\documentclass[12pt]{article}
\pdfoutput=1
\usepackage[nosort]{cite}
\usepackage[height=8.85in,width=6.45in]{geometry}
\usepackage{times}

\usepackage[utf8]{inputenc}
\usepackage{amsmath}
\usepackage{amssymb}
\usepackage{mathtools}
\numberwithin{equation}{section}
\usepackage{slashed}
\usepackage{braket}
\usepackage[svgnames,psnames]{xcolor}
\usepackage[colorlinks,citecolor=DarkGreen,linkcolor=FireBrick,linktocpage,pagebackref]{hyperref}
\usepackage{cite}
\usepackage{graphicx}
\usepackage{tikz}
\usepackage{tikz-cd}

\usepackage{courier}
\usepackage{bm}
\usepackage{dashbox}
\usepackage{subcaption}
\usepackage{enumitem}

\usepackage{mdframed}

\usepackage{blkarray}
\usepackage{arydshln}
\usepackage{dsfont}
\usetikzlibrary{patterns}
\usetikzlibrary{arrows.meta}

\usepackage{calc}

\usepackage{accents}
\newcommand{\dbtilde}[1]{\accentset{\approx}{#1}}

\newcommand{\ie}{\begin{equation}\begin{aligned}}
\newcommand{\fe}{\end{aligned}\end{equation}}

\renewcommand{\title}[1]{\vbox{\center\LARGE{#1}}\vspace{5mm}}
\renewcommand{\author}[1]{\vbox{\center#1}\vspace{5mm}}
\newcommand{\address}[1]{\vbox{\center\em#1}}

\begin{document}
 
\begin{titlepage}
 	\hfill YITP-SB-2022-16, MIT/CTP-5423\\

\title{Non-invertible Condensation, Duality, and Triality Defects in 3+1 Dimensions}

\author{Yichul Choi${}^{1,2}$, Clay C\'ordova${}^3$, Po-Shen Hsin${}^4$, Ho Tat Lam${}^5$, Shu-Heng Shao${}^1$}

		\address{${}^{1}$C.\ N.\ Yang Institute for Theoretical Physics, Stony Brook University\\
        ${}^{2}$Simons Center for Geometry and Physics, Stony Brook University\\
		${}^{3}$Enrico Fermi Institute $\&$ Kadanoff Center for Theoretical Physics,
University of Chicago
		\\
		${}^{4}$Mani L. Bhaumik Institute for Theoretical Physics,   UCLA		\\
		${}^{5}$Center for Theoretical Physics, Massachusetts Institute of Technology
		}

\abstract

We discuss a variety of codimension-one, non-invertible topological defects in general 3+1d QFTs with a discrete one-form global symmetry. 
These include condensation defects from higher gauging of the one-form symmetries on a codimension-one manifold, each labeled by a discrete torsion class, and duality and triality defects from gauging in half of spacetime.  
The universal fusion rules between these non-invertible topological defects and the one-form symmetry surface defects are determined. 
Interestingly, the fusion coefficients are generally not numbers, but 2+1d TQFTs, such as invertible SPT phases,   $\mathbb{Z}_N$ gauge theories, and  $U(1)_N$ Chern-Simons theories. 
The associativity of these algebras over TQFT coefficients relies on nontrivial facts about 2+1d TQFTs. 
We further prove that some of these non-invertible symmetries are intrinsically incompatible with a trivially gapped phase, leading to  nontrivial constraints on renormalization group flows.  
Duality and triality defects are realized in many familiar gauge theories, including free Maxwell theory, non-abelian gauge theories with orthogonal gauge groups, ${\cal N}=1,$ and  ${\cal N}=4$ super Yang-Mills theories.

\end{titlepage}

\eject

\tableofcontents

\section{Introduction} \label{Sec:intro}

The modern characterization of a global symmetry is in terms of its conserved charge or symmetry operator, rather than any specific realization by Lagrangians and  fields \cite{Gaiotto:2014kfa}.  For an ordinary global symmetry $G$, the associated symmetry operators $U_g$ with $g\in G$ are supported on all of space and act on the Hilbert space at a fixed time. Since it is a symmetry, this operator is conserved under time evolution.  In relativistic quantum field theory (QFT), space and time are on equal footing, and hence a symmetry operator/defect $U_g(M)$ may be placed on a general codimension-one manifold $M$ in spacetime.\footnote{Throughout out this paper, we will focus on relativistic QFT in Euclidean signature, and we will use the terms ``operator" and ``defect" interchangeably.}    In this general setting, the conservation under time evolution is upgraded to the statement that the operator $U_g(M)$ depends on $M$ only topologically.  All invariant properties of a global symmetry are completely captured by the codimension-one topological defect $U_g(M)$.

These ideas lead to a natural link between global symmetries and topological defects.  However, not every topological defect is associated with an ordinary global symmetry.  This is made precise by analyzing the fusion of topological defects.  In general, the fusion of two codimenion-one defects, say $U$ and $U'$, is defined by placing them parallel to each other and then allowing them to collide.  This defines a new topological defect denoted as $U \times U'$.  A topological defect $U$ is called 
\textit{invertible} if there exists an inverse topological operator $U^{-1}$ such that:
\begin{equation}
U(M)\times U^{-1}(M)=U^{-1}(M)\times U(M) =1~,
\end{equation}
where the right-hand side denotes the identity defect.  Mirroring the algebraic fact that every element of $g\in G$ has an inverse, it follows that every ordinary global symmetry is associated to an invertible topological operator.  Converseley, a topological operator $U$ is called \emph{non-invertible} if there does not exist an inverse topological operator.  These non-invertible operators and the rich algebraic structure they encode are the focus of this paper.

Building on earlier work of \cite{Verlinde:1988sn,Petkova:2000ip,Fuchs:2002cm,Frohlich:2004ef,Frohlich:2006ch,Frohlich:2009gb}, it has been advocated that non-invertible topological defects should be viewed as generalizations of ordinary global symmetries \cite{Bhardwaj:2017xup,Chang:2018iay}. In particular, many operations and properties familiar from the study of ordinary symmetries have analogs for non-invertible defects as well.  For instance, it is sometimes possible to gauge a collection of  non-invertible topological operators by summing over suitable networks of defects \cite{Frohlich:2009gb,Carqueville:2012dk,Brunner:2013xna,Bhardwaj:2017xup,Komargodski:2020mxz,Gaiotto:2020iye,Huang:2021zvu,Kaidi:2021gbs,Buican:2021axn,Yu:2021zmu,Benini:2022hzx,Roumpedakis:2022aik}.  Furthermore, when there is an obstruction to gauging, the generalized 't Hooft anomaly matching conditions lead to surprising constraints on renormalization group flows \cite{Chang:2018iay,Thorngren:2019iar,Komargodski:2020mxz,Thorngren:2021yso,Choi:2021kmx}. 
In the context of quantum gravity, it has been argued that the no global symmetry conjecture should be extended to the absence of non-invertible topological defects \cite{Rudelius:2020orz,Heidenreich:2021xpr,McNamara:2021cuo, Cordova:2022rer, Arias-Tamargo:2022nlf}. 
For this reason, we will refer to them as \textit{non-invertible symmetries}.

In 1+1d, non-invertible topological defect lines have been studied extensively in the context of conformal field theory \cite{Verlinde:1988sn,Petkova:2000ip,Fuchs:2002cm,Frohlich:2004ef,Frohlich:2006ch,Frohlich:2009gb} (see also \cite{Bachas:2004sy,Fuchs:2007tx,Bachas:2009mc,Ji:2019ugf,Lin:2019hks,Gaiotto:2020fdr,Gaiotto:2020iye,Komargodski:2020mxz,Gaiotto:2020dhf,Lin:2021udi,Thorngren:2021yso,Burbano:2021loy}), and
 related  lattice models \cite{Grimm:1992ni,Feiguin:2006ydp,Hauru:2015abi,Aasen:2016dop,Buican:2017rxc,Aasen:2020jwb,Inamura:2021szw,Koide:2021zxj,Huang:2021nvb,Vanhove:2021zop,Liu:2022qwn}. 
 (See also \cite{Nguyen:2021yld} for a construction in 2+1d.) 
Recently, non-invertible duality defects have also been constructed in various 3+1d continuum and lattice gauge theories \cite{Koide:2021zxj,Choi:2021kmx,Kaidi:2021xfk}, which have led to non-trivial dynamical consequences on the phase diagram of gauge theories \cite{Choi:2021kmx, Kaidi:2021xfk}.  
In fact, in general spacetime dimensions,  it was demonstrated that non-invertible symmetries are almost as common as the higher-form symmetries via higher gauging \cite{Roumpedakis:2022aik}, which we will discuss below.  

In this paper we continue exploring a variety of codimension-one, non-invertible symmetries in 3+1d QFT with a one-form global symmetry, focusing mostly on the case of a $\mathbb{Z}_N^{(1)}$ one-form symmetry. Below we give an overview of these non-invertible symmetries that we will encounter and their fusion rules. As we discuss in examples, the fusion rules are generally non-commutative.  In our examples, the operator products that we derive take the simple form:
\begin{equation} \label{OPEintro}
    \cal{N}(M) \times \cal{N}'(M) = \mathcal{T}(M) \cal{N}''(M) ~,
\end{equation}
where $\cal{N}, \cal{N}', \cal{N}''$ are non-invertible topological defects and $\mathcal{T}$ is the fusion ``coefficient'': a 2+1d Topological Field Theory (TQFT).  
More precisely, the coefficient $\mathcal{T}(M)$ in the parallel fusion rule is the value of the partition function of the theory $\mathcal{T}$ on the closed manifold $M$. 
Similar fusion rules have also recently been emphasized in \cite{Roumpedakis:2022aik} and echo the fusion categories characterizing topological lines in 1+1d as well as the operator product expansions of supersymmetric line defects explored for instance in \cite{Gaiotto:2010be, Cordova:2013bza}.  We anticipate that fusion algebras such as \eqref{OPEintro} form a natural component of the algebraic structure of a higher fusion category \cite{Kong:2014qka,douglas2018fusion,Gaiotto:2019xmp,Kong:2020cie,Johnson-Freyd:2020usu,Johnson-Freyd:2020ivj,Kong:2020wmn,Kong:2020iek,Kong:2021ups,Kong:2022hjj, Bhardwaj:2022yxj}, generalizing the simpler higher algebraic structures defining higher-form and higher group global symmetry \cite{Gaiotto:2014kfa, Cordova:2018cvg, Benini:2018reh}.

\subsection{Condensation Defects}

The most basic and universal non-invertible symmetries in a QFT with a higher-form symmetry are condensation defects.

Given a higher-form global symmetry, it is common to gauge it in the whole spacetime to map one QFT to another. 
It is also common to gauge it in a codimension-zero region in spacetime to produce an interface between two different QFTs, or a boundary condition for the original QFT (see, for example, \cite{Kaidi:2021gbs,Choi:2021kmx}, for some recent discussions).  

Recently, ordinary gauging of a higher-form symmetry was generalized to \textit{higher gauging} \cite{Roumpedakis:2022aik}. 
In general, $p$-gauging  is defined by gauging a discrete $q$-form symmetry in a codimension-$p$ submanifold in spacetime. 
A $q$-form symmetry is $p$-gaugeable if it can be gauged on a codimension-$p$ submanifold, otherwise it is $p$-anomalous. 
The higher gauging does not change the bulk QFT, but inserts a codimension-$p$ topological defect in the same bulk QFT. 
The resulting topological defect is called the \textit{condensation defect}, which is generally non-invertible.\footnote{All the condensation defects we will encounter in this paper are non-invertible. More generally, an anomalous higher-form symmetry can give rise to an invertible condensation defect. For example, a one-gaugeable fermion line in 2+1d leads to a $\mathbb{Z}_2$ condensation surface under one-gauging \cite{Roumpedakis:2022aik}   (see also \cite{Kong:2020cie,Johnson-Freyd:2020twl}). }

The condensation defects have been studied in the condensed matter physics literature \cite{Kong:2014qka,Else:2017yqj,Gaiotto:2019xmp,Kong:2020cie,Johnson-Freyd:2020twl}, and the higher gauging gives a realization of them from the perspective of generalized global symmetries.
The condensation surface defects in 2+1d QFT with a one-form global symmetry are systematically analyzed in \cite{Roumpedakis:2022aik}, which generalizes and unifies various earlier results  in \cite{Fuchs:2002cm,Kapustin:2010if}.

In this paper, we will mostly focus on 3+1d  QFT $\mathcal{Q}$ with a $\mathbb{Z}_N^{(1)}$ one-form symmetry and consider the codimension-one condensation defects from one-gauging.
The resulting condensation defect is defined by summing over all possible insertions of the $\mathbb{Z}_N^{(1)}$ one-form symmetry defects along nontrivial two-cycles of a three-dimensional submanifold $M$. 
Similar to the ordinary gauging, one can  add a discrete torsion in the higher gauging on the three-manifold.
Different choices of the discrete torsion class  $ k \in H^3(B\mathbb{Z}_N;U(1))\cong \mathbb{Z}_N$  lead to distinct types of condensation defects denoted by ${\cal C}_k$.  
Examples of this class of condensation defects played a prominent role in \cite{Choi:2021kmx,Kaidi:2021xfk}, and our analysis generalizes these constructions.

Of special importance are the condensation defects that are orientation reversal invariant. 
These are the condensation defects that will participate in the fusion rules involving the duality and the triality defects discussed below.
For odd $N$ it is ${\cal C}_0$ while for even $N$ they are ${\cal C}_0$ and ${\cal C}_{N\over2}$.  
We determine the universal fusion rules between these codimension-one, non-invertible condensation defects:
\ie\label{intro:condensationfusion}
   \mathcal{C}_{\frac{N\ell_1}{2}} \times \mathcal{C}_{\frac{N\ell_2}{2}} &= (\mathcal{Z}_N)_{N(\ell_1 + \ell_2)} \, \mathcal{C}_{\frac{N\ell_1}{2}} =(\mathcal{Z}_N)_{N(\ell_1 + \ell_2)} \, \mathcal{C}_{\frac{N\ell_2}{2}}\,,
   \fe
   where $\ell=0$ for odd $N$ and $\ell=0,1$ mod 2 for even $N$.  
   Here  $({\cal Z}_N)_K$ stands for the 2+1d $\mathbb{Z}_N$ gauge theory with a level $K$ Dijkgraaf-Witten twist \cite{Dijkgraaf:1989pz} (see \eqref{Eq:ZN_level_K}).  
   For example, the 3+1d $\mathbb{Z}_N$ gauge theory (which is the low energy limit of the $\mathbb{Z}_N$ toric code) has a $\mathbb{Z}_N^{(1)}$ one-form symmetry and therefore realizes these codimension-one condensation defects ${\cal C}_k$. 
   Importantly,  the fusion ``coefficients" of these three-dimensional topological defects are 2+1d TQFTs, rather than numbers.\footnote{For defects of any dimensionality, when the fusion ``coefficient" is  a $\mathbb{Z}_N$-valued topological scalar  theory that gives $N$ trivial vacua on any manifold, it corresponds to a fusion multiplicity $N$. In general, we can decorate a simple object $x(\Sigma)$ that supports on submanifold $\Sigma$  by a lower dimensional TQFT supported on $\Sigma$. If such lower-dimensional TQFT has a non-trivial topological domain wall, then $\text{Hom}(x,x)\not\cong\mathbb{C}$ and the decoration produces a non-simple object.
} 
This generalizes the fusion rule of the condensation surface defects in 2+1d spacetime \cite{Roumpedakis:2022aik}.

\subsection{Duality Defects}

Given any QFT with a $\mathbb{Z}_N^{(1)}$ one-form symmetry, one can create a topological interface between the QFT  $\cal Q$ and   $S{\cal Q}$, where $S$ denotes the gauging of the one-form symmetry \eqref{eq:S_and_T_operations}. 
This is achieved by gauging the $\mathbb{Z}_N^{(1)}$ symmetry only in half of the spacetime manifold and then imposing the topological  Dirichlet boundary condition for the $\mathbb{Z}_N^{(1)}$ two-form gauge field along the interface.

If the QFT $\cal Q$ is invariant under $S$ (in the sense of \eqref{Eq:invariance_under_S}), then the topological interface becomes a topological defect inside the same theory $\mathcal{Q}$ which we call the duality defect $\mathcal{D}_2$ \cite{Choi:2021kmx,Kaidi:2021xfk}.  
In this paper we determine the full set of parallel fusion rules between the duality defect ${\cal D}_2$, the condensation defect ${\cal C}_0$, the charge conjugation defect ${\cal U}$, and the one-form symmetry surface defect $\eta$ (see \eqref{Eq:fusion_duality_summary} and \eqref{Eq:fusion_eta_duality}):
\begin{align} 
\begin{split}
&    \overline{\mathcal{D}}_2 \times \mathcal{D}_2 =\mathcal{D}_2 \times \overline{\mathcal{D}}_2 = \mathcal{C}_0  \,,\\
&    \mathcal{D}_2 \times \mathcal{D}_2 = \mathcal{U} \times \mathcal{C}_0 = \mathcal{C}_0 \times \mathcal{U} \,, \\
&\overline{\mathcal{D}}_2 = \mathcal{U} \times \mathcal{D}_2 = \mathcal{D}_2 \times \mathcal{U}\,,\\
&    \mathcal{D}_2 \times \mathcal{C}_0 = \mathcal{C}_0 \times \mathcal{D}_2 = (\mathcal{Z}_N)_0 \,\mathcal{D}_2  \,, \\
&    \mathcal{C}_0 \times \mathcal{C}_0 = (\mathcal{Z}_N)_0 \, \mathcal{C}_0 \,,\\
&      \mathcal{D}_2 \times \eta =
    \eta\times \mathcal{D}_2= \mathcal{D}_2\,, \\
&    \overline{\mathcal{D}}_2  \times \eta =
    \eta \times \overline{\mathcal{D}}_2  = \overline{\mathcal{D}}_2  \,, \\
&    \mathcal{C}_0  \times \eta =
    \eta \times \mathcal{C}_0  \,,\\
&    \eta^N=1\,.
\end{split}
\end{align}
Here $\overline{\cal D}_2$ denotes the orientation-reversal of ${\cal D}_2$ (see \eqref{Eq:general_orientation_reversal}).  
The other fusion rules involving $\overline{\cal D}_2$ can be obtained from \eqref{Eq:orientation_reversal_fusion}.  
Some of the above fusion rules have  been reported in \cite{Choi:2021kmx,Kaidi:2021xfk}.  

Intuitively, the duality defect ${\cal D}_2$ (and similarly for the triality defect below) can be thought of as the ``square root" of the condensation defect in the sense that ${\cal D}_2\times \overline{\cal D}_2= {\cal C}_0$. Thus, these defects may be viewed as 3+1d analog of the Tambara-Yamagami fusion category of 1+1d QFTs \cite{TY}.  (See \cite{Thorngren:2019iar, Thorngren:2021yso} for a physical perspective on 1+1d defects analogous to that presented here.)

\subsection{Triality Defects}

Analogously, one can always form a topological interface between the two QFTs $\mathcal{Q}$ and $ST\mathcal{Q}$, where $T$ stands for the operation of stacking a 3+1d invertible field theory \eqref{eq:S_and_T_operations}. 
If the QFT is invariant under the $ST$ twisted gauging (or more precisely, \eqref{Eq:invariance_under_ST}), then the interface becomes a topological defect in the same theory $\mathcal{Q}$, and we call this defect the \emph{triality defect} $\mathcal{D}_3$.
The name ``triality'' defect is inspired by the fact that the $ST$ operation is of order 3 (modulo an invertible theory).\footnote{On the other hand, to be precise, the duality defect ${\cal D}_2$ should really be called a ``quadrality defect" since $S^2=C$ and $S^4=1$, where $C$ is the charge conjugation.} 

The fusion rules of the condensation defects ${\cal C}_{N\ell\over2}$, the triality defect ${\cal D}_3$, and the one-form symmetry surface defect are summarized below. 
When $N$ is even, the fusion rule is non-invertible and non-commutative (see \eqref{Eq:fusion_triality_summary} and \eqref{Eq:fusion_eta_triality}):
\ie
&  \overline{\mathcal{D}}_3 \times \mathcal{D}_3 = \mathcal{C}_0 \,,\\
&    \mathcal{D}_3 \times \overline{\mathcal{D}}_3 = \mathcal{C}_{\frac{N}{2}} \,,\\
&    \mathcal{D}_3 \times \mathcal{D}_3 = U(1)_{N} \,\overline{\mathcal{D}}_3 \,,\\
&    \mathcal{D}_3 \times \mathcal{C}_0 = \mathcal{C}_{\frac{N}{2}} \times \mathcal{D}_3 = (\mathcal{Z}_N)_0 \,\mathcal{D}_3 \,,\\
&    \mathcal{D}_3 \times \mathcal{C}_{\frac{N}{2}} = \mathcal{C}_0 \times \mathcal{D}_3 
    = (\mathcal{Z}_N)_N \,\mathcal{D}_3  \,, \\
&    \mathcal{C}_{\frac{N\ell_1}{2}} \times \mathcal{C}_{\frac{N\ell_2}{2}} = (\mathcal{Z}_N)_{N(\ell_1 + \ell_2)} \, \mathcal{C}_{\frac{N\ell_1}{2}} =(\mathcal{Z}_N)_{N(\ell_1 + \ell_2)} \, \mathcal{C}_{\frac{N\ell_2}{2}} \,,\\
&     \mathcal{D}_3 \times \eta = \mathcal{D}_3\,, ~~~~~~~~~~~~~~~\eta \times \mathcal{D}_3 = (-1)^{Q}\, \mathcal{D}_3 \,, \\
&    \overline{\mathcal{D}}_3  \times \eta =
    (-1)^{Q} \,\overline{\mathcal{D}}_3  \,, ~~~~    \eta \times \overline{\mathcal{D}}_3  =    \overline{\mathcal{D}}_3  \,,\\
&    \mathcal{C}_\frac{N\ell}{2}  \times \eta =  \eta \times \mathcal{C}_\frac{N\ell}{2}
    = (-1)^{\ell Q} \,\mathcal{C}_\frac{N\ell}{2} \,, \\
&\eta^N=1\,.
    \fe
The sign $(-1)^Q$ requires some explanation. 
The quantity $Q$ is defined as ${1\over N}\int_M \text{PD}(\Sigma)\cup \delta \text{PD}(\Sigma)$, where $\Sigma$ is the two-manifold where $\eta$ is supported on and $M$ is a three-manifold where the codimension-one defect is supported on. Here PD$(\Sigma)$ stands for the Poincar\'e dual of $\Sigma$ in $M$.  
The sign $(-1)^{Q(M,\Sigma)}$ can be viewed as a 2+1d $\mathbb{Z}_N^{(0)}$ Symmetry Protected Topological (SPT) phase of the background one-form gauge field PD$(\Sigma)$ living in $M$.  
The associativity of this algebra relies on several nontrivial facts about 2+1d TQFTs and this SPT, as we will discuss in Section \ref{Sec:fusion_triality_codim_1}.

When $N$ is odd, the fusion rule is non-invertible but commutative (see \eqref{Eq:fusion_triality_summary_odd_N} and \eqref{Eq:eta_fusion_triality_odd_N}):
\ie
&    \overline{\mathcal{D}}_3 \times \mathcal{D}_3 =
    \mathcal{D}_3 \times \overline{\mathcal{D}}_3 = \mathcal{C}_{0} \,,\\
&    \mathcal{D}_3 \times \mathcal{D}_3 = SU(N)_{-1} \,\overline{\mathcal{D}}_3 \,,\\
&    \mathcal{D}_3 \times \mathcal{C}_0 = \mathcal{C}_{0} \times \mathcal{D}_3 = (\mathcal{Z}_N)_0 \,\mathcal{D}_3 \,,\\
&    \mathcal{C}_0 \times \mathcal{C}_0 = (\mathcal{Z}_N)_0 \,\mathcal{C}_0\,,\\
&\mathcal{D}_3  \times \eta =
    \eta \times \mathcal{D}_3 = \mathcal{D}_3 \,, \\
&    \overline{\mathcal{D}}_3  \times \eta =
    \eta \times \overline{\mathcal{D}}_3  = \overline{\mathcal{D}}_3  \,, \\
    &\mathcal{C}_0 \times \eta = \eta \times \mathcal{C}_0 = \mathcal{C}_0\,,\\
&    \eta^N=1\,.
\fe
The other fusion rules involving $\overline{\cal D}_3$ can be obtained from \eqref{Eq:orientation_reversal_fusion}. 
The triality fusion rules above can be viewed as the generalization of those in 1+1d \cite{Thorngren:2019iar,Thorngren:2021yso}. Examples of similar triality defect fusion rules have recently been explored in $U(1)$ lattice gauge theory \cite{Hayashi:2022fkw}.

Interestingly, the fusion ``coefficients" between these topological defects can either be a 2+1d TQFT such as the twisted $\mathbb{Z}_N$ gauge theory $(\mathcal{Z}_N)_K$ or the $SU(N)_{-1}$ Chern-Simons theory, or a 2+1d SPT such as $(-1)^{Q(M,\Sigma)}$.\footnote{In some of the examples we have analyzed in this paper, the 2+1d TQFT coefficients  arise as the boundary theories of 3+1d invertible TQFTs. In these cases we have implicitly chosen a preferred framing for the 2+1d worldvolume of the defect coming from the embedding into the 3+1d bulk. } 
In general we expect, the fusion ``coefficients" to be a TQFT decorated by the insertion of additional topological defects (see \cite{Roumpedakis:2022aik} for a 2+1d discussion). Such a structure is also known in condensed matter as an Symmetry Enriched Topological (SET) phase.  As emphasized around \eqref{OPEintro}, these fusion coefficients presumably are a core ingredient in a fusion $n$-category characterizing the topological operators.

\subsection{Dynamical Consequences in Gauge Theories}

Some of the  duality and triality defects are not compatible with a trivially gapped phase.  
In Section \ref{sec:dynamical}, we classify non-invertible topological defects from gauging, which include the duality and the triality defects as special cases, that are incompatible with a trivially gapped phase.  
The presence of these non-invertible symmetries in the UV imply that the IR phase cannot be trivial, which is reminiscent of the presence of 't Hooft anomalies.  
In the case of gauge theories, these constraints provide an analytic obstruction to a trivially confining gapped phase. These results are analogous to those shown for duality defects in 3+1d in \cite{Choi:2021kmx}.

Specifically, for theories with $\mathbb{Z}_N^{(1)}$ one-form symmetry, we show that

\paragraph{Theorem} {\it Let $\cal Q$ be a 3+1d QFT that is invariant under the $ST$ gauging of the $\mathbb{Z}_N^{(1)}$ one-form symmetry in the sense of \eqref{Eq:invariance_under_ST}, {\it i.e.} ${\cal Q}$ has a triality defect ${\cal D}_3$. Then $\cal Q$ cannot flow to a trivially gapped phase with a unique ground state if one of the following is true:
\begin{itemize}
\item $3|N$, and there exists a prime factor of $N/3$ that is not one modulo $3$.
\item $3\nmid N$, and there exists a prime factor of $N$ that is not one modulo $3$.
\end{itemize}
}
For example, the triality defects of all even $N$, and of $N=5,9,11,\cdots$, are not compatible with a trivially gapped phase. 

Similarly, for theories with $\mathbb{Z}_N^{(1)}\times \mathbb{Z}_N^{(1)}$ one-form symmetry with even $N$, we show that
\paragraph{Theorem}
{\it If a 3+1d QFT is invariant under gauging $\mathbb{Z}_N^{(1)}\times\mathbb{Z}_N^{(1)}$ one-form symmetry for even $N$ with the minimal mixed counterterm that couples the gauge fields of the two $\mathbb{Z}_N^{(1)}$ one-form symmetry,\footnote{
Denote the two-form $\mathbb{Z}_N$ gauge fields for the $\mathbb{Z}_N^{(1)}$ one-form symmetries by $B,B'$, then this minimal mixed local counterterm is $\frac{2\pi i}{N}\int B\cup B'$. 
} {\it i.e.} the theory has a triality defect, then the theory cannot flow to a gapped phase with a unique ground state.}

In Section \ref{sec:example}, we will give several familiar gauge theory examples realizing triality defects. These include free Maxwell gauge theory at $\tau = e^{2\pi i /3}N$,  $SO(8)$ Yang-Mills theory, ${\cal N}=1$ and ${\cal N}=4$ super Yang-Mills theories.   Additional investigation of some of these examples will appear in \cite{Kaidi:2022uux}.

\section{Definition of Non-invertible Defects} \label{Sec:define_defects}

We consider a general  3+1-dimensional quantum field theory $\cal Q$ with a non-anomalous $\mathbb{Z}^{(1)}_N$ one-form symmetry.\footnote{The $\mathbb{Z}^{(1)}_N$ one-form symmetry in a 3+1d theory is potentially anomalous as can be seen, for instance, from the nontrivial bordism group $\Omega^{\text{SO}}_5(B^2 \mathbb{Z}_2) \cong \mathbb{Z}_2 \times \mathbb{Z}_2$ \cite{Wan:2018bns} where one of the $\mathbb{Z}_2$ factors corresponds to a possible anomaly for the $\mathbb{Z}^{(1)}_2$ one-form symmetry (the other $\mathbb{Z}_2$ factor corresponds to the $w_2 w_3$ gravitational anomaly).
This particular anomaly vanishes for spin theories since $\Omega^{\text{Spin}}_5(B^2 \mathbb{Z}_2)$ is trivial.} 
Let $Z_{\mathcal{Q}}[B]$ be the partition function on a closed spacetime manifold $X$ in the presence of a two-form background gauge field $B \in H^2(X;\mathbb{Z}_N)$ for the $\mathbb{Z}_N^{(1)}$ symmetry.\footnote{Throughout the paper (with an exception in Section \ref{sec:Maxwell}), we use the   upper case letters to denote classical background fields, and the lower case letters to denote dynamical fields.\label{fn:case}}
Following \cite{Gaiotto:2014kfa, Bhardwaj:2020ymp} (which generalize the earlier work of  \cite{Witten:2003ya}), we define $S$ and $T$ operations which  map the QFT $\mathcal{Q}$ to   $S\mathcal{Q}$ and $T\mathcal{Q}$, respectively: 
\begin{align} \label{eq:S_and_T_operations}
\begin{split}
     & S:\quad Z_{S\mathcal{Q}}[B]= \frac{1}{|H^2(X;\mathbb{Z}_N)|^{1/2}}\sum_{b\in H^2(X;\mathbb{Z}_N)} Z_{\mathcal{Q}}[b] \text{exp}\left(\frac{2\pi i}{N} \int_X b \cup B \right) \,, \\
& T:\quad Z_{T\mathcal{Q}}[B]=  
\begin{dcases}
Z_{\mathcal{Q}}[B] \text{exp}\left(\frac{2\pi i}{N} \int_X \frac{1}{2} q(B)\right) & \text{if $N$ is even} \,, \\
Z_{\mathcal{Q}}[B] \text{exp}\left(\frac{2\pi i}{N} \int_X \frac{N+1}{2} B \cup B \right) & \text{if $N$ is odd} \,.
\end{dcases}
\end{split}
\end{align}
That is, the $S$ operation corresponds to gauging the $\mathbb{Z}_N^{(1)}$ symmetry without any twist, the $T$ operation corresponds to stacking an invertible field theory, which is  $e^{\frac{2\pi i}{2N} \int_X q(B)}$ or $e^{\frac{2\pi i}{N} \int_X \frac{N+1}{2} B^2}$ depending on whether $N$ is even or odd, and the $ST$ operation corresponds to a particular twisted gauging of the $\mathbb{Z}_N^{(1)}$ symmetry, etc.
For the even $N$ case, $q: H^2(X;\mathbb{Z}_N) \rightarrow H^4(X;\mathbb{Z}_{2N})$ denotes the Pontryagin square operation.
In terms of an integral lift of a particular cocycle representation, it is given by $q(B) = B \cup B - B \cup_1 \delta B$.

When $N$ is even, the two operations define a projective $SL(2,\mathbb{Z}_{2N})$ action on the space of QFTs with a $\mathbb{Z}_N^{(1)}$ one-form symmetry: 
\ie
\text{even} ~N:~~S^2 = C\,,\quad T^{2N} = 1\,,\quad (ST)^3 = Y\,,
\fe
 where $C$ is the charge conjugation, i.e., $Z_{C\mathcal{Q}}[B] = Z_{\mathcal{Q}}[-B]$, and $Y$ is an invertible field theory \cite{Bhardwaj:2020ymp,Hsin:2021qiy}.
For the case of odd $N$, we have 
\ie
\text{odd} ~N:~~S^2 = C\,,\quad  T^{N} = 1\,,\quad (ST)^3 = Y\,.
\fe
The partition function of the invertible theory $Y$ on a closed manifold $X$ is given by\footnote{In both cases the partition function is equal to $e^{2\pi i \sigma(X) /8}$, modulo Euler counterterms, where $\sigma(X)$ is the signature of the manifold $X$ if we assume that the integral cohomology of the spacetime manifold is torsionless (see, for example, \cite{Morita:1971,Gaiotto:2014kfa,Bhardwaj:2020ymp}).} 
\begin{equation} \label{Eq:invertible_Y}
    Z_Y = 
\begin{dcases}
    \frac{1}{|H^2(X;\mathbb{Z}_N)|^{1/2}}\sum_{b\in H^2(X;\mathbb{Z}_N)} \text{exp}\left(\frac{2\pi i}{N} \int_X \frac{1}{2} q(b)\right) & \text{if $N$ is even} \,, \\
    \frac{1}{|H^2(X;\mathbb{Z}_N)|^{1/2}}\sum_{b\in H^2(X;\mathbb{Z}_N)} \text{exp}\left(\frac{2\pi i}{N} \int_X \frac{N+1}{2} b \cup b \right) & \text{if $N$ is odd} \,.
\end{dcases}
\end{equation}

In the following subsections, we will define three classes of non-invertible codimension-one topological defects:
\begin{description}[style=unboxed,leftmargin=0cm,font=\it\textbullet\space]
\item [Condensation Defects]
 For any  QFT $\mathcal{Q}$ with a $\mathbb{Z}_N^{(1)}$ one-form symmetry, one can gauge the symmetry on  a codimension-one submanifold.  This is known as the higher condensation \cite{Kong:2014qka,Else:2017yqj,Gaiotto:2019xmp,Kong:2020cie,Johnson-Freyd:2020twl} or higher gauging \cite{Roumpedakis:2022aik}. 
The resulting condensation defect is defined by summing over all possible insertions of the $\mathbb{Z}_N^{(1)}$ one-form symmetry defects along nontrivial two-cycles of a three-dimensional submanifold,   twisted by a discrete torsion class in $H^3(B\mathbb{Z}_N;U(1))$.  
\item [Duality Defects]
In general, one can form a topological interface between the theories $\mathcal{Q}$ and $S\mathcal{Q}$, by gauging the $\mathbb{Z}_N^{(1)}$ symmetry only in half of the spacetime manifold and then imposing the Dirichlet boundary condition for the $\mathbb{Z}_N^{(1)}$ gauge field on the interface.
If the theory $\mathcal{Q}$ is invariant under the $S$ operation, that is, $S\mathcal{Q} \cong \mathcal{Q}$ (or more precisely, \eqref{Eq:invariance_under_S}), then the topological interface becomes a topological defect inside the theory $\mathcal{Q}$ which we call the duality defect $\mathcal{D}_2$ \cite{Choi:2021kmx,Kaidi:2021xfk}.  
\item [Triality Defects]
Analogously, one can always form a topological interface between $\mathcal{Q}$ and $ST\mathcal{Q}$ where the dynamical $\mathbb{Z}_N^{(1)}$ gauge field  only lives in half of the spacetime  and  obeys the Dirichlet boundary condition along the interface.
If $ST\mathcal{Q} \cong \mathcal{Q}$, (or more precisely, \eqref{Eq:invariance_under_ST}) then the interface becomes a topological defect in the theory $\mathcal{Q}$, and we call this defect the \emph{triality defect} $\mathcal{D}_3$.

\end{description}

The subscripts of ${\cal D}_2$ and ${\cal D}_3$ are to remind us that these are duality and triality defects, respectively, rather than their dimensionalities. 
We will  denote the codimension-one submanifold on which these defects are supported as $M$.
For simplicity, we assume that both the spacetime manifold $X$ and $M \subset X$ are oriented.
It then follows that $M$ has a neighborhood inside $X$ which is topologically $M \times I$ where $I$ is a small interval.
This allows us to easily visualize the paralell fusion of defects which will be discussed in Sections \ref{Sec:fusion_codim_one} and \ref{Sec:fusion_involving_eta}.

\subsection{Condensation Defects} \label{Sec:define_cond_defects}

In \cite{Roumpedakis:2022aik}, condensation defects are defined by the higher gauging of a higher-form symmetry, i.e., gauging the higher-form symmetry  only along a higher codimensional submanifold in spacetime.
Here we consider condensation defects given by the one-gauging of the $\mathbb{Z}_N^{(1)}$ one-form symmetry along a codimension-one manifold $M$ in the 3+1d spacetime $X$. 

Just like the ordinary gauging, the one-gauging can be twisted by a choice of the discrete torsion class. 
A $\mathbb{Z}_N^{(1)}$ symmetry surface  defect wrapping around a nontrivial two-cycle in $M$ can be represented by a $\mathbb{Z}_N$-valued one-form gauge field through the Poincar\'e duality in $M$.
Possible discrete torsions for a $\mathbb{Z}_N$ one-form gauge field living in $M$ are  classified by $H^3(B\mathbb{Z}_N;U(1))$.
Therefore, we obtain a family  of  condensation defects labeled by this group cohomology.

Let $\eta(\Sigma)$ be the $\mathbb{Z}^{(1)}_N$ one-form symmetry surface defect wrapping around a two-cycle $\Sigma$.
Then these condensation defects on a three-manifold $M$ are defined as:
\begin{equation} \label{Eq:condensation_defect_general}
    \mathcal{C}_k (M) = \frac{1}{|H^0(M;\mathbb{Z}_N)|} \sum_{\Sigma \in H_2(M;\mathbb{Z}_N)} e^{\frac{2\pi i k}{N} Q(M,\Sigma)} \eta(\Sigma) \,.
\end{equation}
As usual, the overall normalization of the condensation defect is subject to the Euler counterterm ambiguity. 
Here we have made a particular choice for later convenience. 
Here $k = 0, \cdots, N-1$ is an integer modulo $N$ which labels the discrete torsion class $H^3(B\mathbb{Z}_N;U(1)) \cong \mathbb{Z}_N$, which we now discuss in more details.

The discrete torsion phase $Q(M,\Sigma)$ is defined as:\footnote{The number $Q(M,\Sigma)$ is non-trivial only if $\Sigma$ is given by a discrete $\mathbb{Z}_N$ cycle, i.e., $\partial \Sigma=N\gamma$ for some one-cycle $\gamma$, where $\beta(a)=\text{PD}(\gamma)$ mod $N$. Then $Q(M,\Sigma)$ equals the intersection number of $\Sigma$ and $\gamma$, or $N$ times the linking form evaluated on the torsion cycle $\gamma$. When $N=2$, using $\beta(a)=a^2$ one finds the number equals to the triple intersection number of $\Sigma$.
}
\begin{equation}
    Q(M,\Sigma) = \int_{M} a \cup \beta(a)\,,
\end{equation}
where   $a = \text{PD}(\Sigma) \in H^1(M;\mathbb{Z}_N)$   is the Poincar\'e dual of $\Sigma$ in $M$. 
Here $\beta : H^1(M;\mathbb{Z}_N) \rightarrow H^2(M;\mathbb{Z}_N)$ is the Bockstein homomorphism associated to the short exact sequence $1 \rightarrow \mathbb{Z}_N \rightarrow \mathbb{Z}_{N^2} \rightarrow \mathbb{Z}_N \rightarrow 1$.
Using an integral lift of a particular cocycle representation of $a$, we may also write it as 
\ie
Q(M,\Sigma) = \frac{1}{N} \int_{M} a \cup \delta a\,.
\fe
This is the familiar Dijkgraaf-Witten term for a $\mathbb{Z}_N$ gauge field $a$  in 2+1d \cite{Dijkgraaf:1989pz}.

When $N$ is even, if we perform the $T$ operation given in Eq. (\ref{eq:S_and_T_operations}), then the condensation defect $\mathcal{C}_k$ in the theory $\mathcal{Q}$ gets mapped to the condensation defect $\mathcal{C}_{k+\frac{N}{2}}$ in the theory $T\mathcal{Q}$.

Two condensation defects $\mathcal{C}_k$ and $\mathcal{C}_{-k}$ are related by the orientation reversal.
For a general defect $\mathcal{N}$ supported on a manifold $M$, its orientation reversal $\overline{\mathcal{N}}$ is defined by
\begin{equation} \label{Eq:general_orientation_reversal}
    \mathcal{N}(M) = \overline{\mathcal{N}}(\overline{M})
\end{equation}
where $\overline{M}$ denotes the orientation reversal of the manifold $M$.\footnote{We thank Sahand Seifnashri for the discussions related to the orientation reversal of defects. See \cite{Roumpedakis:2022aik} for more discussions.}
For our condensation defects, we see that under the orientation reversal of the support $M$, we get $Q(\overline{M},\Sigma) = -Q(M,\Sigma)$, and thus
\begin{equation} \label{Eq:cond_orientation_reversal}
    \overline{\mathcal{C}}_k(M) = \mathcal{C}_{-k}(M) \,.
\end{equation}
In particular, $\mathcal{C}_0$ is its own  orientation reversal. 
Furthermore, when $N$ is even since the discrete torsion class $k$ is defined modulo $N$, $\mathcal{C}_{\frac{N}{2}}$ is also  its own orientation reversal.  
To summarize:
\ie\label{Eq:orientation_invariant}
\overline{ {\cal C}_{N\ell\over2}}(M) =  {\cal C}_{N\ell\over2}(M)~~~~,~~~~
 \ell =
    \begin{cases}
    0\text{ or }1\,,~~~\text{even $N$}\,,\\
    0\,,~~~~~~~~~~\text{odd $N$}\,.
    \end{cases}
\fe

\begin{figure}[!h]
    \centering
    \includegraphics[width=0.8\textwidth]{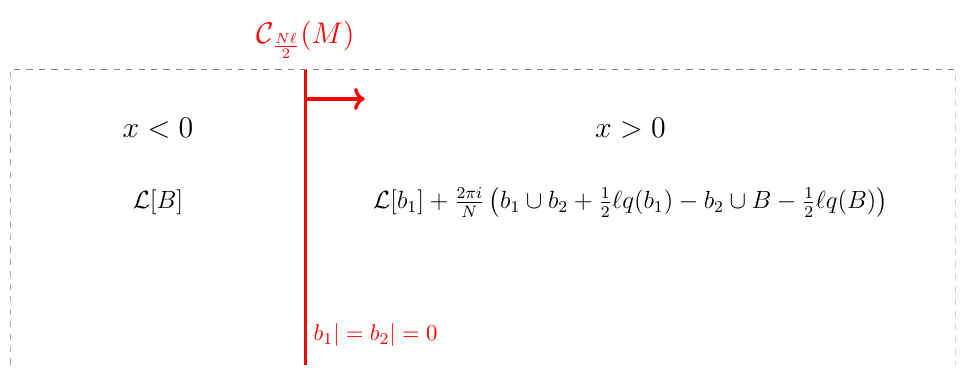}
    \caption{Definition of the condensation defect $\mathcal{C}_{\frac{N \ell}{2}}(M)$.
    The vertical red line corresponds to the codimension-one manifold $M$ on which the condensation defect is supported ($x=0$), and the arrow indicates the orientation of $M$.
    Here $\ell = 0,1$ for even $N$, and $\ell = 0$ for odd $N$.}
    \label{Fig:condensation_defect}
\end{figure}

 \subsubsection*{An Alternative Definition}

For those orientation-reversal-invariant condensation defects, i.e., ${\cal C}_{N\ell\over2}$ with $\ell=0$ for odd $N$ and $\ell=0,1$ mod 2 for even $N$, there is an  alternative definition which we will use more frequently in this paper.\footnote{This is related to the fact that we can take the ``square root'' of the orientation-reversal-invariant condensation defects, but not those that are not invariant under the orientation reversal. We will discuss this more in later sections.}

One way  to understand why ${\cal C}_{N\ell\over2}$  are distinguished among the other condensation defects ${\cal C}_k$ is because the discrete torsion class $(-1)^{\ell Q(M,\Sigma)}\in H^3(B\mathbb{Z}_N;U(1))$ used in the higher gauging comes from the image of an 3+1d $\mathbb{Z}_N^{(1)}$-SPT for a one-form symmetry. 
More precisely, there is a map from the 3+1d $\mathbb{Z}_N^{(1)}$-SPT  to the 2+1d $\mathbb{Z}_N^{(0)}$-SPT \cite{Roumpedakis:2022aik} (with the $N=2$ case studied in \cite{Kaidi:2021xfk})
\ie\label{Eq:map}
H^4(B^2\mathbb{Z}_N;U(1))\to H^3(B\mathbb{Z}_N;U(1)) \,.
\fe
For even $N$, this maps the generator to the order 2 element of the latter, while for odd $N$, it maps everything to the trivial element.  
Precisely when $k={N\ell\over2}$, those discrete torsion classes in $H^3(B\mathbb{Z}_N;U(1))$ are in the image of this map. 
For these values of $k$, the corresponding condensation defect ${\cal C}_k$ can be built by gauging in a thin slab in the 3+1d spacetime, and then collapse the thin slab to a three-manifold $M$.

Below we realize these condensation defects following the picture above. 
Consider the setup  in Figure \ref{Fig:condensation_defect}.
Locally, we denote the coordinate along the direction transverse to the defect as $x$, and let $x=0$ be the three-manifold $M$ on which  the defect is supported.
We choose the orientation of $M$ to be pointing  toward the positive $x$ direction.
Let $B$ be the background $\mathbb{Z}_N$ two-form gauge field  for the $\mathbb{Z}_N^{(1)}$ symmetry.

On one side of the defect  where $x<0$, we have our 3+1d QFT $\cal Q$ described by the Lagrangian $\mathcal{L}[B]$ coupled to the background gauge field $B$.
On the other side of the defect where $x>0$, we have the following Lagrangian:
\begin{align}
   & x>0 \colon \mathcal{L}[b_1] + \frac{2\pi i}{N} \left(
        b_1 \cup b_2 + \frac{1}{2} \ell q(b_1) - b_2 \cup B
        -\frac{1}{2} \ell q(B)
    \right) \,, \notag\\
    &\ell =
     \begin{cases}
  &  0\text{ or }1\,~~~\text{even $N$}\,,\\
    &0\,,~~~~~~~\,~\text{odd $N$}\,.
    \end{cases}
\end{align} 
Here $b_1$ and $b_2$ are dynamical $\mathbb{Z}_N^{(1)}$  two-form gauge fields that live only on  one side of the defect where $x>0$, and  are subject to the Dirichlet boundary condition 
\ie
b_1| = b_2| = 0\,,
\fe
where $|$ denotes the restriction of the dynamical gauge fields to $x=0$.
When $N$ is even, $\ell = 0,1$ is an integer which is defined modulo 2, i.e.,  $\ell \sim \ell+2$.
(This identification  requires some explanation and we will come back to this issue at the end of this subsection.) 
When $N$ is odd, we simply set $\ell = 0$.

We claim that this configuration in Figure \ref{Fig:condensation_defect} gives us the condensation defect $\mathcal{C}_{\frac{N \ell}{2}}$.
More precisely, setting $B=0$ will give us the stand-alone condensation defect $\mathcal{C}_{\frac{N \ell}{2}}$, and the definition given in Figure \ref{Fig:condensation_defect} for a non-vanishing value of $B$ contains slightly more information than the other definition (\ref{Eq:condensation_defect_general}) which enables us to derive various fusion rules as we will explain in later sections.

To verify the claim, first observe that integrating out $b_2$ inside the bulk of the region $x>0$ sets $b_1 = B$, and this leaves us with the original Lagrangian $\mathcal{L}[B]$.
However, due to the Dirichlet boundary condition, this doesn't completely fix the value of $b_1$ near the defect.
Thus, the configuration in Figure \ref{Fig:condensation_defect} defines a codimension-one defect at $x=0$ in the theory given by the Lagrangian $\mathcal{L}[B]$.

\begin{figure}[!h]
    \centering
    \includegraphics[width=0.3\textwidth]{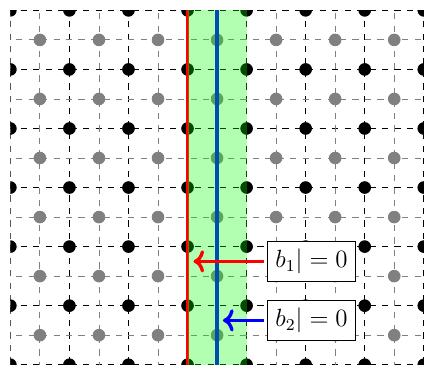}
    \caption{Lattice regularization for the condensation defect.
    The $b_1$ gauge field lives on the right of the red vertical line, and the $b_2$ gauge field lives on the right of the blue vertical line.
    Integrating out $b_2$ sets $b_1=0$ almost everywhere except for the interior of the green shaded region.}
    \label{Fig:condensation_lattice}
\end{figure}
To see that this defect is indeed $\mathcal{C}_{\frac{N \ell}{2}}$, consider the lattice regularization shown in Figure \ref{Fig:condensation_lattice}.\footnote{To be more precise, we use the integer BF lattice presentation of discrete higher-form gauge theories in \cite{Sulejmanpasic:2019ytl,Gorantla:2021svj} (see also \cite{Choi:2021kmx}).}
We set $B=0$, which amounts to not inserting additional symmetry defects.
We have a hypercubic lattice depicted by the black dots and the black dashed lines.
We place the $b_1$ gauge fields  on the 2-cells of this black lattice, and it vanishes on and to the left of the red vertical line.
On the other hand, the $b_2$ gauge fields are placed  on the 2-cells of the dual hypercubic lattice, which are shown in gray color.
$b_2$ vanishes on and to the left of the blue vertical line.
Now, since the $b_2$ gauge field does not couple to the matter fields (which live on the black dots), we can easily integrate it out.
Doing so sets $b_1=0$ almost everywhere, but due to the Dirichlet boundary condition for $b_2$ on the blue vertical line, the $b_1$ gauge fields inside the green shaded region remains to be free.

In the continuum limit, this leaves us with the summation over the two-form gauge fields inside the thin slab of the green shaded region relative to its boundary which are valued in $H^2(M\times I, \partial(M\times I);\mathbb{Z}_N)$, where $I$ is an infinitesimal interval.   
This precisely corresponds to the one-gauging of $\mathbb{Z}_N^{(1)}$ one-form symmetry on a codimension-one manifold as $I$ is taken to zero, which defines the condensation defect.
When $N$ is even, the 3+1d term $\mathbb{Z}_N^{(1)}$-SPT term $\text{exp}\left(\frac{2\pi i}{2N} \ell \int q(b_1) \right)$ inside the slab becomes the discrete torsion phase  \cite{Kaidi:2021xfk}:
\ie
\text{exp}\left(\frac{2\pi i}{2N} \ell \int_{M\times I} q(b_1) \right) = (-1)^{\ell Q(M,\Sigma)}\,,
\fe
which gives the map \eqref{Eq:map}. 
This confirms our claim that the configuration in Figure \ref{Fig:condensation_defect} defines the condensation defect $\mathcal{C}_{\frac{N \ell}{2}}$.

We now come back to the identification $\ell\sim \ell+2$ in Figure \ref{Fig:condensation_defect} for the even $N$ case.
This is obvious from the first definition for the condensation defect given in \eqref{Eq:condensation_defect_general}, which follows directly from the   fact that $k$ is defined modulo $N$.
We will verify the same fact from the alternative definition given in Figure \ref{Fig:condensation_defect} as follows.
To begin with, we first perform a field redefinition $b'_2 = b_2 + b_1 + B$ in the $x>0$ region.
Then, since $b_1 \cup b_2 + \frac{\ell+2}{2} q(b_1) - b_2 \cup B -\frac{\ell+2}{2} q(B) = b_1 \cup b'_2 + \frac{\ell}{2} q(b_1) - b'_2 \cup B -\frac{\ell}{2} q(B)$, the identification $\ell \sim \ell+2$ naively seems to follow.
In general, however, such a field redefinition does not preserve the boundary condition, i.e., $b'_2| =B$ at $x=0$.
Therefore, the two defects defined by $\ell$ and $\ell+2$ are not exactly identical in the presence of nonzero $B$ on the defect (and since the definition (\ref{Eq:condensation_defect_general}) corresponds to setting $B=0$ everywhere in Figure \ref{Fig:condensation_defect}, indeed such a subtlety does not arise in that presentation of the condensation defect).

However, the difference between the two is very mild.
Recall that any nonzero value of $B$ corresponds to inserting some symmetry defects through the Poincar\'e duality.
There are two possible situations where $B$ is nonzero on the defect.
The first case is when there is a symmetry defect which is completely embedded inside $M$. 
This corresponds to the parallel fusion between the symmetry defect and the condensation defect which we will study in Section \ref{Sec:fusion_involving_eta}, and the two defects defined by $\ell$ and $\ell+2$ can't be distinguished by such a parallel fusion as will become clear later.
The second case is when there is a symmetry defect which transversely intersects with the condensation defect along a one-dimensional junction in $M$, and in this case $\ell$ and $\ell +2$ can indeed be distinguished.
Put differently, as long as there is no symmetry defect which transversely intersects with the condensation defect, we can identify $\ell \sim \ell+2$.\footnote{Generally, any two defects which are different only in the presence of such transverse junctions are considered as equivalent defects, i.e., they are isomorphic but not identical.
This is expected to be the higher categorical analogue of the familiar fact in the 1+1d fusion category symmetry case where one has the freedom to choose basis vectors for junction vector spaces.}
In this work, we will focus only on the parallel fusions between defects, and therefore we identify $\ell$ with $\ell + 2$ in Figure \ref{Fig:condensation_defect}.
We will also encounter similar situations in the derivation of the fusion rules later, and we always identify defects up to factors associated to transverse junctions.

\subsection{Duality Defects and Charge Conjugation} \label{Sec:define_duality_defects}

\begin{figure}[!h]
    \centering
    \includegraphics[width=0.8\textwidth]{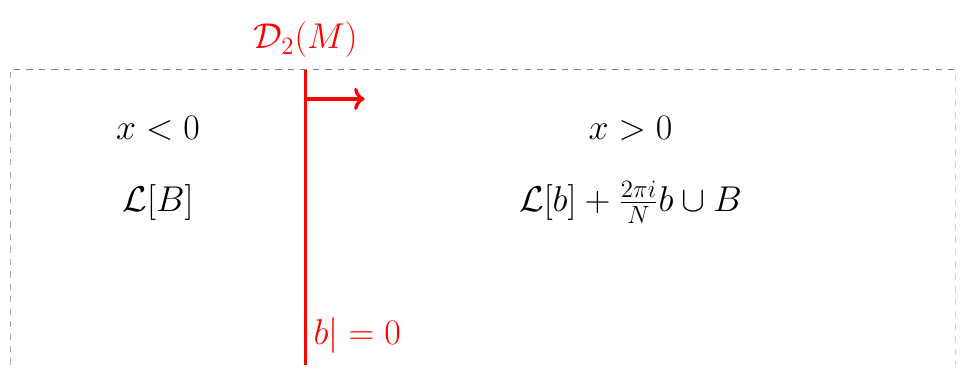}
    \caption{The duality defect $\mathcal{D}_2(M)$ defined by gauging the $\mathbb{Z}_N^{(1)}$ symmetry only for $x>0$.
    Again, the red arrow indicates the orientation of $M$.}
    \label{Fig:duality_defect}
\end{figure}

Now, consider a QFT $\mathcal{Q}$ which is invariant under the $S$ gauging operation. 
More precisely, by $S{\cal Q} \cong {\cal Q}$ we mean the following:
\begin{equation} \label{Eq:invariance_under_S}
    Z_{\mathcal{Q}}[B] = \frac{1}{|H^2(X;\mathbb{Z}_N)|^{1/2}}
    \sum_{b \in H^2(X;\mathbb{Z}_N)} Z_{\mathcal{Q}}[b] e^{\frac{2\pi i}{N} \int_X b \cup B} \,.
\end{equation}
We require the equality to hold modulo counterterms that are independent of the $\mathbb{Z}_N^{(1)}$ symmetry background $B$. 
For such a QFT $\mathcal{Q}$, one can form the duality defect $\mathcal{D}_2$ by gauging only in half of spacetime while imposing the Dirichlet boundary condition for the gauge field $b$ on the defect \cite{Choi:2021kmx,Kaidi:2021xfk}.
This is shown in Figure \ref{Fig:duality_defect}.

In \eqref{Eq:invariance_under_S}, we have fixed a specific meaning of the self-duality $S\mathcal{Q} \cong \mathcal{Q}$ by choosing an isomorphism between the $\mathbb{Z}_N^{(1)}$ one-form symmetries before and after the gauging.  
More generally, one can consider other isomorphisms by replacing $b \cup B$ with other bicharacters. 
This generalization would lead to other types of duality defects. 
We leave the study of these other duality defects for the future. 
A similar comment applies to the triality defect as well.

\begin{figure}[!h]
    \centering
    \includegraphics[width=0.8\textwidth]{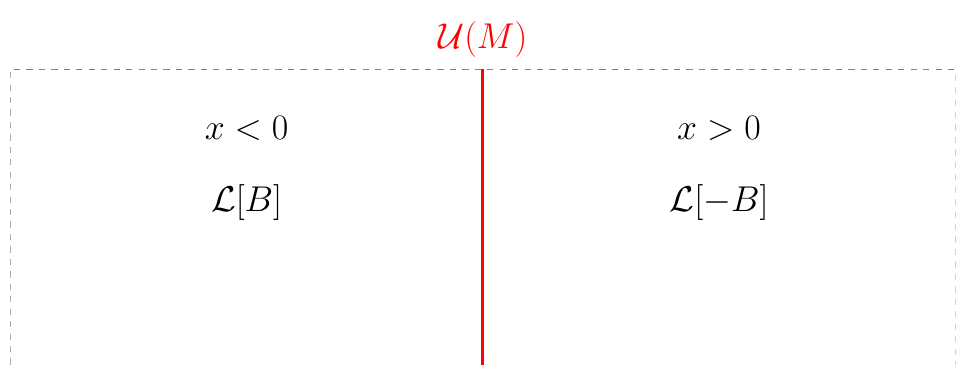}
    \caption{The charge conjugation defect $\mathcal{U}(M)$, which does not depend on the orientation of $M$.}
    \label{Fig:charge_conjugation}
\end{figure}
If the theory is self-dual under the $S$ operation, then it necessarily realizes a ``charge conjugation'' defect since $S^2 = C$.
We denote the charge conjugation defect as $\mathcal{U}$.
This is shown in Figure \ref{Fig:charge_conjugation}.
The charge conjugation defect does not carry an orientation, and it becomes trivial in the case of $N=2$.

\begin{figure}[!h]
    \centering
    \includegraphics[width=0.8\textwidth]{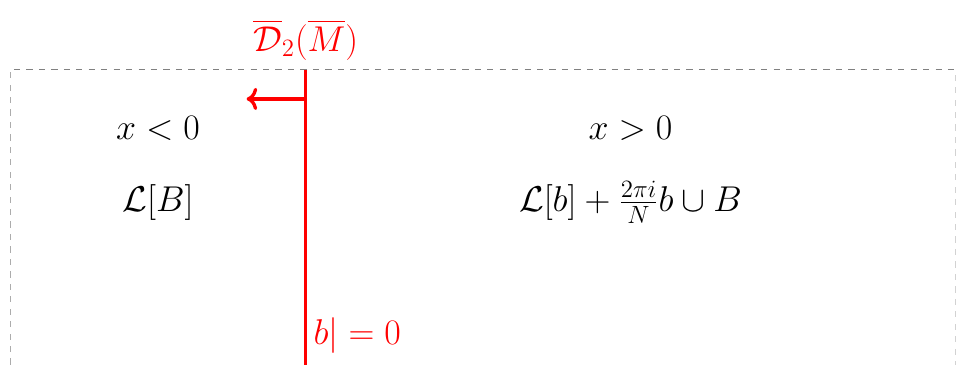}
    \caption{The orientation reversal $\overline{\mathcal{D}}_2(\overline{M})$ of the duality defect.
    The arrow shows that the orientation of $\overline{M}$ is opposite to that of $M$.}
    \label{Fig:duality_defect_rev}
\end{figure}

The orientation reversal $\overline{\mathcal{D}}_2$ is defined by the equation $\mathcal{D}_2 (M) = \overline{\mathcal{D}}_2 (\overline{M})$ according to Eq. (\ref{Eq:general_orientation_reversal}).
Thus, reversing the direction of the arrow in Figure \ref{Fig:duality_defect} gives Figure \ref{Fig:duality_defect_rev}.

\begin{figure}[!h]
    \centering
    \includegraphics[width=0.8\textwidth]{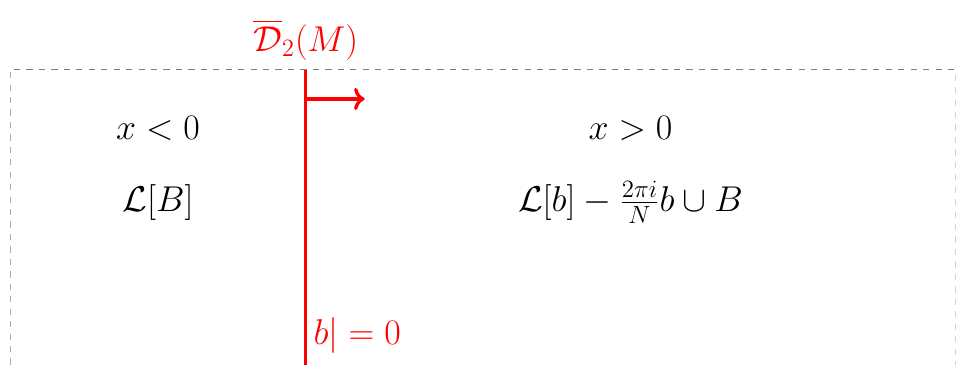}
    \caption{An equivalent definition of $\overline{\mathcal{D}}_2(M)$.}
    \label{Fig:duality_defect_rev_2}
\end{figure}

For later convenience, we will use an alternative definition for $\overline{\mathcal{D}}_2$ which we derive below. 
Since the theory is invariant under the $S$ gauging and the charge conjugation, we can replace the Lagrangians $\cal L$ on both sides  of Figure \ref{Fig:duality_defect_rev} by ${\cal L}[b'] - {2\pi i \over N} b'\cup B$, where $b'$ is a dynamical field living in the whole bulk. 
Next, we integrate out $b$ on half of the spacetime. 
Following a similar lattice discussion in Section \ref{Sec:define_cond_defects}, we arrive at the alternative definition for  $\overline{\mathcal{D}}_2$ in Figure \ref{Fig:duality_defect_rev_2}. 
(We have  renamed $b'$ as the new $b$ and  rotated the figure by 180 degree.)

\subsection{Triality Defects} \label{Sec:define_triality_defects}

\begin{figure}[!h]
    \centering
    \includegraphics[width=0.8\textwidth]{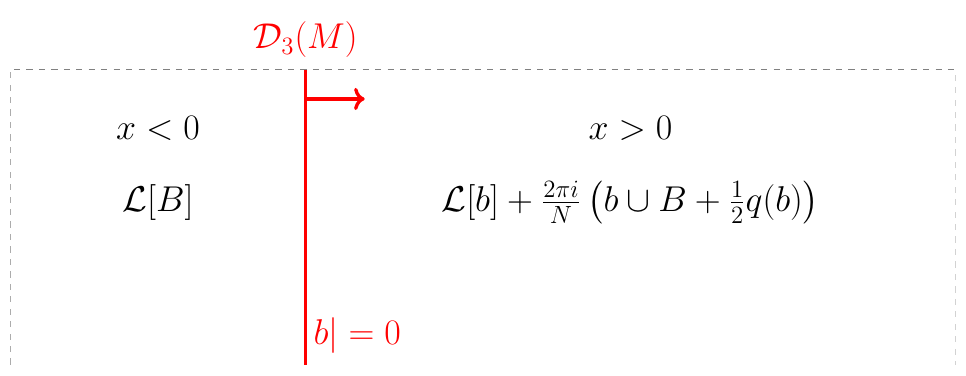}
    \caption{The triality defect $\mathcal{D}_3(M)$ for even $N$.
    For odd $N$, $\frac{1}{2}q(b)$ should be replaced by $\frac{N+1}{2}b\cup b$.}
    \label{Fig:triality_defect}
\end{figure}

\begin{figure}[!h]
    \centering
    \includegraphics[width=0.8\textwidth]{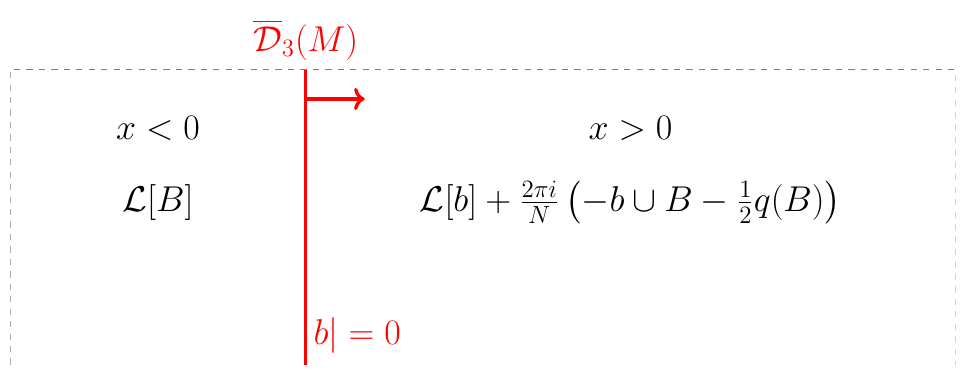}
    \caption{The orientation reversal $\overline{\mathcal{D}}_3(M)$ of the triality defect for even $N$.
    For odd $N$, $\frac{1}{2}q(B)$ should be replaced by $\frac{N+1}{2}B\cup B$.}
    \label{Fig:triality_defect_rev}
\end{figure}

Suppose now we have a QFT $\mathcal{Q}$ which is invariant under the twisted $ST$ gauging. More precisely, by $ST{\cal Q}\cong {\cal Q}$ we mean the following:
\begin{equation} \label{Eq:invariance_under_ST}
    Z_{\mathcal{Q}}[B] = 
    \begin{dcases}
    \frac{1}{|H^2(X;\mathbb{Z}_N)|^{1/2}}
    \sum_{b \in H^2(X;\mathbb{Z}_N)} Z_{\mathcal{Q}}[b] e^{\frac{2\pi i}{N} \int_X \left(b \cup B + \frac{1}{2} q(b) \right)} & \text{for even $N$} \,, \\
    \frac{1}{|H^2(X;\mathbb{Z}_N)|^{1/2}}
    \sum_{b \in H^2(X;\mathbb{Z}_N)} Z_{\mathcal{Q}}[b] e^{\frac{2\pi i}{N} \int_X \left(b \cup B + \frac{N+1}{2} b\cup b \right)} & \text{for odd $N$}  \,.
    \end{dcases}
\end{equation}
Similar to before, we require the equality to hold up to counterterms that are independent of the $\mathbb{Z}_N^{(1)}$ symmetry background $B$.
Then we can construct the \emph{triality defect} $\mathcal{D}_3$ and its orientational reversal $\overline{\mathcal{D}}_3$ as in Figure \ref{Fig:triality_defect}
and Figure \ref{Fig:triality_defect_rev}, respectively.  
Here $\mathcal{D}_3$ is defined by performing the $ST$ operation only to the one side of the defect, while imposing the Dirichlet boundary condition for the dynamical gauge field $b| =0$ at the defect location $x=0$.
We have used a similar argument in Section \ref{Sec:define_duality_defects} to derive the presentation of $\overline{\cal D}_3$ in Figure \ref{Fig:triality_defect_rev} from \eqref{Eq:general_orientation_reversal}.

\section{Fusion of Codimension-One Defects} \label{Sec:fusion_codim_one}

The fusion of two codimension-one defects, say $\mathcal{N}$ and $\mathcal{N}'$, is defined by placing them  parallel to each other and separated by a small distance $\epsilon$, and then taking the $\epsilon \to 0$ limit.
This defines a new codimension-one defect denoted as $\mathcal{N} \times \mathcal{N}'$.
The fusion is generally non-commutative. 
All the fusion rules between codimension-one defects that we derive in this work take the following form:
\begin{equation}
    \mathcal{N}(M) \times \mathcal{N}'(M)
    = \mathcal{T}(M) \mathcal{N}''(M) \,.
\end{equation}
Here $M$ is a codimension-one manifold in spacetime, $\mathcal{N}''$ is a codimension-one defect, and $\mathcal{T}$ is the fusion ``coefficient''.
Interestingly, similar to the observation in 2+1d in \cite{Roumpedakis:2022aik}, we find that the fusion coefficient in general is not a number, but a 2+1d TQFT. 
More precisely, the coefficient $\mathcal{T}(M)$ in the parallel fusion rule is the value of the partition function of the theory $\mathcal{T}$ on the closed manifold $M$.

Using the definition of the orientation-reversal of a defect \eqref{Eq:general_orientation_reversal}, we have
\begin{align}\label{Eq:orientation_reversal_fusion}
 \overline{\mathcal{N}}'(  M) \times \overline{\cal N} ( M)   = \overline{\cal T}(  M)  \,\overline{{\cal N}}''(  M)\,,
\end{align}
where $\overline{\cal T}$ is the orientation reversal of the 2+1d TQFT ${\cal T}$.

 In the previous section, we have defined the codimension-one defects by writing down the Lagrangians on the two sides of the defect.  
In the definition for each one of them, we  have $\mathcal{L}[B]$ of the original theory for the $x<0$ region, and different defects are completely characterized by the Lagrangians in the $x>0$ region.
In this section, we will thus use shorthand notations such as
\begin{equation}
    \mathcal{D}_2 \colon \mathcal{L}[b] + \frac{2\pi i}{N} b \cup B
\end{equation}
to stand for the definition of ${\cal D}_2$ in Figure \ref{Fig:duality_defect}. 
This notation will be useful in deriving the fusion rules systematically.

\begin{figure}[!h]
    \centering
    \includegraphics[width=0.8\textwidth]{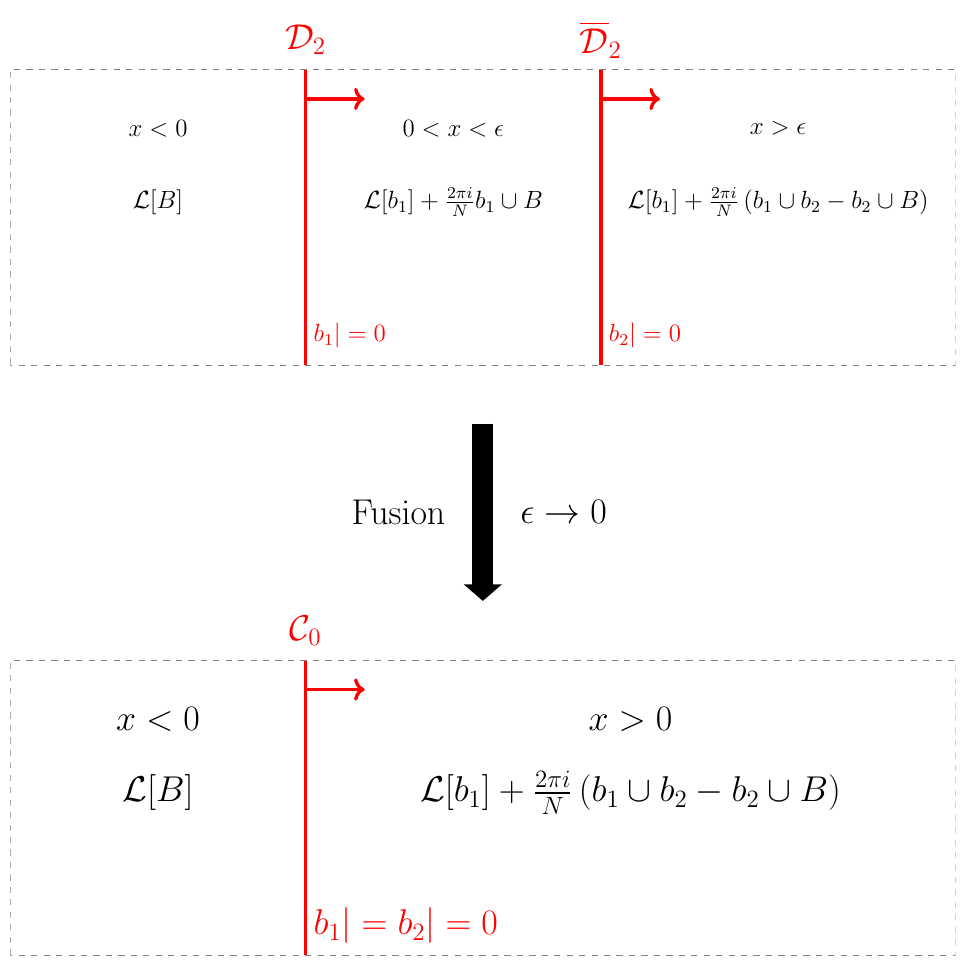}
\caption{Derivation of the fusion rule $\mathcal{D}_2 \times \overline{\mathcal{D}}_2 = \mathcal{C}_0$.}
    \label{Fig:fusion}
\end{figure}
Let us demonstrate the general procedure to determine the fusion of two codimension-one defects using the example of $\mathcal{D}_2 (M) \times \overline{\mathcal{D}}_2 (M)$.
Consider placing these two defects in parallel such that they are separated by a small distance $\epsilon$, as shown in Figure \ref{Fig:fusion}.
In the $x<0$ region, the bulk QFT is described by the Lagrangian $\mathcal{L}[B]$. 
Between the two defects, the bulk Lagrangian is $\mathcal{L}[b_1] + \frac{2\pi i}{N} b_1 \cup B$ which is the expression that defines $\mathcal{D}_2$.
Then, if we let $\mathcal{L}'[B] \equiv \mathcal{L}[b_1] + \frac{2\pi i}{N} b_1 \cup B$, the expression that should appear for the $x > \epsilon$ region is $\mathcal{L}'[b_2] - \frac{2\pi i}{N} b_2 \cup B = \mathcal{L}[b_1]
+ \frac{2\pi i}{N} \left( b_1 \cup b_2 - b_2 \cup B \right)$ using the definition of $\overline{\mathcal{D}}_2$ given in Figure \ref{Fig:duality_defect_rev_2}.
Thus, we arrive at the upper picture of Figure \ref{Fig:fusion}.
Now, fusing the two defects amounts to taking the limit $\epsilon \to 0$.
This brings us to the lower picture of Figure \ref{Fig:fusion}, and we can easily identify the resulting defect as the condensation defect $\mathcal{C}_0$ from Figure \ref{Fig:condensation_defect}.
The fusion rule $\mathcal{D}_2 \times \overline{\mathcal{D}}_2 = \mathcal{C}_0$ indeed confirms the result from \cite{Choi:2021kmx,Kaidi:2021xfk}.

Follwing this general procedure, we will now derive the fusion rules between various codimension one defects.
In Section \ref{Sec:fusion_duality_codim_1}, we will first derive the fusion rule involving the duality defect $\mathcal{D}_2$ in a theory which is invariant under the $S$ operation.
Some of these fusion rules were already derived in \cite{Kaidi:2021xfk,Choi:2021kmx}.
In Sections \ref{Sec:fusion_triality_codim_1} and  \ref{Sec:fusion_triality_codim_1_odd_N}, we present the fusion rule involving the triality defect $\mathcal{D}_3$, which is realized in a theory invariant under the $ST$ operation. 
The derivation of the triality fusion rules is in Appendix \ref{app:triality}. 
The fusion ``coefficients" in these fusion rules are generally nontrivial 2+1d TQFTs.

\subsection{Duality Fusion Rules} \label{Sec:fusion_duality_codim_1}

Given any 3+1d QFT which is invariant under the $S$ operation, $\mathcal{Q} \cong S\mathcal{Q}$, as in Eq. (\ref{Eq:invariance_under_S}), we will derive the following fusion rule between the duality defect ${\cal D}_2$, the condensation defect ${\cal C}_0$, and the charge conjugation defect $\cal U$:
\begin{align} \label{Eq:fusion_duality_summary}
\begin{split}
    \overline{\mathcal{D}}_2 \times \mathcal{D}_2 &=\mathcal{D}_2 \times \overline{\mathcal{D}}_2 = \mathcal{C}_0  \,,\\
    \mathcal{D}_2 \times \mathcal{D}_2 &= \mathcal{U} \times \mathcal{C}_0 = \mathcal{C}_0 \times \mathcal{U} \,, \\
    \mathcal{D}_2 \times \mathcal{C}_0 &= \mathcal{C}_0 \times \mathcal{D}_2 = (\mathcal{Z}_N)_0 \,\mathcal{D}_2  \,, \\
    \mathcal{C}_0 \times \mathcal{C}_0 &= (\mathcal{Z}_N)_0 \, \mathcal{C}_0 \,.
\end{split}
\end{align}
We also have $\overline{\mathcal{D}}_2 = \mathcal{U} \times \mathcal{D}_2 = \mathcal{D}_2 \times \mathcal{U}$.
All the defects and the ``coefficient" TQFTs are supported on the same codimension-one submanifold $M$ and we will not explicitly write the dependence. 
Other fusion rules involving $\overline{\cal D}_2$ can be obtained using \eqref{Eq:orientation_reversal_fusion}. 
$(\mathcal{Z}_N)_K$ denotes the 2+1d $\mathbb{Z}_N$ gauge theory at level $K$ following the convention from \cite{Hsin:2018vcg}.
It can be represented by the action
\begin{equation} \label{Eq:ZN_level_K}
    (\mathcal{Z}_N)_K \colon \quad
    \int_M \left(
        \frac{iK}{4\pi} xdx + \frac{iN}{2\pi} xdy
    \right)
\end{equation}
where $x$ and $y$ are dynamical $U(1)$ gauge fields living on $M$.
The level $K$ is defined modulo $2N$.
On a nonspin manifold, it can take any even integer values, whereas on a spin manifold, it can take any integer values.

Below we present the derivation of these fusion rules:

\begin{description}[style=unboxed,leftmargin=0cm,font=\normalfont\textbullet\space]
\item [$\overline{\mathcal{D}}_2 \times \mathcal{D}_2$ and $\mathcal{D}_2 \times \overline{\mathcal{D}}_2$ :]

We have already illustrated how to obtain the fusion rule $\mathcal{D}_2 \times \overline{\mathcal{D}}_2 = \mathcal{C}_0$ above.
Here we derive the fusion  $\overline{\mathcal{D}}_2 \times \mathcal{D}_2$ in the opposite order.
As $\epsilon \to 0$, we  have the following Lagrangian in the $x>0$ region 
\begin{equation}
    \overline{\mathcal{D}}_2 \times \mathcal{D}_2 \colon
    \mathcal{L}[b_1] + \frac{2\pi i}{N} \left(
        -b_1 \cup b_2 + b_2 \cup B
    \right) \,.
\end{equation}
Next, we perform the field redefinition $b_2 \rightarrow -b_2$ and see that this again corresponds to the condensation defect $\mathcal{C}_0$.
Thus, $\overline{\mathcal{D}}_2 \times \mathcal{D}_2 =\mathcal{D}_2 \times \overline{\mathcal{D}}_2 = \mathcal{C}_0$.

\item[$\mathcal{D}_2 \times \mathcal{D}_2$ :]

We have
\begin{align}
\begin{split}
    \mathcal{D}_2 \times \mathcal{D}_2 \colon
    & \mathcal{L}[b_1] + \frac{2\pi i}{N}
    \left(
        b_1 \cup b_2 + b_2 \cup B
    \right)
    \\ 
    = & \mathcal{L}[-b_1] + \frac{2\pi i}{N}
    \left(
        b_1 \cup b_2 - b_2 \cup B
    \right)
    \,.
\end{split}
\end{align}
In the second line, we have flipped the signs of both $b_1$ and $b_2$.
We see that $\mathcal{D}_2 \times \mathcal{D}_2 = \mathcal{U} \times \mathcal{C}_0 = \mathcal{C}_0 \times \mathcal{U}$.

\item[$\mathcal{D}_2 \times \mathcal{C}_0$ and $\mathcal{C}_0 \times \mathcal{D}_2$ :]

We have
\begin{align}
\begin{split}
    \mathcal{D}_2 \times \mathcal{C}_0 \colon
    &
    \mathcal{L}[b_1] + \frac{2\pi i}{N} \left(
        b_1 \cup b_2 + b_2 \cup b_3 - b_3 \cup B
    \right)
    \\
    =
    &\left(
    \frac{2\pi i}{N} (b_2 - B) \cup b'_3
    \right)
    +
    \left(
    \mathcal{L}[b] + \frac{2\pi i}{N} b\cup B
    \right)
\end{split}
\end{align}
where we have renamed fields as $b \equiv b_1$ and $b'_3 \equiv b_3 + b_1$.
The first term gives  a  2+1d $\mathbb{Z}_N$ gauge theory $(\mathcal{Z}_N)_0$ living on the defect \cite{Hsin:2018vcg}.
This can be seen as follows.\footnote{We will set $B=0$ since the coupling to $B$ contains  information about the junction between the $(\mathcal{Z}_N)_0$ theory and the symmetry defect, which we do not discuss for parallel fusion.} 
As explained in Section \ref{Sec:define_cond_defects} and in Figure \ref{Fig:condensation_lattice}, when we integrate out $b_2$, the remaining gauge field $b'_3$ is set to zero except inside a thin slab $M \times I$, where $M$ is the codimension-one submanifold on which the defect is supported and $I$ is a small interval (the green shaded region in Figure \ref{Fig:condensation_lattice}).
Under the isomorphism $H^2(M\times I,\partial(M\times I);\mathbb{Z}_N) \xrightarrow{\cong} H^1(M;\mathbb{Z}_N)$, the bulk two-form gauge field $b'_3 \in H^2(M\times I,\partial(M\times I);\mathbb{Z}_N)$ inside the slab $M \times I$ is mapped to a dynamical one-form gauge field $a \in H^1(M;\mathbb{Z}_N)$ living on the defect $M$.  
We have thus obtained a 2+1d $(\mathcal{Z}_N)_0$ gauge theory of $a$ living on $M$.

The remaining terms define the defect $\mathcal{D}_2$.  We have thus derived $\mathcal{D}_2 \times \mathcal{C}_0 = (\mathcal{Z}_N)_0 \,\mathcal{D}_2$.

Finally, the fusion $\mathcal{C}_0 \times \mathcal{D}_2 $ is identical to $\mathcal{D}_2 \times \mathcal{C}_0$ by flipping the sign of $b_3$.
To conclude, we obtain $\mathcal{D}_2 \times \mathcal{C}_0 = \mathcal{C}_0 \times \mathcal{D}_2 = (\mathcal{Z}_N)_0 \,\mathcal{D}_2$.

\item[$\mathcal{C}_0 \times \mathcal{C}_0$ :]

We have
\begin{align}
    \begin{split}
        \mathcal{C}_0 \times \mathcal{C}_0 \colon
        &
        \mathcal{L}[b_1] + \frac{2\pi i}{N} \left(
            b_1 \cup b_2 - b_2 \cup b_3 
            +b_3 \cup b_4 - b_4 \cup B
        \right)
        \\
        =
        &
        -\frac{2\pi i}{N} b'_2 \cup b'_3
                +
        \left(
        \mathcal{L}[b_1] + \frac{2\pi i}{N} \left(
        b_1 \cup b_4 - b_4 \cup B
        \right)
        \right)
    \end{split}
\end{align}
where we have renamed $b'_2 \equiv b_2 - b_4$ and $b'_3 \equiv b_3 - b_1$.
The first term again gives us a decoupled 3d $(\mathcal{Z}_N)_0$ gauge theory, and the rest defines $\mathcal{C}_0$.\footnote{To be precise, this $(\mathcal{Z}_N)_0$ gauge theory differs from the previous one in the presence of a transverse junction with $\eta$, which we ignore as we explained at the end of Section \ref{Sec:define_cond_defects}.}
Thus, we obtain $\mathcal{C}_0 \times \mathcal{C}_0=(\mathcal{Z}_N)_0 \,\mathcal{C}_0$.

This fusion rule between the codimension-one condensation  defect ${\cal C}_0$ can, for example, be realized in the 3+1d $\mathbb{Z}_N$ gauge theory, which is the low energy limit of the $\mathbb{Z}_N$ toric code. 
It takes a similar form as the fusion rule  for the ``Cheshire strings'' \cite{Else:2017yqj,Johnson-Freyd:2020twl} (see also \cite{Alford:1990mk,Bucher:1991bc,Alford:1992yx} for earlier papers), which are  condensation defects from the two-gauging of two-form symmetry lines on a two-dimensional surface. 
Here,  our condensation defects arise from the one-gauging of the one-form symmetry surfaces on a three-dimensional submanifold in 3+1d.

\end{description}

\subsection{Triality Fusion Rules for Even $N$} \label{Sec:fusion_triality_codim_1}

We now consider a general 3+1d QFT $\cal Q$ which is invariant under the $ST$ operation, $  ST\mathcal{Q}\cong \mathcal{Q}$ in the sense of Eq. (\ref{Eq:invariance_under_ST}). 
We will see examples of such QFTs in later sections. 
Such a QFT has a triality defect defined in Figure \ref{Fig:triality_defect}.
The fusion rules for even and odd $N$ are qualitatively different, and will be discussed separately. 
We start with the even $N$ case.

The two condensation defects ${\cal C}_0$ and ${\cal C}_{N\over2}$ are distinguished compared to the more general ${\cal C}_k$ in that they are their own orientation reversal \eqref{Eq:orientation_invariant}. 
Here we present the fusion rules of the triality defect $\mathcal{D}_3$ and the condensation defects ${\cal C}_{N\ell\over2}$ for even $N$ (here $\ell =0,1$ mod 2), while the derivation is given in Appendix \ref{app:evenN}:
\begin{align} \label{Eq:fusion_triality_summary}
\begin{split}
    \overline{\mathcal{D}}_3 \times \mathcal{D}_3 &= \mathcal{C}_0 \,,\\
    \mathcal{D}_3 \times \overline{\mathcal{D}}_3 &= \mathcal{C}_{\frac{N}{2}} \,,\\
    \mathcal{D}_3 \times \mathcal{D}_3 &= U(1)_{N} \,\overline{\mathcal{D}}_3 \,,\\
    \mathcal{D}_3 \times \mathcal{C}_0 &= \mathcal{C}_{\frac{N}{2}} \times \mathcal{D}_3 = (\mathcal{Z}_N)_0 \,\mathcal{D}_3 \,,\\
    \mathcal{D}_3 \times \mathcal{C}_{\frac{N}{2}} &= \mathcal{C}_0 \times \mathcal{D}_3 
    = (\mathcal{Z}_N)_N \,\mathcal{D}_3  \,, \\
    \mathcal{C}_{\frac{N\ell_1}{2}} \times \mathcal{C}_{\frac{N\ell_2}{2}} &= (\mathcal{Z}_N)_{N(\ell_1 + \ell_2)} \, \mathcal{C}_{\frac{N\ell_1}{2}} =(\mathcal{Z}_N)_{N(\ell_1 + \ell_2)} \, \mathcal{C}_{\frac{N\ell_2}{2}} \,.
\end{split}
\end{align}
Other fusion rules involving $\overline{\cal D}_3$ can be obtained using \eqref{Eq:orientation_reversal_fusion}. 
Again, all the defects and TQFTs are supported on the same codimension-one manifold $M$. 
Here, $U(1)_{N}$ is the 2+1d Chern-Simons theory living on $M$ given by the action $\frac{iN}{4\pi} \int_M ada$ where $a$ is a $U(1)$ gauge field in $M$, and $(\mathcal{Z}_N)_K$ again denotes the 2+1d $\mathbb{Z}_N$ gauge theory at level $K$ (\ref{Eq:ZN_level_K}).
Even though some of these fusion rules can be obtained from other fusion rules,  we will derive all of them independently for consistent checks.

The fusion between the defects has to be associative. 
Unlike the duality case, now the associativity of the fusion rules (\ref{Eq:fusion_triality_summary}) is not trivial.
For instance, we can compute $\mathcal{D}_3 \times \mathcal{D}_3 \times \mathcal{D}_3$ in two different ways, which gives
\begin{equation} \label{Eq:DDD_triality_fusion_rule}
        \mathcal{D}_3 \times \mathcal{D}_3 \times \mathcal{D}_3
    =  U(1)_{N} \, \mathcal{C}_0 =  U(1)_{N} \, \mathcal{C}_{\frac{N}{2}} \,. 
\end{equation}
Naively, this seems to be a contradiction.
However, there is an interesting resolution.
As discussed in Section \ref{Sec:define_cond_defects},  the two condensation defects $\mathcal{C}_0$ and $\mathcal{C}_{\frac{N}{2}}$ differ by the Dijkgraaf-Witten twist $(-1)^{Q(M,\Sigma)}$. 
It turns out that this sign, which can be viewed as a 2+1d invertible field theory of PD$(\Sigma)$ labeled by the order 2 element of $H^3(B\mathbb{Z}_N;U(1))$, can be  absorbed  by the $U(1)_{N}$ Chern-Simons theory, 
\ie\label{Eq:signU1}
(-1)^{Q(M,\Sigma)}Z_{U(1)_{N}}[M] = Z_{U(1)_{N}}[M]\,,
\fe
where $Z_{U(1)_N}[M]$ denotes the partition function of $U(1)_N$ on $M$. 
(Similar examples of invertible phases trivialized in the presence of TQFTs were also discussed in \cite{Hsin:2019fhf}.) 
That is, for every two-cycle $\Sigma$ in a  closed three-manifold $M$ such that with $Q(M,\Sigma) =1$ mod 2, the partition function of $U(1)_{N}$ on $M$  vanishes.
We prove this fact in Appendix \ref{App:U1_N_times_SPT}.
It follows that  $U(1)_{N} \, \mathcal{C}_0 =  U(1)_{N} \, \mathcal{C}_{\frac{N}{2}}$, which is essential for the fusion rule to be associative. 

Similarly, \eqref{Eq:signU1}  also explains why the last equation in (\ref{Eq:fusion_triality_summary}) is consistent.
There, if we set $\ell_1 =0$ and $\ell_2 =1$, then  $(\mathcal{Z}_N)_N\,\mathcal{C}_0 = (\mathcal{Z}_N)_N\,\mathcal{C}_{\frac{N}{2}}$. 
This equation is again consistent since $(\mathcal{Z}_N)_N = U(1)_N \times U(1)_{-N}$ also absorbs the phase $(-1)^{Q(M,\Sigma)}$.
Using these facts, one can check that the set of fusion rules (\ref{Eq:fusion_triality_summary}) are associative and consistent.

The coefficient TQFT $U(1)_{N}$ appearing in Eq. (\ref{Eq:DDD_triality_fusion_rule}) can be understood as the boundary of the invertible theory $Y$ in Eq. (\ref{Eq:invertible_Y}).
The fusion $\mathcal{D}_3 \times \mathcal{D}_3 \times \mathcal{D}_3$ corresponds to applying $(ST)^3 = Y$ operation only in half of the spacetime, and the Dirichlet boundary condition imposed for the invertible theory $Y$ corresponds to $U(1)_{N}$ Chern-Simons theory living on the defect \cite{Gaiotto:2014kfa,Hsin:2018vcg}.
In the notation of \cite{Hsin:2018vcg}, $U(1)_{N}$ is the minimal TQFT $\mathcal{A}^{N,1}$ which has $p=1$ anomaly for the $\mathbb{Z}_N^{(1)}$ one-form symmetry (see also \cite{Moore:1988qv,Bonderson:2007ci,Barkeshli:2014cna} for earlier discussions).

Finally, the fusion rules tell us that the quantum dimension of the triality defect $\mathcal{D}_3$ on $S^3$ is equal to $\langle \mathcal{D}_3 \rangle_{S^3} = 1/\sqrt{N}$, which is the same as that of the duality defect $\mathcal{D}_2$.

\subsection{Triality Fusion Rules for Odd $N$} \label{Sec:fusion_triality_codim_1_odd_N}

We now move on to the triality fusion rule for odd $N$.  
Here we present the fusion rule between the triality defect ${\cal D}_3$ and the condensation defect ${\cal C}_0$, while the derivation  is given in Appendix \ref{app:oddN}:
\begin{align} \label{Eq:fusion_triality_summary_odd_N}
\begin{split}
    \overline{\mathcal{D}}_3 \times \mathcal{D}_3 &=
    \mathcal{D}_3 \times \overline{\mathcal{D}}_3 = \mathcal{C}_{0} \,,\\
    \mathcal{D}_3 \times \mathcal{D}_3 &= SU(N)_{-1} \,\overline{\mathcal{D}}_3 \,,\\
    \mathcal{D}_3 \times \mathcal{C}_0 &= \mathcal{C}_{0} \times \mathcal{D}_3 = (\mathcal{Z}_N)_0 \,\mathcal{D}_3 \,,\\
    \mathcal{C}_0 \times \mathcal{C}_0 &= (\mathcal{Z}_N)_0 \,\mathcal{C}_0\,.
\end{split}
\end{align}
This fusion rule apply to any 3+1d QFT $\mathcal{Q}$ with a $\mathbb{Z}_N^{(1)}$ one-form symmetry such that $\mathcal{Q} \cong ST\mathcal{Q}$ in the sense of Eq. (\ref{Eq:invariance_under_ST}). 
Other fusion rules involving $\overline{\mathcal{D}}_3$ can be obtained using Eq. (\ref{Eq:orientation_reversal_fusion}). 
In particular, we have
\begin{equation} \label{Eq:DDD_fusion_odd_N}
    \mathcal{D}_3 \times \mathcal{D}_3 \times \mathcal{D}_3
    = SU(N)_{-1} \,\mathcal{C}_0 \,.
\end{equation}
Similar to the even $N$ case, the 2+1d $SU(N)_{-1}$ Chern-Simons theory which appears as the fusion  ``coefficient" can  be understood as the boundary of the 3+1d invertible theory $Y$ in Eq. (\ref{Eq:invertible_Y}) for odd $N$ with the Dirichlet boundary condition imposed.
It's the minimal TQFT $\mathcal{A}^{N,N+1}$ in the notation of \cite{Hsin:2018vcg}.
If the three-manifold $M$ on which the defects are supported is a spin manifold, the $SU(N)_{-1}$ Chern-Simons theory is dual to the $U(1)_{N}$ Chern-Simons theory  \cite{Hsin:2016blu}.\footnote{Since we  assume that $M$ has a neighborhood $M \times I$ inside the spacetime manifold $X$, $M$ is  spin as long as  $X$ is spin.}

Similar to the even $N$ case, the fusion rule implies that the quantum dimension of the triality defect $\mathcal{D}_3$ on $S^3$ is given by $\langle \mathcal{D}_3 \rangle_{S^3} = 1/\sqrt{N}$.

\section{Fusion Involving Symmetry Surface Defects} \label{Sec:fusion_involving_eta}

Having derived the fusion rules between the codimension-one defects, we now turn to the fusion between a codimension-one defect and the symmetry surface defect $\eta$, which is of codimension-two.
The fusion is defined in a similar fashion as before.
A codimension-one defect $\mathcal{N}(M)$ is supported on a three-dimensional submanifold $M$, around which we have a neighborhood that is topologically $M \times I$.
Then, we place the symmetry surface defect $\eta(\Sigma)$   on a two-dimensional surface $\Sigma \subset M \times I$ while being parallel to $M$ but separated with it by a small distance $\epsilon$ in the interval $I$.
When we take the limit $\epsilon \to 0$, the surface $\Sigma$ is embedded into the three-manifold $M$, and this defines a topological surface operator living inside the three-dimensional defect $\mathcal{N}$.\footnote{In general, two defects of the same dimensionalities can be fused in the presence of such higher codimensional topological operators living inside them \cite{Kaidi:2021xfk,Choi:2021kmx,Chen:2021xuc,Roumpedakis:2022aik}. 
From this more general perspective, the fusion between $\mathcal{N}$ and $\eta$ can be thought of as fusing the trivial surface operator inside $\mathcal{N}$ with a nontrivial surface operator given by $\eta$ inside a trivial three-dimensional defect.
A simple analogy can be made in a 1+1d QFT having a non-invertible symmetry described by a fusion category.
Consider two topological line defects $\mathcal{L}_1$ and $\mathcal{L}_2$ in a 1+1d theory which are not necessarily simple, and also the morphisms $\alpha_1 : \mathcal{L}_1 \rightarrow \mathcal{L}_1$ and $\alpha_2 : \mathcal{L}_2 \rightarrow \mathcal{L}_2$ inside these lines, which are topological point operators living on these lines.
The morphisms $\alpha_1$ and $\alpha_2$ can be represented by finite-dimensional matrices, and the fusion $\alpha_1 \times \alpha_2 : \mathcal{L}_1 \times \mathcal{L}_2 \rightarrow \mathcal{L}_1 \times \mathcal{L}_2$ is simply given by the tensor product of two matrices, and it defines a topological point operator living on the line $\mathcal{L}_1 \times \mathcal{L}_2$.
If we let $\alpha_1 = \text{Id}_{\mathcal{L}_1}$ to be the trivial point operator and $\mathcal{L}_2 = 1$ to be the trivial line defect, then the fusion can be thought as the fusion between the topological line defect $\mathcal{L}_1$ and the topological local operator $\alpha_2$.
We thank Kantaro Ohmori for discussions on this point. 
}
The fusion rules that we encounter in this work always take the following form:
\begin{align} \label{Eq:fusion_eta_general}
\begin{split}
    \mathcal{N}(M) \times \eta(\Sigma)
    &= \mathcal{I}(M,\Sigma) \mathcal{N}(M) \,, \\
    \eta(\Sigma) \times \mathcal{N}(M) &= \mathcal{I}'(M,\Sigma) \mathcal{N}(M) \,.
\end{split}
\end{align}
Here, $\mathcal{I}(M,\Sigma)$ (and similarly $\mathcal{I}'(M,\Sigma)$) denotes the partition function of a 2+1d $\mathbb{Z}_N^{(0)}$-SPT on the three-manifold $M$ of the background $\mathbb{Z}_N^{(0)}$ one-form gauge field PD$(\Sigma)$, which is the Poincar\'e dual of the homology class of the surface $\Sigma \in H_2(M;\mathbb{Z}_N)$ in $M$.
The fusion is generally noncommutative, that is, $\mathcal{I}$ and $\mathcal{I}'$ can be different.
The form of the fusion rule (\ref{Eq:fusion_eta_general}) implies that when $\eta$ is brought inside the codimension-one defect $\mathcal{N}$, it becomes a trivial surface operator living on $\mathcal{N}$ which is completely characterized by a local counterterm, namely the 2+1d SPT $\mathcal{I}$ or $\mathcal{I}'$.

Below, we first discuss such fusion rules involving the duality defect in Section \ref{Sec:eta_fusion_duality}.
In that case, there is no nontrivial 2+1d SPT appearing in the fusion rules.
Then in Section \ref{Sec:eta_fusion_triality} and Section \ref{Sec:eta_fusion_triality_odd_N}, we discuss the triality defect case.
There, we will encounter a nontrivial 2+1d SPT in some of the fusion rules when $N$ is even.

\subsection{Duality Fusion Rules} \label{Sec:eta_fusion_duality}

We first consider the case where the QFT $\mathcal{Q}$ is invariant under the $S$ operation, $\mathcal{Q} \cong S\mathcal{Q}$ as in Eq. (\ref{Eq:invariance_under_S}), so that the theory realizes the duality defect $\mathcal{D}_2$.
The fusion rule of the codimension-one defects as summarized in Eq. (\ref{Eq:fusion_duality_summary}).
The fusion rule with the symmetry surface defects are given as follows \cite{Choi:2021kmx,Kaidi:2021xfk}:
\begin{align} \label{Eq:fusion_eta_duality}
\begin{split}
    \mathcal{D}_2 (M) \times \eta(\Sigma) &=
    \eta(\Sigma) \times \mathcal{D}_2(M) = \mathcal{D}_2(M) \,, \\
    \overline{\mathcal{D}}_2 (M) \times \eta(\Sigma) &=
    \eta(\Sigma) \times \overline{\mathcal{D}}_2 (M) = \overline{\mathcal{D}}_2 (M) \,, \\
    \mathcal{C}_0 (M) \times \eta(\Sigma) &=
    \eta(\Sigma) \times \mathcal{C}_0 (M) \,.
\end{split}
\end{align}
We see that the symmetry surface $\eta$ is always ``absorbed" by the codimension-one defect under fusion. 
They can be easily derived by recalling that inserting a symmetry defect $\eta(\Sigma)$ wrapping around a $\mathbb{Z}_N$ two-cycle $\Sigma$ is equivalent to turning on a two-form background gauge field $B$  which is Poincar\'e dual to  $\Sigma$ in the 3+1d spacetime $X$.

For instance, consider the fusion $\mathcal{D}_2 \times \eta$.
From the definition of $\mathcal{D}_2$ given in Figure \ref{Fig:duality_defect}, we see that in the region $x>0$, we have
\begin{equation}
    \eta(\Sigma) = \text{exp} \left(
        \frac{2\pi i}{N} \int_{X:~x>0} b \cup B
    \right) 
    =
    \text{exp} \left(
        \frac{2\pi i}{N} \oint_\Sigma b
    \right) \,.
\end{equation}
Bringing $\eta(\Sigma)$ into $\mathcal{D}_2(M)$ from the right, due to the Dirichlet boundary condition $b| = 0$ on $M$, the symmetry defect $\eta(\Sigma)$ just gets absorbed, and we conclude that $\mathcal{D}_2 \times \eta = \mathcal{D}_2$.
Exactly the same argument applies also to $\overline{\mathcal{D}}_2 \times \eta = \overline{\mathcal{D}}_2$ and $\mathcal{C}_0 \times \eta = \mathcal{C}_0$.
Combined with orientation reversed versions of these fusion rules using Eq. (\ref{Eq:orientation_reversal_fusion}), we arrive at Eq. (\ref{Eq:fusion_eta_duality}).

\subsection{Triality Fusion Rules for Even $N$} \label{Sec:eta_fusion_triality}

Now, we turn to the case where we have a theory $\mathcal{Q}$ invariant under the $ST$ operation as in Eq. (\ref{Eq:invariance_under_ST}), so that the theory realizes the triality defect $\mathcal{D}_3$.
The fusion rule for the codimension-one defects is given in Eq. (\ref{Eq:fusion_triality_summary}) for the even $N$ case and in Eq.(\ref{Eq:fusion_triality_summary_odd_N}) for the odd $N$ case.

When $N$ is even, the fusion rule between the codimension-one defects and the symmetry surface defect $\eta$ is given by:
\begin{align} \label{Eq:fusion_eta_triality}
\begin{split}
    \mathcal{D}_3(M) \times \eta(\Sigma) &= \mathcal{D}_3(M)\,, \\
    \eta(\Sigma) \times \mathcal{D}_3(M) &= (-1)^{Q(M,\Sigma)}\mathcal{D}_3(M) \,, \\
    \overline{\mathcal{D}}_3 (M) \times \eta(\Sigma) &=
    (-1)^{Q(M,\Sigma)} \overline{\mathcal{D}}_3 (M) \,, \\
    \eta(\Sigma) \times \overline{\mathcal{D}}_3 (M) &=
    \overline{\mathcal{D}}_3 (M) \,,\\
    \mathcal{C}_\frac{N\ell}{2} (M) \times \eta(\Sigma) 
    &=  \eta(\Sigma) \times \mathcal{C}_\frac{N\ell}{2}(M)
    = (-1)^{\ell Q(M,\Sigma)} \mathcal{C}_\frac{N\ell}{2} (M)\,. \\
\end{split} 
\end{align}
(Recall that $\ell=0,1$ mod 2.) 
Here, we see that the partition function of a nontrivial 2+1d SPT $(-1)^{Q(M,\Sigma)}$ appears in some of the fusion rules.

The way that these fusion rules are derived is completely analogous to the duality case.
The only additional subtlety is that now to the right of a codimension-one defect, we sometimes have another $B$ dependent term $\text{exp}\left( \pm \frac{2\pi i}{2N} \int_{X:~x>0} q(B) \right)$.
For instance, consider the fusion $\overline{\mathcal{D}}_3 (M) \times \eta(\Sigma)$.
From Figure \ref{Fig:triality_defect_rev}, we see that
\begin{align}
\begin{split}
    \eta(\Sigma) &= \text{exp}\left(
        -\frac{2\pi i}{N} \int_{X:~x>0} b \cup B
        -\frac{2\pi i}{2N} \int_{X:~x>0} q(B)
        \right) \\
        &= \text{exp} \left(
            -\frac{2\pi i}{N} \oint_\Sigma b
        \right) \times
        \text{exp}\left(
        -\frac{2\pi i}{2N} \int_{X:~x>0} q(B)
        \right)
\end{split}
\end{align}
in the $x>0$ region, that is, when $\Sigma$ is to the right of $M$.
As we bring $\eta$ to $\overline{\mathcal{D}}_3 (M)$,  due to the Dirichlet boundary condition $b| = 0$ at $x=0$, the symmetry defect gets absorbed.
However, now there is an extra phase factor $\text{exp}\left(-\frac{2\pi i}{2N} \int_{X:~x>0} q(B) \right)$.
If we denote the Poincar\'e dual of $\Sigma$ in $M$ as $A = \text{PD}(\Sigma)$, then we have
\begin{equation} \label{Eq:Pontryagin_square_slab}
    \text{exp}\left(-\frac{2\pi i}{2N} \int_{X:~x>0} q(B) \right)
    = \text{exp}\left(\pi i \int_M A \cup \beta(A) \right) =(-1)^{Q(M,\Sigma)}\,.
\end{equation}
Let us explain \eqref{Eq:Pontryagin_square_slab} below.
First, since $\Sigma \subset M$ defines a class in $H_2(M;\mathbb{Z}_N)$, we can use the Lefschetz duality $H_2(M;\mathbb{Z}_N) \xrightarrow{\cong} H^2(M\times I, \partial(M\times I);\mathbb{Z}_N)$ to express the symmetry defect in terms of the two-form background gauge field $B \in H^2(M\times I, \partial(M\times I);\mathbb{Z}_N)$ which lives inside the thin slab $M\times I$ near the defect, with the Dirichlet boundary condition imposed on the boundary of the slab $\partial(M\times I)$.
Then, as we explained in Section \ref{Sec:define_cond_defects} and as was derived in \cite{Kaidi:2021xfk}, under the isomorphism
$H^4(M\times I, \partial(M\times I);\mathbb{Z}_{2N}) \xrightarrow{\cong} H^3(M;\mathbb{Z}_{2N})$, the phase factor $\text{exp}\left(-\frac{2\pi i}{2N} \int_{X:~x>0} q(B) \right)$ is mapped to $(-1)^{Q(M,\Sigma)}$.
Thus, we conclude $\overline{\mathcal{D}}_3 (M) \times \eta(\Sigma)= (-1)^{Q(M,\Sigma)} \overline{\mathcal{D}}_3 (M)$.

Mathematically, Eq. (\ref{Eq:Pontryagin_square_slab}) is a map between the cohomology groups \eqref{Eq:map} \cite{Roumpedakis:2022aik}. 
It is realized as a dimensional reduction of a 3+1d $\mathbb{Z}_N^{(1)}$-SPT, where one obtains a 2+1d $\mathbb{Z}_N^{(0)}$-SPT by putting the 3+1d $\mathbb{Z}_N^{(1)}$-SPT on a thin slab while imposing the Dirichlet boundary condition for the background gauge field.\footnote{In \cite{Roumpedakis:2022aik}, Eq. \eqref{Eq:Pontryagin_square_slab} was interpreted as a map from the zero-anomaly to the one-anomaly of  a $\mathbb{Z}_N^{(1)}$ one-form symmetry in 2+1d.}

The remaining fusion rules in Eq. (\ref{Eq:fusion_eta_triality}) are derived in the same way when $\eta$ is to the right of a codimension-one defect. 
When $\eta$ is to the left of a codimension-one defect, the fusion rule is derived by taking the   orientation reversal of the previous ones using Eq. (\ref{Eq:orientation_reversal_fusion}).\footnote{For the fusion with the condensation defect $\mathcal{C}_{\frac{N\ell}{2}}$, one can also use the definition (\ref{Eq:condensation_defect_general}) to obtain the same results.}

Finally, recall that at the end of Section \ref{Sec:define_cond_defects}, we argued that the two condensation defects $\mathcal{C}_{\frac{N\ell}{2}}$ and $\mathcal{C}_{\frac{N(\ell+2)}{2}}$ defined in Figure \ref{Fig:condensation_defect} obey the same parallel fusion rule with the symmetry defect $\eta$.
One can now see that this is indeed the case.
For instance, the difference between the two fusion rules $\mathcal{C}_{\frac{N\ell}{2}} \times \eta$ and $\mathcal{C}_{\frac{N(\ell+2)}{2}} \times \eta$ would be a phase factor $\text{exp}\left(\frac{2\pi i}{N} \int_{X:~x>0} q(B) \right) = \text{exp}\left(\frac{2\pi i}{N} \int_{X:~x>0} B\cup B \right)$.
However, this phase factor is trivial as it is equal to $(-1)^{2 Q(M,\Sigma)} = 1$.
Therefore, $\mathcal{C}_{\frac{N\ell}{2}}$ and $\mathcal{C}_{\frac{N(\ell+2)}{2}}$ indeed can't be distinguished by the parallel  fusion with $\eta$ as claimed.

\subsection{Triality Fusion Rules for Odd $N$} \label{Sec:eta_fusion_triality_odd_N}

For the odd $N$ case, we have the following fusion rules:
\begin{align} \label{Eq:eta_fusion_triality_odd_N}
\begin{split}
\mathcal{D}_3 (M) \times \eta(\Sigma) &=
    \eta(\Sigma) \times \mathcal{D}_3(M) = \mathcal{D}_3(M) \,, \\
    \overline{\mathcal{D}}_3 (M) \times \eta(\Sigma) &=
    \eta(\Sigma) \times \overline{\mathcal{D}}_3 (M) = \overline{\mathcal{D}}_3 (M) \,, \\
    \mathcal{C}_0 (M)\times \eta(\Sigma) &= \eta(\Sigma) \times \mathcal{C}_0(M) = \mathcal{C}_0(M)\,.
\end{split}
\end{align}

First, to the right of the defect $\mathcal{D}_3 (M)$, that is, $x>0$ region in Figure \ref{Fig:triality_defect}, we again have $\eta(\Sigma) = \text{exp}\left({\frac{2\pi i}{N} \oint_{\Sigma} b} \right)$.
The Dirichlet boundary condition $b|=0$ tells us $\mathcal{D}_3 (M) \times \eta(\Sigma) = \mathcal{D}_3 (M)$. 
  Taking the orientation reversal following Eq. (\ref{Eq:orientation_reversal_fusion}), we obtain $\eta(\Sigma) \times \overline{\mathcal{D}}_3 (M) = \overline{\mathcal{D}}_3 (M)$.
  
Next, if $\eta$ is to the right of the defect $\overline{\mathcal{D}}_3 (M)$, we have
\begin{equation}
    \eta(\Sigma) = \text{exp} \left(
            -\frac{2\pi i}{N} \oint_\Sigma b
        \right) \times
        \text{exp}\left(
        -\frac{2\pi i}{N} \int_{X:~x>0} \frac{N+1}{2} B \cup B
        \right)
\end{equation}
where $B$ is the background two-form gauge field dual to the insertion of the surface defect $\eta(\Sigma)$.
Similar to the even $N$ case, we use the fact that when $\Sigma$ is embedded inside $M$, we can represent $B$ as an element in the relative cohomology group $H^2(M\times I, \partial(M\times I);\mathbb{Z}_N)$.
Then, under the isomorphism $H^4(M\times I, \partial(M\times I);\mathbb{Z}_{N}) \xrightarrow{\cong} H^3(M;\mathbb{Z}_{N})$ the cup product $B \cup B$  maps to the trivial element, and thus the phase factor $\text{exp}\left(-\frac{2\pi i}{N} \int_{X:~x>0} \frac{N+1}{2} B \cup B \right)$ in such a topology is actually trivial.
Put differently, for odd $N$,  the map \eqref{Eq:map} 
takes every element to the trivial element in $H^3 (B\mathbb{Z}_N; U(1))$. 
The map is again given by the dimensional reduction of a 3+1d $\mathbb{Z}_N^{(1)}$-SPT on a slab with the Dirichlet boundary condition which gives us a 2+1d $\mathbb{Z}_N^{(0)}$-SPT.
Therefore, we get $\overline{\mathcal{D}}_3 (M) \times \eta(\Sigma) = \overline{\mathcal{D}}_3 (M)$.
Taking the orientation reversal, we obtain $\eta(\Sigma) \times \mathcal{D}_3 (M) = \mathcal{D}_3 (M)$.
Combing the above, we arrive at the fusion rules (\ref{Eq:eta_fusion_triality_odd_N}).

\section{Dynamical Consequences}\label{sec:dynamical}

One of the key applications of global symmetries is to constrain renormalization group flows via anomaly matching.  In this section we investigate an analog of this question for the non-invertible symmetries.  Specifically, we determine whether there exist symmetry protected topological phases that are invariant under gauging either $\mathbb{Z}_N^{(1)}$ or $\mathbb{Z}_N^{(1)}\times\mathbb{Z}_N^{(1)}$ one-form symmetries, with local counterterms. If there does not exist such an SPT phase, then any theory invariant under gauging the symmetries are necessarily non-trivial at all energy scales, i.e.\ cannot be trivially gapped. 

Our analysis in this section apply not only to the duality and triality defects discussed in the earlier sections associated with a $\mathbb{Z}_N^{(1)}$ one-form symmetry, but also to more general $N$-ality defects and defects associated with a $\mathbb{Z}_N^{(1)}\times\mathbb{Z}_N^{(1)}$ symmetry. 

\subsection{Gauging $\mathbb{Z}_N^{(1)}$ One-Form Symmetry}

\subsubsection{Even $N$}

The most general $\mathbb{Z}_N^{(1)}$ one-form symmetric SPT phase $\mathcal{Q}_{p'}$ takes the form:
\begin{equation}\label{genspteven}
Z_{\mathcal{Q}_{p'}}=\exp\left(\frac{2\pi i p'}{2N} \int_{X} q(B)\right)~.
\end{equation}
Here $q$ is the Pontryagin square operation introduced below equation \eqref{eq:S_and_T_operations}, while $p'$ characterizes the SPT phase.  For bosonic theories $p'\in \mathbb{Z}_{2N}$, while for fermionic theories (formulated on spin manifolds) $q(B)$ is even and the phase is characterized  by $p'\in \mathbb{Z}_{N}$.

Now we consider gauging the one-form symmetry in the SPT \eqref{genspteven}.  Prior to gauging we add to the action an additional local counterterm $p$.  In the language of the $S$ and $T$ operations discussed in \eqref{eq:S_and_T_operations} we are thus considering the operation $S T^{p}(\mathcal{Q}_{p'})$.  We ask when the theory is left invariant.  The special case $p=0$ corresponds to the existence of a duality defect as discussed in \cite{Choi:2021kmx} while the case $p=1$ would imply the existence of a triality defect.

After gauging, the theory becomes
\begin{equation}
Z_{S T^{p}\mathcal{Q}_{p'}}= \frac{1}{|H^2(X;\mathbb{Z}_N)|^{1/2}}\sum_{b\in H^2(X;\mathbb{Z}_N)}\exp\left(\frac{2\pi i(p'+p)}{2N}\int_{X} q(b)+\frac{2\pi i}{N}\int_{X} b\cup B\right)~.
\end{equation}
If $\gcd(p+p',N)=1$, then the equation of motion of $b$ trivializes the dynamical gauge field and the theory becomes invertible.  Integrating out then gives a gravitational term and a new SPT
\begin{equation}
Z_{S T^{p}\mathcal{Q}_{p'}}=\exp\left(-\frac{2\pi i(p+p')^{-1}}{2N}\int_{X} q(B)\right )~.
\end{equation}
In other words under the condition that $\gcd(p+p',N)=1$ we have derived the relationship:
\begin{equation}
S T^{p}\mathcal{Q}_{p'} \cong \mathcal{Q}_{-(p+p')^{-1}}~.
\end{equation}

Thus, for the bosonic SPT $\mathcal{Q}_{p'}$ to be invariant under $S T^{p},$ we require
\begin{equation}
p'(p+p')=-1\text{ mod }2N~.
\end{equation}
When $p$ is odd, the left hand side is always even and thus the equation does not have a solution.
Taking mod 4 on both sides and use $p'^2=0,1$ mod 4 for even/odd $p'$ we find that there can be solution only if $p'$ is odd and $p=2$ mod 4.
In the cases where there is no solution, there is correspondingly no possible SPT phase realizing these defects.  Hence we have proved the following theorem.

\paragraph{Theorem} Let $\cal{Q}$ be bosonic 3+1d QFT with a $\mathbb{Z}_N^{(1)}$ one-form symmetry for even $N$.  If $\cal{Q}$ is invariant under $S T^{p}$ with $p\neq 2$ mod $4$ then $\cal{Q}$ cannot flow to a gapped phase with a unique vacuum.  In particular, this includes the case of a theory with a triality defect $p=1.$
\\

For fermionic theories the condition that the SPT phase $\mathcal{Q}_{p'}$ is invariant under $S T^{p},$ is instead
\begin{equation}
p'(p+p')=-1\text{ mod }N~.
\end{equation}
For odd $p$ the left hand side is even, and thus the equation does not have any solution for $p'$.  Meanwhile, when $N=0$ mod 4, the same analysis as above shows that for even $p\neq 2$ mod 4 there also does not exist any solution for $p'$ and hence no possible SPT.  Thus in this case we conclude the following. 

\paragraph{Theorem} Let $\cal{Q}$ be a fermionic 3+1d QFT with a $\mathbb{Z}_N^{(1)}$ one-form symmetry for even $N$. If $\cal{Q}$ is invariant under $S T^{p}$ with 
$p$ odd, or for $N=0$ mod 4 and $p\neq 2$ mod 4, then $\cal{Q}$ cannot flow to a gapped phase with a unique vacuum.  In particular, this includes the case of a theory with a triality defect $p=1.$

\subsubsection{Odd $N$}

In this case the most general $\mathbb{Z}_N^{(1)}$ one-form symmetric SPT phase takes the form:
\begin{equation}\label{gensptodd}
Z_{\mathcal{Q}_{p'}}=\exp\left(\frac{2\pi i p'}{N} \int_{X} B\cup B\right)~,
\end{equation}
where now $p' \in \mathbb{Z}_{N}$.  We consider the action of the operation $ST^{2p}$ on the above.  This yields:
\begin{equation}
Z_{S T^{2p}\mathcal{Q}_{p'}}= \frac{1}{|H^2(X;\mathbb{Z}_N)|^{1/2}}\sum_{b\in H^2(X;\mathbb{Z}_N)}\exp\left(\frac{2\pi i (p'+p)}{N}\int_{X} b\cup b+\frac{2\pi i}{N}\int_{X} b\cup B\right)~.
\end{equation}
The resulting theory is invertible if $\gcd(p+p',N)=1$.  In that case, integrating out $b$ gives a gravitational term and
\begin{equation}
Z_{S T^{2p}\mathcal{Q}_{p'}}=\exp\left(-2\pi i \frac{2^{-1}(2(p+p'))^{-1}}{N}\int B\cup B\right)~,
\end{equation}
where $2^{-1},(2(p+p'))^{-1}$ are well defined in $\mathbb{Z}_N$.  Hence under these conditions we determine:
\begin{equation}
S T^{2p}\mathcal{Q}_{p'} \cong \mathcal{Q}_{-2^{-1}(2(p+p'))^{-1}}~.
\end{equation}
Thus for the theory to be invariant under gauging the one-form symmetry, we require
\begin{equation}\label{eqn:SPToddNZN}
4p'(p+p')=-1\text{ mod }N~.
\end{equation}

The case $p=0$ of the equation above was derived in \cite{Choi:2021kmx} and corresponds to the existence of a duality defect.  Next, let us take $p=1$. Then the condition is $(2p'+1)^2=0$ mod $N$. The equation always has a solution $p'=(N-1)/2$. Thus a quantum system invariant under gauging $\mathbb{Z}_N^{(1)}$ one-form symmetry with odd $N$ and additional local counterterm $p=1$ can flow to a trivially gapped phase.

Finally, let us take $p=\frac{N+1}{2}$, which corresponds to the triality defect. 
The equation (\ref{eqn:SPToddNZN}) reduces to $m^2+m+1=0$ mod $N$, where we denote $m=2p'$. 
Multiplying the equation by 4, we get $(2m+1)^2=-3$ mod $N$.
The solution exists if and only if $(-3)$ is a quadratic residue of $N$ i.e. there is a $y$ such that $y^2=-3$ mod $N$. The solution is $p'=4^{-1}(-1\pm y)$ mod $N$. Since $\gcd(p',N)=1$, at least one of $y+1,y-1$ must be coprime with $N$.
Hence, the solution exists if and only if $N$ lives in the set
\begin{equation}\label{eqn:solsetX}
{\cal X}\equiv\{N:\, \exists y\in\mathbb{N}, y^2 =-3\text{ mod }N,\; \gcd(y+1,y-1,N)=1\}
=
\{3,7,13,19,21,\cdots\}~.
\end{equation}
When $N$ does not belong to the set ${\cal X}$, such as when $(-3)$ is not a quadratic residue of $N$, the triality defect cannot be realized by an invertible phase. 
For $(-3)$ to be a quadratic residue of $N$, the prime factorization of $N$ contains $3^r$ with $r=0,1$, and all other prime factors must be one modulo $3$. 
Thus we have the following theorem:

\paragraph{Theorem} Let $\cal{Q}$ be a  3+1d QFT with a $\mathbb{Z}_N^{(1)}$ one-form symmetry for odd $N$.  If $\cal{Q}$ is invariant under $S T$, i.e.\ if $\cal{Q}$ has the triality defect, and one of the following conditions is true:
\begin{itemize}
\item if $3|N$, and there exits a prime factor of $N/3$ that is not one modulo 3~,
\item if $3\nmid N$, and there exists a prime factor of $N$ that is not one modulo 3~,
\end{itemize}
then $\cal{Q}$ cannot flow to a gapped phase with a unique vacuum.
\\

In fact, the theorem also applies to the previous case when $N$ is even, where $N$ (and $N/3$ if $3|N$) contain the prime factor 2 that is not one modulo 3.

\subsection{Gauging $\mathbb{Z}_N^{(1)}\times\mathbb{Z}_N^{(1)}$ One-Form Symmetry}\label{sec:znzn}

Although our analysis in previous sections has focused on defects that arise in theories with $\mathbb{Z}_{N}^{(1)}$ symmetry, there are also interesting non-invertible defects that arise in theories with product group one form symmetry.  Here we focus on a particular triality defect that arises in theories with $\mathbb{Z}_N^{(1)}\times\mathbb{Z}_N^{(1)}$ one-form symmetry leaving more complete investigations for the future.  We will see a realization of this defect in section \ref{secso8}.

For a theory $\mathcal{Q}$ with $\mathbb{Z}_N^{(1)}\times\mathbb{Z}_N^{(1)}$ symmetry we can define a new theory by a twisted gauging with an off-diagonal counterterm.  We denote this operation as $ST_{12}$:
\begin{equation}
Z_{ST_{12}{\cal Q}}[B_{1},B_{2}]=\frac{1}{|H^2(X;\mathbb{Z}_N)|}\sum_{b_{1}, b_{2}\in H^2(X;\mathbb{Z}_N)}
Z_{\cal Q}[b_1,b_2]
\text{exp}\left(\frac{2\pi i}{N} \int_X b_{1}\cup b_{2}+b_{1} \cup B_{2}+b_{2} \cup B_{1} \right) ~.
\end{equation}
Note that the operation $ST_{12}$ is order three.  Indeed, neglecting the overall normalization for notational convenience, the partition function of the theory $(ST_{12})^{3}{\cal Q}$ is:
\ie
 \,&Z_{(ST_{12})^{3} \cal{Q}}[B_{1}, B_{2}] = \sum_{a_{i}, b_{i}, c_{i}}Z_{Q}[a_1,a_2]\exp\left(\frac{2\pi i}{N}\int_{X} \left( a_1\cup a_2+ a_1\cup b_2+a_2\cup b_1\right) \right.\\
& +  \left. \left( b_1\cup b_2+ b_1\cup c_2+b_2\cup c_1\right) +\left( c_1\cup c_2+ c_1\cup B_2+c_2\cup B_1\right) \phantom{\int}\hspace{-.2in}\right) \nonumber \\
&= \sum_{a_i,b_i}Z_{{\cal Q}}[a_1,a_2] e^{{2\pi i\over N}\int_X (a_1-B_1)\cup b_2 +(a_2-B_2)\cup b_1+a_1\cup a_2-B_1\cup B_2} \left(\sum_{c_i'\equiv c_i+b_i+B_i} e^{{2\pi i\over N}\int_X c_1'\cup c_2'}\right)\cr
& = Z_{\cal{Q}}[B_{1},B_{2}]~.
\fe
Thus, as in our general analysis in section \ref{Sec:define_duality_defects}, a theory $\cal{Q}$ invariant under the operation $ST_{12}$ admits a triality defect.

Motivated by this result, as a final example of anomaly type analysis, let us investigate invertible phases invariant under gauging $\mathbb{Z}_N^{(1)}\times\mathbb{Z}_N^{(1)}$ one-form symmetry. We will focus on the case of even $N$.

A general bosonic invertible phase with $\mathbb{Z}_N^{(1)}\times\mathbb{Z}_N^{(1)}$ one-form symmetry can be labeled by $p_1,p_2\in\mathbb{Z}_{2N}$ as before, and $p_{12}\in\mathbb{Z}_N$ for the mixed term.  It takes the general form:
\begin{equation}
Z_{\mathcal{Q}_{p_{1}, p_{2}, p_{12}}}=\exp\left(\frac{2\pi i p_{1}}{2N} \int_{X} q(B_{1})+\frac{2\pi i p_{2}}{2N} \int_{X} q(B_{2})+\frac{2\pi i p_{12}}{N} \int_{X} B_{1}\cup B_{2}\right)~.
\end{equation}

Let us gauge the one-form symmetry in the invertible phase labeled by $p_1',p_2',p_{12}'$ with additional local counterterms with coefficients $p_1,p_2,p_{12}$. The resulting theory has the action:\footnote{
Gauging the $\mathbb{Z}_N^{(1)}\times\mathbb{Z}_N^{(1)}$ one-form symmetry produces a dual $\mathbb{Z}_N^{(1)}\times\mathbb{Z}_N^{(1)}$ one-form symmetry, and we turn on background $B^1,B^2$. In the following we choose the generators for the dual one-form symmetry to be $\oint b^2,\oint b^1$, but the conclusion does not change if we choose the generators to be $\oint b^1,\oint b^2$.  This corresponds to choosing a different bicharacter under the gauging (see e.g.\ the discussion below \eqref{Eq:invariance_under_S}). Our results below imply that the triality defect arising from invariance under twisted gauging with this different bicharacter also obstructs a trivially gapped phase.}
\begin{equation}
 2\pi i\int_{X} \frac{p_1'+p_1}{2N}q(b^1)+\frac{p_2'+p_2}{2N}q(b^2)+\frac{p_{12}'+p_{12}}{N}b^1\cup b^2+\frac{1}{N}b^1\cup B^2+\frac{1}{N}b^2\cup B^1~,
\end{equation}
where $b^i$ are dynamical two-form $\mathbb{Z}_N$ gauge fields and $B^i$ are background two-form gauge field for the dual $\mathbb{Z}_N^{(1)}\times\mathbb{Z}_N^{(1)}$ one-form symmetry. The equations of motion for fields $b^1$ and $b^{2}$ give
\begin{equation}
(p_1+p_1')b^1+(p_{12}+p_{12}')b^2+B^2=0~,\quad (p_{12}+p_{12}')b^1+(p_{2}+p_2')b^2+B^1=0~.
\end{equation}
For the theory to be invertible after gauging the symmetry, the dynamical gauge fields $b^i$ need to be fixed by the background gauge field $B^i$, and this requires $\gcd(P,N)=1$, where
\begin{equation}
P=(p_1+p_1')(p_2+p_2')-(p_{12}+p_{12}')^2~.
\end{equation}
When this holds, we can integrate out $b^i$, which gives a decoupled gravitational term, and the a new SPT with partition function
\begin{equation}
Z=\exp\left(2\pi i \int_{X} \frac{-(p_2+p_2')/P}{2N}q(B^{2})+\frac{-(p_1+p_1')/P}{2N}q(B^1)
+\frac{(p_{12}'+p_{12})/P}{N}B^1\cup B^2\right)~.
\end{equation}
Thus the condition that the theory is invariant under gauging the symmetry is
\begin{equation}
-(p_2+p_2')=p_2'P\text{ mod }2N~,\quad -(p_1+p_1')=p_1'P\text{ mod }2N~,\quad 
p_{12}+p_{12}'=p_{12}'P\text{ mod }N~.
\end{equation}
(When the theory is fermionic, the first two equations are mod $N$ instead of mod $2N$, while the last equation remains the same.)

Let us focus on the case with odd $p_{12}$. This includes $p_1=p_2=0,p_{12}=1$ corresponding to the operation $ST_{12}$ studied above. 
In the last equation $p_{12}'P=p_{12}'+p_{12}$ mod $N$, $P$ is odd (since $\gcd(P,N)=1$ and $N$ is even), and thus the two sides of the equation have different even/odd parity, and the equation does not have a solution. The conclusion is the same also for fermionic theories, since the last equation is the same for bosonic and fermionic theories.  Therefore we have proved the following.

\paragraph{Theorem}
Let $\cal{Q}$ be a  3+1d QFT with a $\mathbb{Z}_N^{(1)}\times\mathbb{Z}_N^{(1)}$ one-form symmetry for even $N$.  If $\cal{Q}$ is invariant under gauging $\mathbb{Z}_N^{(1)}\times\mathbb{Z}_N^{(1)}$ with an additional counterterm with odd $p_{12}$, then $\cal{Q}$ cannot flow to a gapped phase with a unique ground state.  In particular this includes the case of a triality defect arising from invariance under $ST_{12}$.

\section{Examples}\label{sec:example}

In this section, we discuss concrete examples of the various non-invertible defects defined in Section \ref{Sec:define_defects}. Each example provides an independent check to the fusion rule derived in Section \ref{Sec:fusion_codim_one} and \ref{Sec:fusion_involving_eta}.

\subsection{3+1d Maxwell Theory}\label{sec:Maxwell}

All the defects defined in Section \ref{Sec:define_defects} can be realized explicitly in the 3+1d Maxwell theory:
\ie
\mathcal{L}_{\text{bulk}}[A]=\frac{1}{2e^2}F\wedge \star F+\frac{i\theta}{8\pi^2}F\wedge F~,
\fe
where $A$ is the dynamical bulk $U(1)$ one-form gauge field and $F=dA$ is the field strength.\footnote{In contrast to the convention elsewhere in this paper (see footnote \ref{fn:case}), in this subsection we use upper case letters for the dynamical bulk gauge fields, and lower case letters for dynamical gauge fields living on the defects. }
The theory is parametrized by a complex coupling
\ie
\tau=\frac{\theta}{2\pi}+\frac{2\pi i}{e^2}~.
\fe
It has two $U(1)^{(1)}$ one-form symmetries \cite{Gaiotto:2014kfa}:
\begin{itemize}
\item
An electric $U(1)^{(1)}$ one-form symmetry. The symmetry shifts the $U(1)$ gauge field $A$ by a flat connection. The charged operators are the Wilson lines $W_E=\exp(i\oint A)$, and the charge is $Q_E=\frac{1}{2\pi}\oint(-\frac{2\pi i}{e^2} \star F+\frac{\theta}{2\pi} F)$.

\item
A magnetic $U(1)^{(1)}$ one-form symmetry. The charged operators are the 't Hooft lines $W_M$, and the charge is $Q_M=\frac{1}{2\pi}\oint F$.

\end{itemize}
Both symmetries are anomaly free but there is a mixed anomaly between them.

The theory enjoys an electromagnetic duality, which will be crucial for establishing the duality and triality defects. The duality group depends on which manifolds the theory is placed on. 

On spin manifolds, the duality group is $SL(2,\mathbb{Z})$. It is generated by the $\mathbb{S}$ and $\mathbb{T}$ duality\footnote{We use a different font here to distinguish between $\mathbb{S},\mathbb{T}$ electromagnetic duality of the Maxwell theory  and the $S,T$ operations defined in \eqref{eq:S_and_T_operations}.}
\ie
&\mathbb{S}:\quad \tau\rightarrow -1/\tau~,
\\
&\mathbb{T}:\quad \tau\rightarrow \tau+1~.
\fe
They obey the relation
\ie
\mathbb{S}^2=\mathbb{C}~,\quad (\mathbb{S}\mathbb{T})^3=1~,
\fe
where $\mathbb{C}$ is the charge conjugation symmetry: $A\rightarrow -A$. 

The $\mathbb{S}$ duality relates the dynamical gauge field $A$ and its dual gauge field $\tilde A$. Their field strength are related by
\ie
\tilde F=\frac{2\pi i}{e^2} \star F-\frac{\theta}{2\pi}F~.
\fe
Under the duality, the Wilson line $W_E=\exp(i\oint A)$ and 't Hooft line $W_M=\exp(i\oint \tilde A)$ transform as
\ie
(\widetilde W_E,\widetilde W_M)=(W_M, W_E^\dagger)~.
\fe
The electric and magnetic charges transform as
\ie
(\widetilde Q_E,\widetilde Q_M)=(Q_M,-Q_E)~.
\fe
Here, we use variables with a tilde to denote the observables in the $\mathbb{S}$-dual frame.

Under the $\mathbb{T}$ transformation, the Wilson and 't Hooft lines transform as
\ie
(W_E',W_M')=(W_E,W_MW_E^\dag)~.
\fe
The electric and magnetic charges transform as
\ie
(Q_E',Q_M')=(Q_E+Q_M,Q_M)~.
\fe
Here, we use variables with a prime to denote the observables in the duality frame after the $\mathbb{T}$ transformation.

On non-spin manifolds, $\mathbb{T}$ transformation is no longer a duality due to the half-instantons. The duality group is instead generated by $\mathbb{S}$ and $\mathbb{T}^2$ transformations. It is sometimes denoted as $\Gamma_\theta$.

In the rest of this subsection, we will discuss how various defects defined in Section \ref{Sec:define_defects} are realized in the Maxwell theory, and provide explicit worldvolume Lagrangian descriptions for all of them. We will denote a codimension-one defect $\mathcal{N}$ by its worldvolume Lagrangian $\mathcal{L}_{\mathcal{N}}[A_L,A_R]$, where $A_L, A_R$ are the restriction of the bulk $U(1)$ gauge fields from the left and right, respectively.\footnote{Generally, the worldvolume Lagrangian ${\cal L}_{\cal N}[A_L,A_R]$ also depend on other dynamical fields that only live on the defect. We will not write the dependence on the defect fields explicitly.}

Using the worldvolume Lagrangian, we can explicitly derive the fusion rules between the codimension-one defects. 
We prepare two parallel codimension-one defects $\cal N$ and ${\cal N}'$ at $x=0$ and $x=\epsilon$, respectively, and bring them close to each other. 
The action for this configuration is given by
\ie
&\int_{x<0}  {\cal L}_{\text{bulk}}[A_L] +\int_{0<x<\epsilon}  {\cal L}_{\text{bulk}}[A_I]+\int_{x>0}  {\cal L}_{\text{bulk}}[A_R]\\
&+\int_{x=0} {\cal L}_{\cal N}[A_L,A_I]+\int_{x=\epsilon} {\cal L}_{{\cal N}'}[A_I,A_R]\,,
\fe
where $A_L, A_I, A_R$ are the dynamical bulk one-form gauge fields living in the regions $x<0$, $0<x<\epsilon$, and $x>0$, respectively. 
As we bring the two defects close to each other, i.e., $\epsilon\to0$, the field $A_I$ becomes a defect field that only lives on $x=0$. The worldvolume Lagrangian of the fused defect is then 
\ie
\mathcal{N}\times\mathcal{N}': ~\mathcal{L}_{\mathcal{N}}[A_L,A_I]+\mathcal{L}_{\mathcal{N}'}[A_I,A_R]\,.
\fe

The worldvolume Lagrangian also allows us to derive the fusion between the codimension-one defects $\mathcal{N}(M)$ and the $\mathbb{Z}_N^{(1)}$ one-form symmetry defect $\eta(\Sigma)$. Fusing $\eta(\Sigma)$ with $\mathcal{N}(M)$ from the left/right amounts to shifting $A_L/A_R$ in the worldvolume Lagrangian $\mathcal{L}_\mathcal{N}[A_L,A_R]$ of $\mathcal{N}(M)$ by $\text{PD}(\Sigma)$, the Poincar\'e dual of $\Sigma$ in $M$.

Some of the codimension-one defects of the Maxwell theory can be described by 
\begin{itemize}
\item On the defect, there is a $U(1)_L\times U(1)_R$ gauge symmetry whose gauge fields are $A_L,A_R$. This gauge group is Higgsed to a subgroup, which can be thought of as higher condensation \cite{Kong:2014qka,Else:2017yqj,Gaiotto:2019xmp,Kong:2020cie,Johnson-Freyd:2020twl} or higher gauging \cite{Roumpedakis:2022aik}. 
\item A Chern-Simons term for the unbroken gauge group on the defect.
Since the bulk lines can move to the wall, the Chern-Simons terms need to be compatible with the   coupling to the bulk fields.
\end{itemize}
In this section, we will describe condensation defect where the diagonal $U(1)_L\times U(1)_R$ is broken to $\mathbb{Z}_N$, and the duality and triality defects with unbroken $U(1)_L\times U(1)_R$.  

\subsubsection{Condensation Defects}

For any $\tau$, we can define condensation defect $\mathcal{C}_k$ of \eqref{Eq:condensation_defect_general} using the $\mathbb{Z}_N^{(1)}$ subgroup of the $U(1)^{(1)}$ electric one-form symmetry. It can be realized explicitly by the following worldvolume Lagrangian
\ie\label{eq:Maxwell_condensation}
\mathcal{C}_k:\quad \frac{iN}{2\pi}ad(A_L-A_R)+\frac{ik}{2\pi}(A_L-A_R)d(A_L-A_R)~.
\fe
where $a$ is a $U(1)$ gauge field on the worldvolume $M$.

The worldvolume Lagrangian \eqref{eq:Maxwell_condensation} can be understood as follows. Integrating out $a$ constrains $A_L-A_R$ to be a $\mathbb{Z}_N$ gauge field on $M$. It is equivalent to summing over all possible insertions of the electric $\mathbb{Z}_N^{(1)}$ one-form symmetry defect on $M$. This is because inserting a symmetry defect $\eta(\Sigma)$ on $M$ induces a discontinuity $A_L-A_R$ given by $\text{PD}(\Sigma)\in H^1(M,\mathbb{Z}_N)$. The second term in \eqref{eq:Maxwell_condensation} generates a phase in the summation given by $e^{\frac{2\pi i k}{N}Q(M,\Sigma)}$ where $Q(M,\Sigma)=\frac{1}{N}\text{PD}(\Sigma)\cup  \delta \text{PD}(\Sigma)$. This sum over symmetry defects is precisely the condensation defect $\mathcal{C}_k(M)$.

Following similar steps in \cite{Banks:2010zn,Kapustin:2014gua}, the worldvolume Lagrangian  \eqref{eq:Maxwell_condensation} can be dualized to the following Higgs presentation
\ie
\mathcal{C}_k:\quad \frac{i}{2\pi}
H\left[ N(A_L-A_R)-d\phi\right]+\frac{ik}{2\pi}(A_L-A_R)d(A_L-A_R)~,
\fe
where $H$ is Lagrangian multiplier two-form field and $\phi$ is a scalar on $M$.
 The equation of motion of $H$ constrains $N(A_L-A_R)=d\phi$, which means that the $U(1)_L\times U(1)_R$ gauge symmetry  is Higgsed to a $\mathbb{Z}_N$ subgroup on $M$. 
 This can be thought of as a Higgs mechanism  on the defect, where $\phi$ is the angular part of a complex scalar field that condenses. 
 This justifies the name ``codensation defects." 
 See \cite{Roumpedakis:2022aik} for more examples of Higgs Lagrangians for the condensation defects in 2+1d.

\paragraph{Fusion of condensation defects}  We will focus on the fusions involving $\mathcal{C}_{N\ell\over2}$ with $\ell=0$ if $N$ is odd and $\ell=0,1$ mod 2 if $N$ is even. 
These condensation defects are distinguished in that they are orientation reversal invariant and they participate in the duality \eqref{Eq:fusion_duality_summary} and the triality fusion rules \eqref{Eq:fusion_triality_summary} and \eqref{Eq:fusion_triality_summary_odd_N}. 
Fusing two such condensation defects gives the following worldvolume Lagrangian:
\ie
\mathcal{C}_{\frac{N\ell_1}{2}}\times \mathcal{C}_{\frac{N\ell_2}{2}}:~
&\frac{iN}{2\pi}a_1d(A_L-A_I)+\frac{iN\ell_1}{4\pi}(A_L-A_I)d(A_L-A_I)\\
&+\frac{iN}{2\pi}a_2d(A_I-A_R)+\frac{iN\ell_2}{4\pi}(A_I-A_R)d(A_I-A_R)~.
\fe
If we redefine $a\equiv a_2+\ell_2(A_I-A_L)$, $x\equiv A_I-A_L$, $y\equiv a_2-a_1$, the expression becomes
\ie
 \left(\frac{iN}{2\pi}xdy+\frac{iN(\ell_1+\ell_2)}{4\pi} xdx\right)+ \left(\frac{iN}{2\pi}ad(A_L-A_R)+\frac{iN\ell_2}{4\pi}(A_L-A_R)d(A_L-A_R)\right)~.
\fe
The first term is a decoupled $(\mathcal{Z}_N)_{N(\ell_1+\ell_2)}$ theory and the second term is the $\mathcal{C}_{\frac{N \ell_2}{2}}$ defect. Thus, 
\ie
\mathcal{C}_{\frac{N\ell_1}{2}}\times \mathcal{C}_{\frac{N\ell_2}{2}}=(\mathcal{Z}_N)_{N(\ell_1+\ell_2)}\mathcal{C}_{\frac{N \ell_2}{2}}~.
\fe
We can also redefine $a\equiv a_1+\ell_1(A_I-A_R)$, $x\equiv A_I-A_R$, $y\equiv a_2-a_1-2\ell_1(A_L-A_R)$, and the expression becomes
\ie
 \left(\frac{iN}{2\pi}xdy+\frac{iN(\ell_1+\ell_2)}{4\pi} xdx\right)+ \left(\frac{iN}{2\pi}ad(A_L-A_R)+\frac{iN\ell_1}{4\pi}(A_L-A_R)d(A_L-A_R)\right)~.
\fe
The first term is a decoupled $(\mathcal{Z}_N)_{N(\ell_1+\ell_2)}$ theory and the second term is the $\mathcal{C}_{\frac{N \ell_2}{2}}$ defect. Thus, 
\ie
\mathcal{C}_{\frac{N\ell_1}{2}}\times \mathcal{C}_{\frac{N\ell_1}{2}}=(\mathcal{Z}_N)_{N(\ell_1+\ell_2)}\mathcal{C}_{\frac{N \ell_1}{2}}~.
\fe
Both fusion rules are consistent with \eqref{Eq:fusion_duality_summary}, \eqref{Eq:fusion_triality_summary}, and \eqref{Eq:fusion_triality_summary_odd_N}. 

\paragraph{Fusion of condensation defects and symmetry defects} Fusing a symmetry defect $\eta(\Sigma)$ with a condensation defect $\mathcal{C}_k(M)$ from the left changes the worldvolume Lagrangian of $\mathcal{C}_k(M)$ from \eqref{eq:Maxwell_condensation} to
\ie
\frac{iN}{2\pi}ad\left(A_L-A_R+\frac{2\pi}{N}\text{PD}(\Sigma)\right)+\frac{ik}{2\pi}\left(A_L-A_R+\frac{2\pi}{N}\text{PD}(\Sigma)\right)d\left(A_L-A_R+\frac{2\pi}{N}\text{PD}(\Sigma)\right)~.
\fe
Shifting $a\rightarrow a-\frac{4\pi k}{N^2}\text{PD}(\Sigma)$, the expression simplifies to
\ie
\left(\frac{iN}{2\pi}ad\left(A_L-A_R\right)+\frac{ik}{2\pi}\left(A_L-A_R\right)d\left(A_L-A_R\right)\right)+\left(\frac{2\pi i k}{N}Q(M,\Sigma)\right)~.
\fe
Here we have used the $da \,\text{PD}(\Sigma) = 0$ mod $2\pi$.\footnote{Here we use a mixed notation where $a,A_L,A_R$ are represented as $U(1)$ gauge fields and PD$(\Sigma)$ is a $\mathbb{Z}_N$ gauge field. We hope this will not cause too much confusion.} 
The first term is the condensation defect $\mathcal{C}_k(M)$ and the second term is a decoupled 2+1d $\mathbb{Z}_N^{(0)}$-SPT. Thus, \ie
\eta(\Sigma)\times \mathcal{C}_k(M)=e^{\frac{2\pi ik}{N}Q(\Sigma,M)}\mathcal{C}_k(M)~.
\fe 
Similarly, we also have $\mathcal{C}_k(M)\times \eta(\Sigma)=e^{\frac{2\pi ik}{N}Q(\Sigma,M)}\mathcal{C}_k(M)$. The fusion rules are consistent with \eqref{Eq:fusion_eta_triality}.

\subsubsection{Duality Defects} 
The Maxwell theory admits duality defect at \cite{Choi:2021kmx}: 
\ie
\tau=iN~.
\fe
At this coupling, the theory is invariant under the $S$ gauging of the electric $\mathbb{Z}_N^{(1)}$ one-form symmetry:
\ie
Z_{\tau =iN}[B]=\frac{1}{|H^{2}(X;\mathbb{Z}_N)|^{1/2}}\sum_{b\in H^{2}(X;\mathbb{Z}_N)} Z_{\tau=iN}[b]e^{\frac{2\pi i}{N}\int_X b\cup B}~,
\fe
where $B$ is the background gauge field for the $\mathbb{Z}_N^{(1)}$ one-form symmetry.
Gauging the symmetry replaces $A$ by $A/N$, which is equivalent to changing the coupling from $\tau=iN$ to $\tau=i/N$. It also couples $B$ to $Q_M$, the charge of the dual $\mathbb{Z}_N^{(1)}$ magnetic one-form symmetry. Using the $\mathbb{S}$ duality, we recover the original theory at $\tau=iN$ and maps $Q_M$ to $Q_E$. This ensures that $B$ continues coupling to the electric $\mathbb{Z}_N^{(1)}$ one-form symmetry.

The duality defect $\mathcal{D}_2$ at $\tau=iN$ can be realized explicitly by the following worldvolume Lagrangian 
\ie
\mathcal{D}_2:\quad \frac{iN}{2\pi}A_LdA_R~.
\fe
It can be understood by examining the equations of motion. 
The variation  of $A_L,A_R$ gives the following equation of motion on the defect
\ie
F_R=\frac{i}{N}\star(NF_L)~.
\fe
It can be interpreted as first gauging the electric $\mathbb{Z}_N^{(1)}$ one-form symmetry and then perform the $\mathbb{S}$ duality transformation. This is exactly the operations we discussed above. When $N=1$, the duality defect reduces to the $\mathbb{S}$ duality defect at the $\mathbb{S}$ invariant point $\tau=i$ in \cite{Gaiotto:2008ak,Kapustin:2009av}.

As we pull a Wilson line across the duality defect, it becomes an improperly quantized 't Hooft line attached to the surface defect $\eta$ \cite{Choi:2021kmx}.  
This reflects the fact that the duality defect ${\cal D}_2$ is non-invertible. 
If, however, we restrict the action to the local operators, then ${\cal D}_2$ is an order 4 operator. 
In particular, it acts on the field strength $F$ as:
\ie
F\rightarrow i\star F\rightarrow -F \rightarrow -i\star F\rightarrow F\rightarrow \cdots~.
\fe
As far as the local operators are concerned, the square of the generator is the charge conjugation symmetry that maps $F\rightarrow -F$. 
This is consistent with the expected fusion rule $\mathcal{D}_2\times \mathcal{D}_2=\mathcal{U}\times \mathcal{C}_0=\mathcal{C}_0\times \mathcal{U}$ \eqref{Eq:fusion_duality_summary}, where $\mathcal{U}$ is the charge conjugation defect.  
Note that since the condensation defect ${\cal C}_0$ is made out of surfaces, it is ``porous" to the local operators and act trivially on the latter \cite{Roumpedakis:2022aik}. 

 Following the definition in \eqref{Eq:general_orientation_reversal}, the worldvolume Lagrangian  for the orientation reversal $\overline{{\cal D}_2}$ of ${\cal D}_2$ is obtained by swapping $A_L$ and $A_R$ in that of $\mathcal{D}_2$ and adding an overall minus sign due to the change of the orientation. This gives
\ie
\overline{\mathcal{D}}_2:\quad -\frac{iN}{2\pi}A_LdA_R~.
\fe
It is related to the worldvolume Lagrangian of ${\cal D}_2$ by  applying the charge conjugation transformation to either $A_L$ or $A_R$ (but not both). Thus, we have $\overline{\mathcal{D}}_2=\mathcal{U}\times \mathcal{D}_2={\mathcal{D}}_2\times \mathcal{U}$.

\paragraph{Fusion of duality defects} Fusing $\mathcal{D}_2\times \overline{\mathcal{D}}_2$ gives
\ie
\mathcal{D}_2\times \overline{\mathcal{D}}_2:\quad \frac{iN}{2\pi}A_LdA_I-\frac{iN}{2\pi}A_I dA_R=\frac{iN}{2\pi} A_Id(A_L-A_R)~.
\fe
This is the condensation defect $\mathcal{C}_0$. Thus, $\mathcal{D}_2\times \overline{\mathcal{D}}_2=\mathcal{C}_0$. The other fusions involving $\mathcal{D}_2$, $\overline{\mathcal{D}}_2$ can be obtained by using the relation $\overline{\mathcal{D}}_2=\mathcal{U}\times \mathcal{D}_2={\mathcal{D}}_2\times \mathcal{U}$. These fusion rules are consistent with \eqref{Eq:fusion_duality_summary}.

\paragraph{Fusion of duality defects and symmetry defects} Fusing a symmetry defect $\eta(\Sigma)$ with a duality defect $\mathcal{D}_2(M)$ from the left/right leaves the worldvolume Lagrangian of $\mathcal{D}_2(M)$ invariant because  $dA_{R/L} \,\text{PD}(\Sigma) = 0$ mod $2\pi$. Thus, $\eta(\Sigma)\times \mathcal{D}_2(M)=\mathcal{D}_2(M)\times \eta(\Sigma)=\mathcal{D}_2(M)$. It is consistent with \eqref{Eq:fusion_eta_duality}.

\subsubsection{Triality Defect}
The Maxwell theory admits a triality defect at:
\ie
\tau=e^{2\pi i/3}N~.
\fe
At this coupling, the theory is invariant under the $ST$ gauging of the electric $\mathbb{Z}_N^{(1)}$ one-form symmetry. We will divide the discussion into the even and odd $N$ cases.

For even $N$,  we have
\ie
Z_{\tau =e^{2\pi i/3}N}[B]=\frac{1}{|H^{2}(X;\mathbb{Z}_N)|^{1/2}}\sum_{H^{2}(X;\mathbb{Z}_N)} Z_{\tau=e^{2\pi i/3}N}[b]\, e^{\frac{2\pi i}{N}\int_X\left(\frac{1}{2}q(b)+ b\cup B\right)}~,
\fe
where $B$ is the background gauge field for the $\mathbb{Z}_N^{(1)}$ one-form symmetry.
This equality holds on both spin and non-spin manifolds. Gauging the symmetry replaces $A$ by $A/N$ and add a $\theta$-term $\frac{2\pi}{N}\frac{ i}{8\pi^2}F\wedge F$, which is equivalent to changing the coupling from $\tau=e^{2\pi i/3}N$ to $\tau=e^{2\pi i/3}/N + 1/N=e^{\pi i/3}/N$. It also couples $B$ to $Q_M$, the charge of the dual $\mathbb{Z}_N^{(1)}$ magnetic one-form symmetry. Using the $\mathbb{S}$ duality, we recover the original theory at $\tau=e^{2\pi i/3}N$ and maps $Q_M$ to $Q_E$. This ensures that $B$ continues coupling to the electric $\mathbb{Z}_N^{(1)}$ one-form symmetry.

For odd $N$,  we have
\ie\label{eq:Maxwell_triality_condition}
Z_{\tau =e^{2\pi i/3}N}[B]=\frac{1}{|H^{2}(X,\mathbb{Z}_N)|^{1/2}}\sum_{H^{2}(X,\mathbb{Z}_N)} Z_{\tau=e^{2\pi i/3}N}[b]\,e^{\frac{2\pi i}{N}\int_X\left(\frac{N+1}{2}b\cup b+ b\cup B\right)}~,
\fe
where $B$ is the background gauge field for the $\mathbb{Z}_N^{(1)}$ one-form symmetry.
The equality holds only on spin manifolds as we will explain below. Gauging the symmetry replaces $A$ by $A/N$ and add a $\theta$-term $\frac{2\pi(N+1)}{N}\frac{ i}{8\pi^2}F\wedge F$, which is equivalent to changing the coupling from $\tau=e^{2\pi i/3}N$ to $\tau=e^{2\pi i/3}/N + (N+1)/N=1+e^{\pi i/3}/N$. It also couples $B$ to $Q_M$, the charge of the dual $\mathbb{Z}_N^{(1)}$ magnetic one-form symmetry. Using the $\mathbb{S}\mathbb{T}^{-1}$ duality, we recover the original theory at $\tau=e^{2\pi i/3}N$ and maps $Q_M$ to $Q_E$. This ensures that $B$ continues coupling to the electric $\mathbb{Z}_N^{(1)}$ one-form symmetry. Since the $\mathbb{T}$ duality is valid only on spin manifolds, the equality \eqref{eq:Maxwell_triality_condition} holds only on spin manifold.

The triality defect $\mathcal{D}_3$ at $\tau=e^{2\pi i/3}N$ can be realized explicitly by the following worldvolume Lagrangian 
\ie\label{Maxwell_triality_defect_Lag}
\mathcal{D}_3:\quad \frac{iN}{2\pi}A_LdA_R+\frac{iN}{4\pi}A_LdA_L~.
\fe
For odd $N$, the worldvolume Lagrangian has a diagonal Chern-Simons term with an odd level, which can defined only on spin manifolds. This reflects the fact that the equality \eqref{eq:Maxwell_triality_condition} holds only on spin manifolds. When $N=1$, the triality defect reduces to the $\mathbb{S}\mathbb{T}$ triality defect at the $\mathbb{ST}$ invariant point $\tau=e^{2\pi i/3}$ in \cite{Kapustin:2009av}. 

The variation  of $A_L,A_R$ gives the following equation of motion on the defect
\ie\label{eq:triality_field_strength}
F_R=\frac{\sqrt{3}i}{2N}\star (NF_L)-\frac{1}{2N}(NF_L)~,
\fe
or equivalently
\ie
F_L
=-\frac{\sqrt{3}i}{2N}\star(N F_R)-\frac{1}{2N}(NF_R)~.
\fe
It is consistent with the operations discussed above. We first perform the $ST$ operation with the electric $\mathbb{Z}_N^{(1)}$ one-form symmetry. It relates the field strength after gauging $F'$ to $F_L$ by $F'=NF_L$. Then we perform $\mathbb{S}$ or $\mathbb{S}\mathbb{T}^{-1}$ duality transformation depending on whether $N$ is even or odd. In both cases, the transformations on the field strength are the same as $\mathbb{S}$ duality transformation at $\tau=e^{2\pi i/3}/N$ since $\mathbb{T}$ acts trivially on the field strength. It maps $F'$ to $F_R=\frac{\sqrt{3}i}{2N}F'-\frac{1}{2N}F'$, which gives \eqref{eq:triality_field_strength}.

The triality defect acts invertibley on the local operators, such as the field strength $F$, as an order 3 operator:
\ie
F \rightarrow \frac{\sqrt{3}i}{2}\star F-\frac{1}{2}F \rightarrow -\frac{\sqrt{3}i}{2}\star F-\frac{1}{2}F\rightarrow F\rightarrow \cdots~.
\fe
It is consistent with the fusion rule $\mathcal{D}_3\times \mathcal{D}_3\times\mathcal{D}_3=U(1)_{N}\mathcal{C}_0=U(1)_{N}\mathcal{C}_{\frac{N}{2}}$ for even $N$ in \eqref{Eq:DDD_triality_fusion_rule} and $\mathcal{D}_3\times \mathcal{D}_3\times\mathcal{D}_3=SU(N)_{-1}\mathcal{C}_0$ for odd $N$ \eqref{Eq:DDD_fusion_odd_N}.  
Again, since the condensation defect is made out of surfaces, it acts trivially on the local operators. 

The worldvolume Lagrangian for the orientation reversal defect $\overline{\mathcal{D}}_3$ is obtained by swapping $A_L$ and $A_R$ in that of ${\mathcal{D}}_3$ and adding an overall minus sign due to the change of orientation, which gives
\ie
\overline{\mathcal{D}}_3:\quad - \frac{iN}{2\pi}A_LdA_R-\frac{iN}{4\pi}A_RdA_R~.
\fe

\paragraph{Fusion of triality defects} 
The worldvolume Lagrangian for the fused defect $ \overline{\mathcal{D}}_3\times \mathcal{D}_3$ is
\ie
\overline{\mathcal{D}}_3\times\mathcal{D}_3:\quad -\frac{iN}{2\pi}A_LdA_I-\frac{iN}{4\pi}A_IdA_I+\frac{iN}{2\pi}A_IdA_R+\frac{iN}{4\pi}A_RdA_R=\frac{iN}{2\pi}A_Id(A_R-A_L)~.
\fe
Redefining $A_I\rightarrow -A_I$, the expression becomes the worldvolume Lagrangian of the condensation defect $\mathcal{C}_0$. Thus, $\overline{\mathcal{D}}_3\times {\mathcal{D}}_3=\mathcal{C}_0$, which agrees with \eqref{Eq:fusion_triality_summary}.

Fusing $\mathcal{D}_3\times \overline{\mathcal{D}}_3$ gives
\ie
\mathcal{D}_3\times \overline{\mathcal{D}}_3:\quad \frac{iN}{2\pi}A_LdA_I+\frac{iN}{4\pi}A_LdA_L-\frac{iN}{2\pi}A_IdA_R-\frac{iN}{4\pi}A_RdA_R~.
\fe
Let $a\equiv A_I+A_R$, then the expression becomes 
\ie
\overline{\mathcal{D}}_3\times\mathcal{D}_3:\quad \frac{iN}{2\pi}ad(A_L-A_R)+\frac{iN}{4\pi}(A_L-A_R)d(A_L-A_R)~.
\fe
For even $N$, this is the condensation defect $\mathcal{C}_{\frac{N}{2}}$. Thus, ${\mathcal{D}}_3\times \overline{\mathcal{D}}_3=\mathcal{C}_{\frac{N}{2}}$. It agrees with \eqref{Eq:fusion_triality_summary}. For odd $N$, since the triality defects in the Maxwell theory are defined only on spin manifolds, we can use the equivalence $(\mathcal{Z}_N)_N\leftrightarrow (\mathcal{Z}_N)_0$ for spin TQFTs to simplify the worldvolume Lagrangian to
\ie
\overline{\mathcal{D}}_3\times\mathcal{D}_3:\quad \frac{iN}{2\pi}ad(A_L-A_R)~.
\fe
Thus, ${\mathcal{D}}_3\times \overline{\mathcal{D}}_3=\mathcal{C}_{0}$, which agrees with \eqref{Eq:fusion_triality_summary_odd_N}.

Fusing $ {\mathcal{D}}_3\times \mathcal{D}_3$ gives
\ie
{\mathcal{D}}_3\times\mathcal{D}_3:\quad \frac{iN}{2\pi}A_LdA_I+\frac{iN}{4\pi}A_LdA_L+\frac{iN}{2\pi}A_IdA_R+\frac{iN}{4\pi}A_IdA_I~.
\fe
Let $a\equiv A_I+A_L+A_R$, then the expression becomes 
\ie
{\mathcal{D}}_3\times\mathcal{D}_3:\quad \left(\frac{iN}{4\pi}ada\right)-
\left(\frac{iN}{2\pi}A_LdA_R+\frac{iN}{4\pi}A_RdA_R\right)~.
\fe
The first term is a decoupled $U(1)_{N}$ and the second term is $\overline{\mathcal{D}}_3$. Thus, $\mathcal{D}_3\times \mathcal{D}_3=U(1)_{N}\overline{\mathcal{D}}_3$. For even $N$, it agrees with \eqref{Eq:fusion_triality_summary}. For odd $N$, since the defects in the Maxwell theory are defined only on spin manifolds, using the level-rank duality $U(1)_N\leftrightarrow SU(N)_{-1}$ for spin TQFTs \cite{Hsin:2016blu}, we can also write the fusion rule as $\mathcal{D}_3\times \mathcal{D}_3=SU(N)_{-1}\overline{\mathcal{D}}_3$. It agrees with \eqref{Eq:fusion_triality_summary_odd_N} for odd $N$.

Using the above fusion, we can derive all the fusions involving $\mathcal{D}_3,\overline{\mathcal{D}}_3$ and $\mathcal{C}_0,\mathcal{C}_{\frac{N}{2}}$. They agree with  \eqref{Eq:fusion_triality_summary} and \eqref{Eq:fusion_triality_summary_odd_N}.  

For odd $N$,  while the triality defects in the Maxwell theory can only be realized on spin manifolds, the triality fusion rule \eqref{Eq:fusion_triality_summary_odd_N} applies more generally to any non-spin QFT that is invariant under the $ST$ gauging in the sense of \eqref{Eq:invariance_under_ST}.

\paragraph{Fusion of triality defects and symmetry defects} Fusing a symmetry defect $\eta(\Sigma)$ with a triality defect $\mathcal{D}_3(M)$ from the right leaves the worldvolume Lagrangian of $\mathcal{D}_3(M)$ invariant since $dA_{L} \, \text{PD}(\Sigma)=0$ mod $2\pi$. Thus, 
\ie\mathcal{D}_3(M)\times\eta(\Sigma)=\mathcal{D}_3(M)~.
\fe
It agrees with \eqref{Eq:fusion_eta_triality} and \eqref{Eq:eta_fusion_triality_odd_N}.

Fusing a symmetry defect $\eta(\Sigma)$ with a triality defect $\mathcal{D}_3(M)$ from the left changes the worldvolume Lagrangian of $\mathcal{D}_3(M)$ from \eqref{Maxwell_triality_defect_Lag} to
\ie
\frac{iN}{2\pi}\left(A_L+\frac{2\pi}{N}\text{PD}(\Sigma)\right)dA_R+\frac{iN}{4\pi} \left(A_L+\frac{2\pi}{N}\text{PD}(\Sigma)\right)d\left(A_L+\frac{2\pi}{N}\text{PD}(\Sigma)\right)~.
\fe
Again since $dA_{L/R} \, \text{PD}(\Sigma)=0$ mod $2\pi$, we end up with
\ie
\left(\frac{iN}{2\pi}A_LdA_R+\frac{iN}{4\pi}A_LdA_L\right)+i\pi  Q(M,\Sigma)~.
\fe
The first term is $\mathcal{D}_3$ and the second term is a decoupled SPT. For even $N$, the fusion rule is $\eta(\Sigma)\times \mathcal{D}_3(M)=(-1)^{Q(\Sigma,M)}\mathcal{D}_3(M)$. It agrees with \eqref{Eq:fusion_eta_triality}.

For odd $N$, since the triality defects in Maxwell theory can only be defined on spin manifolds, the last term is trivial (see, for example, (B.10) of \cite{Cordova:2017vab}). 
Therefore, for odd $N$,  fusion rule is $\eta(\Sigma)\times \mathcal{D}_3(M)=\mathcal{D}_3(M)$. It agrees with \eqref{Eq:eta_fusion_triality_odd_N}.

\subsection{$N$-ality Defects in ${\cal N}=1$ Super Yang-Mills Theory}

\paragraph{$SU(3)$ super Yang-Mills theory}

When a zero-form symmetry has mixed anomaly with one-form symmetry, gauging the one-form symmetry can sometimes extend the zero-form symmetry to become a non-invertible symmetry \cite{Kaidi:2021xfk}.
Consider as an example, ${\cal N}=1$ super Yang-Mills theory with $SU(3)$ gauge group.  The $\mathbb{Z}_6$ R-symmetry has a mixed anomaly with the $\mathbb{Z}_3^{(1)}$ one-form symmetry \cite{Gaiotto:2014kfa}. The R-symmetry is spontaneously broken to $\mathbb{Z}_2$, resulting in three vacua related by the broken generators, which changes $\theta$ by $2\pi k$ with $k=0,1,2$. 

In the presence of two-form gauge field $B$ for the one-form symmetry, changing $\theta$ by a multiple of $2\pi$ no longer leaves the theory invariant, but generates a local counterterm for $B$. This is a manifestation of the mixed anomaly between the R-symmetry and the one-form symmetry.
The local counterterm is
\begin{equation}
\frac{2\pi k}{3}\int BB,\quad k=0,1,2~.
\end{equation}

\paragraph{$PSU(3)$ super Yang-Mills theory}

Let us gauge the $\mathbb{Z}_3^{(1)}$ one-form symmetry on the wall and in the bulk following \cite{Hsin:2018vcg}, to obtain the domain walls in bulk $PSU(3)$ super Yang-Mills theory.  Due to the mixed anomaly with the $R$ symmetry, the $R$ symmetry is extended to be a non-invertible symmetry. (For a similar discussion in $SU(2)$ and $SO(3)$ super Yang-Mills theories, see \cite{Kaidi:2021xfk}.)

After gauging the one-form symmetry, the domain wall between $k_L,k_R$ depends on the bulk by
\begin{equation}
2\pi\frac{k_L}{3}\int_{x<0} BB+2\pi \frac{k_R}{3}\int_{x>0}BB~,
\end{equation}
where we place the domain wall at $x=0$. Consider the basic domain wall, we can decorate it with $U(1)_{-3}$ to make the domain wall free of gauge anomalies.
Let us denote the resulting decorated domain wall by ${\cal D}_3$.
Similarly, $\overline{{\cal D}}_3$ is decorated with $U(1)_{3}$. This makes the domain wall no longer invertible.
Using the duality \cite{Delmastro:2019vnj}\footnote{
Since the theories involved in the duality are Abelian TQFTs, the duality can be proven by matching the spin and the fusion of the lines.
The duality map is
\begin{equation}
(q_1,q_2,l)\rightarrow (q_1+q_2,q_1+2q_2,l+q_1)\sim (q_1+q_2,q_1-q_2,l+q_1+q_2)~,
\end{equation}
where $q_1,q_2\in\mathbb{Z}_6$ labels the charge for the $U(1)\times U(1)$ gauge groups, and $l\in\mathbb{Z}_2$ labels the transparent fermion line $\psi^l=1,\psi$ for $l=0,1$. We have the identification $(q_1,q_2,l)\sim (q_1+3n,q_2+3m,l+n+m)$, $l\sim l+2$.
}
\begin{equation}
U(1)_{-3}\times U(1)_{-3}\leftrightarrow U(1)_{3}\times U(1)_{3}+8\text{CS}_\text{grav}~,
\end{equation}
we find that (omitting gravitational Chern-Simons terms)
\begin{equation}
{\cal D}_3\times{\cal D}_3=U(1)_{3}\overline{{\cal D}}_3
\end{equation}
or by the level/rank duality \cite{Hsin:2016blu}
\begin{equation}
{\cal D}_3\times{\cal D}_3=SU(3)_{-1}\overline{{\cal D}}_3~.
\end{equation}
This agrees with the fusion algebra (\ref{Eq:fusion_triality_summary_odd_N}) for $N=3$.

\paragraph{$N$-ality domain wall defects for $N>3$}
The discussion can be generalized to $N$-ality defect in ${\cal N}=1$ super Yang-Mills theory with $PSU(N)$ gauge group. 
As before, we start with $SU(N)$ gauge theory and then gauge the $\mathbb{Z}_N^{(1)}$ one-form symmetry.
The theory has $\mathbb{Z}_{2N}$ axial $R$ symmetry, which is broken to $\mathbb{Z}_2$. The $R$ symmetry has mixed anomaly with the $\mathbb{Z}_N$ one-form symmetry: in the presence of background $B$ for the one-form symmetry, the two sides of the $k$-th wall differ by
\begin{equation}
2\pi\frac{k(N-1)}{2N}\int q(B)~, \ \  N~\text{even}~, \hspace{.5in} 2\pi\frac{k(N-1)}{2N}\int B \cup B~, \ \  N ~\text{odd}~.
\end{equation}
When we gauge the one-form symmetry to obtain $PSU(N)$ gauge theory, the basic domain wall $k=1$ needs to be decorated with the minimal Abelian TQFT ${\cal A}^{N,(N-1)}$, which makes the wall non-invertible in the $PSU(N)$ super Yang-Mills theory.

For instance, let us consider odd $N$, and fuse two basic domain walls.  We will use the following duality with $\gcd(N,p_1)=\gcd(N,p_2)=\gcd(N,p_1+p_2)=1$ for $p_1=p_2=(N-1)$
\begin{equation}
{\cal A}^{N,p_1}\times {\cal A}^{N,p_2}\leftrightarrow {\cal A}^{N,p_1^{-1}+p_2^{-1}}\times {\cal A}^{N,p_1+p_2}~,
\end{equation}
where $p_1^{-1},p_2^{-1}$ are their inverse in $\mathbb{Z}_{2N}$ for even $N$, and in $\mathbb{Z}_N$ for odd $N$.
The duality can be proven by noticing that the left hand side has a generator of ${\cal A}^{N,p_1+p_2}$ and using the factorization property of \cite{Hsin:2018vcg}.
\begin{equation}
{\cal A}^{N,p_1}\times {\cal A}^{N,p_2}\leftrightarrow {\cal A}^{N,p_1+p_2}\times \text{Q},\quad \text{Q}={{\cal A}^{N,p_1}\times {\cal A}^{N,p_2}\times {\cal A}^{N,-(p_1+p_2)}\over \mathbb{Z}_N}={\cal A}^{N,p_1^{-1}+p_2^{-1}}~,
\end{equation}
where in the quotient Q we can use the $\mathbb{Z}_N$ identification generated by the line $(1,1,1)$ to parametrize the lines of the quotient theory as $(p_1^{-1}q,p_2^{-1}(-q),0)\in {\cal A}^{N,p_1}\times {\cal A}^{N,p_2}\times {\cal A}^{N,-(p_1+p_2)}$ for $q\in\mathbb{Z}_N$, which has trivial braiding with the line $(1,1,1)$. The line has spin $p'q^2/(2N)$ with $p'=p_1^{-1}+p_2^{-1}$.
Note $\gcd(p_1^{-1}+p_2^{-1},N)=\gcd(p_1^{-1}p_2^{-1}(p_1+p_2),N)=1$ for $\gcd(p_1,N)=\gcd(p_2,N)=\gcd(p_1+p_2,N)=1$.

Thus we find that the $N$-ality domain wall ${\cal D}_N$ in the $PSU(N)$ theory for odd $N$ obeys
\begin{equation}
{\cal D}_N\times {\cal D}_N={\cal A}^{N,-2}({\cal D}_N^2)~,
\end{equation}
where ${\cal D}_N^2$ is the domain wall that comes from the $R$ symmetry element that is twice that of ${\cal D}_N$.

\subsection{${\cal N}=4$ Super Yang-Mills Theories}

Let us discuss non-invertible duality and triality defects in ${\cal N}=4$ super Yang-Mills theory with various gauge groups. We will determine the value of the gauge couplings that have non-invertible defects using the $SL(2,\mathbb{Z})$ duality action discussed in \cite{Aharony:2013hda} for super Yang-Mills theories with various discrete theta angles. 
A  complete analysis of non-invertible symmetries in ${\cal N}=4$ Yang-Mills gauge theories will appear in \cite{Kaidi:2022uux}.
As in our discussion of Maxwell theory in section \ref{sec:Maxwell}, we use $\mathbb{S}$ and $\mathbb{T}$ to denote elements of the electromagnetic duality group $SL(2,\mathbb{Z}).$ Since the theories are fermionic, in this section we will assume the spacetime manifold has a spin structure.

\subsubsection{Gauge algebra ${\frak g}=\mathfrak{su}(N)$}

\paragraph{Gauge group $SU(N)$}

Let us start with $SU(N)$ super Yang-Mills theory. If we gauge the $\mathbb{Z}_N^{(1)}$ one-form symmetry with local counterterm $p$, the theory becomes $PSU(N)$ gauge theory with discrete theta angle $p$. However, according to \cite{Aharony:2013hda}, the same $PSU(N)$ theory can be obtained from $SU(N)$ by acting on the latter with the electromagnetic duality operation $\mathbb{T}^{-p}\mathbb{S}$.  Thus, the theory is invariant under gauging the one-form symmetry with local counterterm $p$ provided the gauge coupling satisfies
\begin{equation}\label{eqn:suninvar}
\tau=-\frac{1}{\tau}-p\Rightarrow \tau=-\frac{p}{2}+ \frac{\sqrt{4-p^2}}{2}i~.
\end{equation}
To obtain a physically well-defined theory we also require $\text{Im}(\tau)>0,$ which restricts the allowed values to $p=0,\pm 1$.  When $p=0$, the theory at $\tau=i$ has a duality defect. Similarly, when $p=\pm 1$, $\tau=\mp\frac{1}{2}+\frac{\sqrt{3}}{2}i$, the theory has a triality defect (we note that the gauge couplings for the two signs are related by $\mathbb{T}$-duality).

\subsubsection{Gauge algebra ${\frak g}=\mathfrak{so}(4n+2)$}

\paragraph{Gauge group $Spin(4n+2)$}

The theory has $\mathbb{Z}_4^{(1)}$ one-form symmetry. Gauging the $\mathbb{Z}_2^{(1)}$ subgroup one form symmetry with local counterterm $p'$ gives $SO(4n+2)$ gauge theory with discrete theta angle $p'$. Gauging the entire $\mathbb{Z}_4^{(1)}$ one-form symmetry with local counterterm $p$ gives $PSO(4n+2)=Spin(4n+2)/\mathbb{Z}_4$ gauge theory with discrete theta angle $p$.

The theory is $\mathbb{S}$ dual to $PSO(4n+2)$ gauge theory with zero discrete theta angle.
Thus at gauge coupling $\tau=i$ the theory is invariant under gauging the $\mathbb{Z}_4^{(1)}$ one-form symmetry, and it has a duality defect for $\mathbb{Z}_4^{(1)}$ one-form symmetry.

The theory is $\mathbb{T}^{\pm1}\mathbb{S}$ dual to $PSO(4n+2)$ theory with discrete theta angle $p=\pm1$ for even $n$ and $p=\mp1$ for odd $n$.
Thus the theory is invariant under gauging the $\mathbb{Z}_4^{(1)}$ one-form symmetry with additional local counterterm $p=\pm 1$, respectively for even and odd $n$, at gauge coupling
\begin{equation}
\tau = -\frac{1}{\tau}\pm 1\Rightarrow \tau=\pm \frac{1}{2} +\frac{\sqrt{3}}{2}i~.
\end{equation}
Thus the theory at this coupling has a triality defect for $\mathbb{Z}_4^{(1)}$ one-form symmetry.

As a consistency check, for $n=1$ it is the $Spin(6)=SU(4)$ gauge theory, and the theory with the above gauge coupling has a duality and a triality defect as in the previous case.

\subsubsection{Gauge algebra ${\frak g}=\mathfrak{so}(4n)$}

\paragraph{Gauge group $Spin(4n)$}

The theory has $\mathbb{Z}_2^{(1)}\times\mathbb{Z}_2^{(1)}$ one-form symmetry. Gauging different $\mathbb{Z}_2^{(1)}$ subgroup one-form symmetry with local counterterm $p$ produces $SO(4n)$, $Sc(4n)$, $Ss(4n)$ gauge theories with discrete theta angle $p$.
Gauging the entire one-form symmetry with local counterterm $n_1,n_2,n_{12}$ gives $PSO(4n)=Spin(4n)/(\mathbb{Z}_2\times\mathbb{Z}_2)$ gauge theory with the corresponding discrete theta angles.\footnote{
The discrete theta angles for $Spin(4n)/(\mathbb{Z}_2\times\mathbb{Z}_2)$ bundle can be specified by $n_{11},n_{22},n_{12}$ that take value in $\mathbb{Z}_2$ on spin manifolds, with the action \cite{Aharony:2013hda}
\begin{equation} 
\frac{i\pi n_{11}}{2}\int q(w_2^{(1)})+\frac{i\pi n_{22}}{2}\int q(w_2^{(2)})+i n_{12}\pi\int w_2^{(1)}\cup w_2^{(2)}~,
\end{equation}
where $w_2^{(1)},w_2^{(2)}$ is the obstruction to lifting the bundle to an $Sc(4n)$, $Ss(4n)$ bundle, respectively.
Shifting the theta angle by $2\pi$ is equivalent to shifting $(n_{11},n_{12},n_{22})\rightarrow (n_{11}+n,n_{12}+n+1,n_{22}+n)$.
For even $n$ this is the theory $PSO(4n)^{n_{12},n_{11}}_{n_{22},n_{12}}$ with $\mathbb{Z}_2$ values $0,1$ denoted by $+,-$, and for odd $n$ this is $PSO(4n)^{n_{11},n_{12}}_{n_{12}
,n_{22}}$. 
}

The theory is $\mathbb{S}$ dual to $PSO(4n)$ theory with zero discrete theta angles, and thus the theory at the gauge coupling $\tau=i$ is invariant under gauging $\mathbb{Z}_2^{(1)}\times\mathbb{Z}_2^{(1)}$ one-form symmetry and has a duality defect.

The theory is $\mathbb{ST}^{\pm 1}$ dual to $PSO(4n)$ gauge theory with discrete theta angle $n_{12}=1,n_{11}=n_{22}=0$ for even $n$ and $n_{12}=0,n_{11}=n_{22}=1$ for odd $n$. Thus the theory at gauge coupling $\tau=(\mp1+\sqrt{3}i)/2$ is invariant under gauging $\mathbb{Z}_2^{(1)}\times\mathbb{Z}_2^{(1)}$ one-form symmetry with the above local counterterms. In particular, for even $n$ it has a triality defect.

\subsection{$SO(8)$ Gauge Theory} \label{secso8}

In this section we consider $SO(8)$ gauge theory without any topological action for the gauge fields. 
As observed in \cite{Choi:2021kmx}, this theory has both duality and triality defects. Here we  discuss these defects and their fusion algebra in more details.

\subsubsection{Partition functions for $Spin(8),SO(8),Sc(8),Ss(8)$}

Let us start with $Spin(8)$ gauge theory, which has $\mathbb{Z}_2^{(1)}\times\mathbb{Z}_2^{(1)}$ center one-form symmetry. The Wilson line in the spinor representation transforms under the first $\mathbb{Z}_2$, the cospinor line transforms under the second $\mathbb{Z}_2$, while the vector line transforms under both $\mathbb{Z}_2$.
Let us denote the partition function of the $Spin(8)$ gauge theory coupled to the background gauge field of the $\mathbb{Z}_2^{(1)}\times\mathbb{Z}_2^{(1)}$ center one-form symmetry by $Z[B_s,B_c]$.
The gauge group has charge conjugation symmetry that exchanges the spinor and cospinor, and thus the symmetry charge for the two one-form symmetries are identical in any correlation functions
\begin{equation}\label{eqn:Zchargeconj}
Z[B_s,B_c]=Z[B_c,B_s]~.
\end{equation}
In addition, the gauge group has triality that permutes the symmetry charges, and thus
\begin{equation}\label{eqn:Ztriality}
Z[B_s,B_c]=Z[B_c,B_s+B_c]=Z[B_s+B_c,B_s]~.
\end{equation}

The $SO(8)$ gauge theory is obtained from $Spin(8)$ gauge theory by gauging the diagonal $\mathbb{Z}_2$ subgroup one-form center symmetry. Similarly, gauging the $\mathbb{Z}_2^{(1)}$ subgroup one-form symmetry given by the first $\mathbb{Z}_2^{(1)}$ in the $\mathbb{Z}_2^{(1)}\times\mathbb{Z}_2^{(1)}$ center one-form symmetry gives the $Sc(8)$ gauge theory, while gauging the second $\mathbb{Z}_2^{(1)}$ gives the $Ss(8)$ gauge theory.
The new theory has $\mathbb{Z}_2^{(1)}$ one-form symmetry that transforms the Wilson line, which we will call the electric symmetry, and a new $\mathbb{Z}_2^{(d-2)}$ quantum $(d-2)$-form symmetry in $d+1$ spacetime dimension, which we will call the magnetic symmetry.
Their partition functions coupled to the backgrounds for these symmetries are
\begin{align}\label{eqn:so8partitionfn}
&Z_{Ss(8)}[B_e,B_m]=\sum_b Z[b,B_e](-1)^{\int b\cup B_m}
=\sum_b Z[b+B_e,B_e](-1)^{\int b\cup B_m+B_e\cup Bm}
\cr 
&Z_{Sc(8)}[B_e,B_m]=\sum_b Z[B_e,b](-1)^{\int b\cup B_m}
=\sum_b Z[B_e,b+B_e](-1)^{\int b\cup B_m +B_e \cup B_m}
\cr 
&Z_{SO(8)}[B_e,B_m]=\sum_b Z[b+B_e,b](-1)^{\int b\cup B_m}
=\sum_b Z[b,b+B_e](-1)^{\int b\cup B_m +B_e \cup B_m}~,
\end{align}
where the second equality in each line is obtained by the redefinition $b\rightarrow b+B_e$.
Applying \eqref{eqn:Zchargeconj} and \eqref{eqn:Ztriality} to the first equality in each line gives
\begin{equation}\label{eqn:so8triality}
Z_{Ss(8)}[B_e,B_m]=Z_{Sc(8)}[B_e,B_m]=Z_{SO(8)}[B_e,B_m]~.
\end{equation}

\subsubsection{Duality Defect for $\mathbb{Z}_2\times\mathbb{Z}_2$ Symmetry}

The $SO(8),Ss(8),Sc(8)$ gauge theories are invariant under the $S_{12}$ operation that simultaneously gauges the $\mathbb{Z}_2^{(1)}$ electric and $\mathbb{Z}_2^{(d-2)}$ magnetic symmetries:
\begin{align}\label{eqn:KWdualitySO8}
&\sum_{b_e,b_m}Z_{SO(8)}[b_e,b_m](-1)^{\int b_e \cup B_m+b_m\cup B_e}=
\sum_{b,b_e,b_m}Z[b+b_e,b](-1)^{\int b\cup b_m+b_e\cup B_m+b_m\cup B_e}\cr 
&=\sum_{b_e} Z[B_e+b_e,B_e](-1)^{\int b_e\cup B_m}=\sum_{b_e}Z[b_e,b_e+B_e](-1)^{\int b_e\cup B_m}=Z_{SO(8)}[B_e,B_m]~,
\end{align}
where the last equality uses the equation (\ref{eqn:Ztriality}) $Z[B_e+b_e,B_e]=Z[b_e,b_e+B_e]$ with $B_c=B_e+b,B_s=b_e$.
Thus the theory is invariant under the $S_{12}$ transformation and has a corresponding duality defect, denoted by ${\cal D}_2$.\footnote{We use the same symbols ${\cal D}_2$ for these duality defects (and similarly for the triality defects ${\cal D}_3$ below) which arise from gauging a $\mathbb{Z}_N^{(1)}\times \mathbb{Z}_N^{(1)}$ symmetry, even though we anticipate that these defects may obey different fusion algebras from those discussed in sections \ref{Sec:fusion_codim_one}-\ref{Sec:fusion_involving_eta}. We hope this will not cause much confusion.  }

\subsubsection{Invertible Symmetry: $T_{12}$ Transformation}

From (\ref{eqn:so8triality}) and the second equality in each line of (\ref{eqn:so8partitionfn}), we also find the theories are invariant under stacking with the SPT phase with effective action $(-1)^{\int B_e\cup B_m}$:\footnote{
For instance, in 3+1d the $SO(8)$ gauge theory flows to the pure $\mathbb{Z}_2$ gauge theory at low energy, which can be represented by a $\mathbb{Z}_2$ two-form gauge theory with gauge field $b$. The coupling to $B_m$ is
\begin{equation}
i\pi\int b\cup B_m~,
\end{equation}
and the theory is invariant under stacking $i \pi\int B_e\cup B_m$ using the field redefinition $b\rightarrow b+B_e$. If we turn on theta angle (which is present in the $Spin(8)$ gauge theory), then for $\theta=2\pi k$ with odd $k$ the action has additional $i\pi\int ( b\cup b+b\cup B_e+B_e\cup B_e)$, which is invariant mod $2\pi$ under the redefinition $b\rightarrow b+B_e$, and thus the redefinition still produces $i\pi\int B_e\cup B_m$ from the total action.
}
\begin{align}
&Z_{SO(8)}[B_e,B_m](-1)^{\int B_e\cup B_m}=Z_{SO(8)}[B_e,B_m],\cr
&Z_{Sc(8)}[B_e,B_m](-1)^{\int B_e\cup B_m}=Z_{Sc(8)}[B_e,B_m],\cr
&Z_{Ss(8)}[B_e,B_m](-1)^{\int B_e\cup B_m}=Z_{Ss(8)}[B_e,B_m]~.
\end{align}
We will call the stacking SPT phase the $T_{12}$ transformation. This is an invertible symmetry, and
we will denote the domain wall that generates the symmetry by ${\cal D}_{T_{12}}$, which satisfies ${\cal D}_{T_{12}}^2=1$, since the SPT phase has order two.

For $SO(8)$ theory, this invertible symmetry is the charge conjugation zero-form symmetry. This is a consequence of the mixed anomaly between charge conjugation zero-form symmetry, electric and magnetic symmetries, as described by the bulk term $\pi \int B_e\cup  B_m\cup B_1$ with background $B_1$ for the charge conjugation symmetry \cite{Hsin:2019fhf,Hsin:2020nts}. The anomaly implies that a charge conjugation transformation $B_1\rightarrow B_1+\delta \lambda$ with $\lambda=0,1$ produces $\pi\int B_e\cup B_m \lambda$, and thus a global charge conjugation transformation $\lambda=1$ stacks the theory with additional SPT phase $\pi\int B_e\cup B_m$ \cite{Hsin:2019fhf}.

\subsubsection{Triality Defects for $\mathbb{Z}_2\times\mathbb{Z}_2$ Symmetry}

Since the $SO(8),Ss(8),Sc(8)$ gauge theories are invariant under $S_{12}$ and $T_{12}$ transformations separately, they are also invariant under the composite operation $ST_{12}\equiv S_{12}T_{12}$. In 3+1d, this is the $\mathbb{Z}_{2}^{(1)}\times \mathbb{Z}_{2}^{(1)}$ version of the general $\mathbb{Z}_{N}^{(1)}\times \mathbb{Z}_{N}^{(1)}$ discussion in section \ref{sec:znzn}.  Thus, the theories also have a triality defect 
\begin{equation}
{\cal D}_3\equiv {\cal D}_2\circ {\cal D}_{T_{12}}~,
\end{equation}
In other words, the partition function obeys
\begin{equation}
\sum_{b_e,b_m} Z_{SO(8)}[b_e,b_m](-1)^{\int b_e\cup b_m+ b_e \cup B_m+b_m\cup B_e}=Z_{SO(8)}[B_e,B_m]~,
\end{equation}
with the same equality holding for $Sc(8)$ and $Ss(8)$.  This defect has order three as follows from the general discussion in section \ref{sec:znzn}.

As a dynamical application, we note that the general constraints on trivially gapped phases realizing duality and triality defects derived in section \ref{sec:dynamical} imply that the $SO(8)$ gauge theory cannot flow to a trivially gapped phase. This is consistent with the mixed anomaly between the $\mathbb{Z}_2^{(1)}\times\mathbb{Z}_2^{(1)}$ one-form symmetries and the charge conjugation zero-form symmetry \cite{Cordova:2017vab,Cordova:2017kue,Hsin:2020nts}, which also implies the theory cannot flow to a gapped phase with a unique vacuum.

\section*{Acknowledgements}

We are grateful to M.\ Cheng, D.\ Freed, J.\ Kaidi, R.\ Kobayashi, K.\ Ohmori, S.\ Seifnashri, Y.\ Zheng for helpful conversations.  
SHS would particularly like to thank S.\ Seifnashri for numerous illuminating discussions   on a related project \cite{Roumpedakis:2022aik}.  
We thank J.\ Kaidi, G.\ Zafrir, and Y. Zheng for comments on the first version of the paper. 
 CC is supported by the US Department of Energy DE-SC0021432 and the Simons Collaboration on Global Categorical Symmetries.  PSH is supported by the Simons Collaboration on Global Categorical Symmetries.  HTL is supported in part by a Croucher fellowship from the Croucher Foundation, the Packard Foundation and the Center for Theoretical Physics at MIT. 
The authors of this paper were ordered alphabetically.

\appendix

\section{$ (-1)^{Q}\times U(1)_N=U(1)_N$} \label{App:U1_N_times_SPT}

 Let $M$ be an oriented three-manifold. 
In this appendix, we prove that, for even $N$, the $U(1)_{N}$ Chern-Simons theory is invariant under stacking the 2+1d $\mathbb{Z}_N^{(0)}$-SPT given by the  Dijkgraaf-Witten term
\ie
(-1)^{Q(M,\Sigma)} = \exp\left(  {i \pi\over N} \int_M A\cup \delta A\right)\,,
\fe
where $A=$PD$(\Sigma)$ is the Poincar\'e dual of the surface $\Sigma$ in $M$.  
This $\mathbb{Z}_N^{(0)}$-SPT   corresponds to the order 2 element of $H^3(B\mathbb{Z}_N;U(1))\cong \mathbb{Z}_N$. 
More precisely, we mean the following
\ie
(-1)^{Q(M,\Sigma)}  Z_{U(1)_N} [M ]=Z_{U(1)_N} [M ]\,,
\fe
where $Z_{U(1)_N}[M]$ is the partition function of $U(1)_N$ on $M$. 
 This  fact was crucial for the triality fusion rule in \eqref{Eq:fusion_triality_summary} to be associative, where both the SPT $(-1)^Q$ and the TQFT $U(1)_N$ appear as the fusion ``coefficients" of the defects.  

We can schematically write the above equality as
\ie
`` \, (-1)^Q \times U(1)_N  = U(1)_N\,."
\fe
Here, $(-1)^Q$, which is an SPT of PD$(\Sigma)$, and $U(1)_N$, which is a TQFT, are both regarded as special cases of an SET of a $\mathbb{Z}_N^{(0)}$ bundle on $M$, and the product above means the tensor product between two SETs. 
Indeed, most generally the fusion ``coefficients" between two  topological defects are SETs (see, for example, \cite{Roumpedakis:2022aik}).

Using the $U(1)$ gauge field notations, the combined system of the Chern-Simons theory and this SPT phase can be expressed by the following Lagrangian:
\begin{equation}
    \frac{iN}{4\pi} ada + \frac{iN}{4\pi} AdA + \frac{iN}{2\pi} A d\tilde{a}
\end{equation}
where $a$ and $\tilde{a}$ are dynamical $U(1)$ gauge fields and $A$ is a classical background $U(1)$ gauge field.
The first term corresponds to the $U(1)_{N}$ Chern-Simons theory.
In the last term, $\tilde{a}$ is a Lagrange multiplier enforcing $A$ to  be a $\mathbb{Z}_N$ gauge field.
The second term is the Dijkgraaf-Witten term $(-1)^{Q(M,\Sigma)}$ for the $\mathbb{Z}_N$ gauge field $A$.

We can rewrite the Lagrangian as
\begin{equation}
    \frac{iN}{4\pi} (a+A)d(a+A)  + \frac{iN}{2\pi} A d(\tilde{a}-a)\,.
\end{equation}
By performing field redefinitions, $a' \equiv a+A$ and $\tilde{a}' \equiv \tilde{a} - a$, we see that this Lagrangian describes $U(1)_{N} \times (\text{trivial SPT}) = U(1)_{N}$.
Thus, the $U(1)_{N}$ Chern-Simons theory is invariant under stacking the $(-1)^{Q(M,\Sigma)}$ SPT phase.
This implies that the partition function of $U(1)_{N}$ on $M$ must vanish if  there exists a two-cycle $\Sigma \subset M$ such that $(-1)^{Q(M,\Sigma)} \neq 1$.

For example, take $N=2$, $M = \mathbb{R}P^3$, and $\Sigma = \mathbb{R}P^2 \subset \mathbb{R}P^3$.
Then, we have $(-1)^{Q(M,\Sigma)} = -1$, and indeed one can verify that the partition function of the $U(1)_{2}$ Chern-Simons theory on $\mathbb{R}P^3$ vanishes.

\section{More on the Triality Fusion Rules}\label{app:triality}
 
 In this appendix we give the detailed derivation of the triality fusion rules presented in Sections \ref{Sec:fusion_triality_codim_1} and \ref{Sec:fusion_triality_codim_1_odd_N}. 
Below all the codimension-one defects are supported on a three-dimensional manifold $M$ at $x=0$.

\subsection{Even $N$}\label{app:evenN}

Here we derive the fusion rule in Section \ref{Sec:fusion_triality_codim_1} for the triality defect ${\cal D}_3$ and the condensation defects ${\cal C}_{N\ell\over2}$ in any 3+1d QFT $\cal Q$ invariant under the $ST$ gauging in the sense of \eqref{Eq:invariance_under_ST} for the case of even $N$.  

\begin{description}[style=unboxed,leftmargin=0cm,font=\normalfont\textbullet\space]
\item[$\overline{\mathcal{D}}_3 \times \mathcal{D}_3$ :]

We have
\begin{align}
\begin{split}
    \overline{\mathcal{D}}_3 \times \mathcal{D}_3 \colon
    & \mathcal{L}[b_1] + \frac{2\pi i}{N}\left(
        -b_1 \cup b_2 - \frac{1}{2} q(b_2)
        + b_2 \cup B + \frac{1}{2} q(b_2)
    \right)
    \\ 
    = & \mathcal{L}[b_1] + \frac{2\pi i}{N} \left(
        -b_1 \cup b_2 + b_2 \cup B
    \right)
        \,.
\end{split}
\end{align}
By flipping the sign of $b_2$, we see that $\overline{\mathcal{D}}_3 \times \mathcal{D}_3 = \mathcal{C}_0$.

\item[$\mathcal{D}_3 \times \overline{\mathcal{D}}_3$ :]

We have
\begin{equation}
    \mathcal{D}_3 \times \overline{\mathcal{D}}_3 \colon
    \mathcal{L}[b_1] +
        \frac{2\pi i}{N} \left(
            b_1 \cup b_2 + \frac{1}{2} q(b_1) - b_2 \cup B - \frac{1}{2} q(B)
        \right)
    \,,
\end{equation}
which is our definition of $\mathcal{C}_{\frac{N}{2}}$.
Thus, $\mathcal{D}_3 \times \overline{\mathcal{D}}_3 = \mathcal{C}_{\frac{N}{2}}$.

\item[$\mathcal{D}_3 \times \mathcal{D}_3$ :]

We have
\begin{equation}
    \mathcal{D}_3 \times \mathcal{D}_3 \colon
    \mathcal{L}[b_1] + \frac{2\pi i}{N}\left(
        b_1 \cup b_2 + \frac{1}{2} q(b_1)
        +b_2 \cup B + \frac{1}{2} q(b_2)
    \right)
     \,.
\end{equation}
If we redefine the dynamical gauge fields as $b \equiv b_1$ and $\tilde{b} \equiv b_1 + b_2$, the expression becomes
\begin{align}
    \mathcal{D}_3 \times \mathcal{D}_3 \colon
    \left(
        \frac{2\pi i}{2N} q(\tilde{b}+B)
    \right)
    +
    \left(
    \mathcal{L}[b] +
        \frac{2\pi i}{N} \left(
            -b \cup B - \frac{1}{2} q(B)
        \right)
    \right) \,.
\end{align}
In the first term we see that the $\tilde{b}$ gauge field is decoupled from other dynamical fields including the matter fields.
The path integral over $\tilde{b}$ defines a 3+1d invertible field theory in the region $x>0$, with the Dirichlet boundary condition $\tilde b|=0$ imposed at $x=0$.  
This gives  the $U(1)_{N}$ Chern-Simons theory living on $M$ at $x=0$ as shown in \cite{Hsin:2018vcg,Gaiotto:2014kfa}.
Its coupling to $B$ only affects transverse junctions.
The remaining terms correspond to the definition of $\overline{\mathcal{D}}_3$.
Thus, we conclude $\mathcal{D}_3 \times \mathcal{D}_3 = U(1)_{N} \,\overline{\mathcal{D}}_3$.

\item[$\mathcal{D}_3 \times \mathcal{C}_{\frac{N\ell}{2}}$ and $\mathcal{C}_{\frac{N\ell}{2}} \times \mathcal{D}_3$ :]

We have
\begin{equation} \label{eq:triality_cond_fusion_1}
    \mathcal{D}_3 \times \mathcal{C}_{\frac{N\ell}{2}} \colon
    \mathcal{L}[b_1] +
        \frac{2\pi i}{N} \left(
            b_1 \cup b_2 + \frac{1}{2} q(b_1) + b_2 \cup b_3 + \frac{1}{2} \ell q(b_2) - b_3 \cup B - \frac{1}{2} \ell q(B)
        \right)
    \,,
\end{equation}
and
\begin{equation} \label{eq:triality_cond_fusion_2}
    \mathcal{C}_{\frac{N\ell}{2}} \times \mathcal{D}_3 \colon
    \mathcal{L}[b_1] +
        \frac{2\pi i}{N} \left(
            b_1 \cup b_2 + \frac{1}{2} \ell q(b_1) - b_2 \cup b_3 - \frac{1}{2}\ell q(b_3) + b_3 \cup B + \frac{1}{2}q(b_3)
        \right)
    \,.
\end{equation}
By setting $\ell=0$ in Eq. (\ref{eq:triality_cond_fusion_1}) and redefining $b_3 \rightarrow -b_3$, we obtain $\mathcal{D}_3 \times \mathcal{C}_0 = \mathcal{C}_{\frac{N}{2}} \times \mathcal{D}_3$.
Furthermore, if we let $b \equiv b_1$ and $b'_3 \equiv b_3 + b_1$, we get
\begin{equation}
    \mathcal{D}_3 \times \mathcal{C}_0 \colon
    \left(
            \frac{2\pi i}{N}(b_2 - B) \cup b'_3
    \right) +
    \left(
    \mathcal{L}[b] +
        \frac{2\pi i}{N} 
            \left(  b \cup B + \frac{1}{2}q(b) \right)
    \right) \,. 
\end{equation}
The first term gives  the 2+1d $(\mathcal{Z}_N)_0$ theory as we explained in Section \ref{Sec:fusion_duality_codim_1}, and the other terms give  $\mathcal{D}_3$.
Thus, $\mathcal{D}_3 \times \mathcal{C}_0 = \mathcal{C}_{\frac{N}{2}} \times \mathcal{D}_3 = (\mathcal{Z}_N)_0 \,\mathcal{D}_3$.

Next, we have
\begin{align} \label{eq:triality_cond_fusion_4}
\begin{split}
    \mathcal{C}_0 \times \mathcal{D}_3 \colon
    &\mathcal{L}[b_1] +
        \frac{2\pi i}{N} \left(
            b_1 \cup b_2 - b_2 \cup b_3 + \frac{1}{2} q(b_3) + b_3 \cup B
        \right)
    \\
    &=
        \frac{2\pi i}{N}
        \left( (-b'_2 + B) \cup b'_3 + \frac{1}{2} q(b'_3)
        \right)
        +
    \left(
    \mathcal{L}[b] +
        \frac{2\pi i}{N} 
            \left(  b \cup B + \frac{1}{2}q(b) \right)
    \right)
\end{split}
\end{align}
where $b \equiv b_1$, $b'_2 \equiv b_2 - b_1$, and $b'_3 \equiv b_3 - b_1$.
The terms inside the first pair of parentheses define a 2+1d $(\mathcal{Z}_N)_N$ gauge theory living on the defect \cite{Hsin:2018vcg}.
To see this, we first set $B=0$ for simplicity, as the coupling  to $B$ only affects the transverse junctions. 
Next, following a similar discussion around Figure \ref{Fig:condensation_lattice}, we  integrate out $b'_2$ to force $b'_3$ to effectively become a gauge field $a \in H^1(M;\mathbb{Z}_N)$ living on $M$. 
The 4d twist term $\frac{2\pi i}{2N} q(b'_3)$ reduces to the level $N$ 3d Dijkgraaf-Witten twist for the gauge field $a$ as shown in \cite{Kaidi:2021xfk}.  
We therefore obtain a 2+1d $(\mathcal{Z}_N)_N$ gauge theory on $M$. 
Finally, remaining terms define the triality defect $\mathcal{D}_3$.
We conclude $\mathcal{C}_0 \times \mathcal{D}_3 = (\mathcal{Z}_N)_N \, \mathcal{D}_3$.

If instead $\ell=1$ in Eq. (\ref{eq:triality_cond_fusion_1}), we redefine the variables as $b_2 \rightarrow -b_3$ and $b_3 \rightarrow b_1 + b_2$ to obtain
\begin{equation} \label{eq:triality_cond_fusion_3}
    \mathcal{D}_3 \times \mathcal{C}_{\frac{N}{2}} \colon
    \mathcal{L}[b_1] +
        \frac{2\pi i}{N} \left(
            b_1 \cup b_2 - b_2 \cup b_3 + \frac{1}{2} q(b_3) + b_2 \cup B - \frac{1}{2}q(B)
        \right)
     \,.
\end{equation}
Let $b \equiv b_1$, $b'_2 \equiv b_2 - b_1$, and $b'_3 \equiv b_3 - b_1$, then we obtain
\begin{equation}
    \mathcal{D}_3 \times \mathcal{C}_{\frac{N}{2}} \colon
        \frac{2\pi i}{N}
        \left( -(b'_2-B) \cup (b'_3 -B) + \frac{1}{2} q(b'_3 -B)
        \right)
    +
    \left(
    \mathcal{L}[b] +
        \frac{2\pi i}{N} 
            \left(  b \cup B + \frac{1}{2}q(b) \right) 
    \right) \,.
\end{equation}
We again see the decoupled 2+1d $(\mathcal{Z}_N)_N$ gauge theory on the defect and also $\mathcal{D}_3$.
The coupling of this $(\mathcal{Z}_N)_N$ to the bulk symmetry background $B$ is different from the previous case, but it only affects transverse junctions.
We conclude that $\mathcal{D}_3 \times \mathcal{C}_{\frac{N}{2}} = (\mathcal{Z}_N)_N \,\mathcal{D}_3$. 

\item[$\mathcal{C}_{\frac{N\ell_1}{2}} \times \mathcal{C}_{\frac{N\ell_2}{2}}$ :] 

We have already derived $\mathcal{C}_0 \times \mathcal{C}_0 = (\mathcal{Z}_N)_0 \,\mathcal{C}_0$ in Section \ref{Sec:fusion_duality_codim_1}.
Now, consider the fusion
\begin{equation}\label{Eq:C0CN2}
    \mathcal{C}_0 \times \mathcal{C}_{\frac{N}{2}} \colon
    \mathcal{L}[b_1]
    +\frac{2\pi i}{N}\left(
        b_1 \cup b_2 - b_2 \cup b_3 + b_3 \cup b_4
        +\frac{1}{2} q(b_3) - b_4 \cup B -
        \frac{1}{2} q(B)
    \right)
    \,.
\end{equation}
Let $b'_4 \equiv b_4 - b_2$, then this becomes
\begin{equation}
    \mathcal{C}_0 \times \mathcal{C}_{\frac{N}{2}} \colon
    \frac{2\pi i}{N}
    \left( (b'_4 +B) \cup (b_3 -B) + \frac{1}{2} q(b_3 -B)
    \right)
+
\left(
\mathcal{L}[b_1] +
    \frac{2\pi i}{N} 
        \left(  b_1 \cup b_2 - b_2 \cup B \right) 
\right) 
\end{equation}
which gives $\mathcal{C}_0 \times \mathcal{C}_{\frac{N}{2}} = (\mathcal{Z}_N)_N \,\mathcal{C}_0$.
Again, the coupling of the coefficient TQFT $(\mathcal{Z}_N)_N$ to the bulk symmetry background $B$ only affects transverse junctions.

We can perform an alternative change of variables $b'_3 \equiv b_3 - b_1$ and $b'_2 \equiv -b_2 + b_1 + b_4$ to rewrite \eqref{Eq:C0CN2} as
\begin{equation}
    \mathcal{C}_0 \times \mathcal{C}_{\frac{N}{2}} \colon
    \frac{2\pi i}{N}
    \left( b'_2 \cup b'_3 + \frac{1}{2} q(b'_3)
    \right)
+
\left(
\mathcal{L}[b_1] +
    \frac{2\pi i}{N} 
        \left(  b_1 \cup b_4 + \frac{1}{2} q(b_1)
        - b_4 \cup B - \frac{1}{2} q(B)
        \right) 
\right) 
\end{equation}
which gives $\mathcal{C}_0 \times \mathcal{C}_{\frac{N}{2}} = (\mathcal{Z}_N)_N \,\mathcal{C}_{\frac{N}{2}}$.
One can also verify that the two condensation defects commutes, and we obtain $\mathcal{C}_0 \times \mathcal{C}_{\frac{N}{2}} = \mathcal{C}_{\frac{N}{2}} \times \mathcal{C}_0
    =(\mathcal{Z}_N)_N \,\mathcal{C}_0
    = (\mathcal{Z}_N)_N \,\mathcal{C}_{\frac{N}{2}}$.

Finally, consider
\begin{equation}
    \mathcal{C}_{\frac{N}{2}} \times \mathcal{C}_{\frac{N}{2}} \colon
    \mathcal{L}[b_1] + \frac{2\pi i}{N} \left(
        b_1 \cup b_2 + \frac{1}{2} q(b_1)
        -b_2 \cup b_3  
        +b_3 \cup b_4          - b_4 \cup B - \frac{1}{2} q(B)
    \right) \,.
\end{equation}
Let $b'_4 \equiv b_4 - b_2$, then this reduces to
\begin{equation}
    \mathcal{C}_{\frac{N}{2}} \times \mathcal{C}_{\frac{N}{2}} \colon
    \frac{2\pi i}{N} b'_4 \cup (b_3 - B)
    +
    \left(
        \mathcal{L}[b_1] + \frac{2\pi i}{N} \left(
            b_1 \cup b_2 + \frac{1}{2} q(b_1)
            -b_2 \cup B -\frac{1}{2} q(B)
        \right)
    \right)
     \,,
\end{equation}
which gives $\mathcal{C}_{\frac{N}{2}} \times \mathcal{C}_{\frac{N}{2}} = (\mathcal{Z}_N)_0 \,\mathcal{C}_{\frac{N}{2}}$.

To summarize, we have derived $\mathcal{C}_{\frac{N\ell_1}{2}} \times \mathcal{C}_{\frac{N\ell_2}{2}} = (\mathcal{Z}_N)_{N(\ell_1 + \ell_2)} \,\mathcal{C}_{\frac{N\ell_1}{2}} =(\mathcal{Z}_N)_{N(\ell_1 + \ell_2)} \,\mathcal{C}_{\frac{N\ell_2}{2}}$ for general $\ell_1$ and $\ell_2$.
The same fusion rules can also be derived from the definition of condensation defects given in Eq. (\ref{Eq:condensation_defect_general}) although we will not do so explicitly.

\item[$\mathcal{D}_3 \times \mathcal{D}_3 \times \mathcal{D}_3$ :] 

We have
\ie
    \mathcal{D}_3 \times \mathcal{D}_3 
    \times \mathcal{D}_3 \colon
    &\mathcal{L}[b_1] +
        \frac{2\pi i}{N} \left(
            b_1 \cup b_2 + \frac{1}{2} q(b_1) 
            + b_2 \cup b_3 + \frac{1}{2} q(b_2) 
            + b_3 \cup B + \frac{1}{2} q(b_3) 
        \right)
    \\
    &=
        \frac{2\pi i}{2N} q(\tilde{b})
        +
    \left(
    \mathcal{L}[b_1] +
        \frac{2\pi i}{N} 
            \left(  b_1 \cup b'_3 - b'_3 \cup B \right)
    \right)\,,
\fe
where we performed field redefinitions $\tilde{b} \equiv b_1 + b_2 + b_3$ and $b'_3 \equiv - b_3$.
The first term reduces to a 2+1d $U(1)_{N}$ Chern-Simons theory on the defect \cite{Hsin:2018vcg,Gaiotto:2014kfa}, and the remaining terms define the condensation defect $\mathcal{C}_0$.
Thus, $\mathcal{D}_3 \times \mathcal{D}_3 \times \mathcal{D}_3
    = U(1)_{N} \,\mathcal{C}_0$.

Alternatively, one can also define $\dbtilde{b} \equiv b_2 + b_3$, in which case we obtain
\begin{equation}
        \mathcal{D}_3 \times \mathcal{D}_3 
        \times \mathcal{D}_3 \colon
                   \frac{2\pi i}{2N} q(\dbtilde{b} + B)
            +
        \left(
        \mathcal{L}[b_1] +
            \frac{2\pi i}{N} 
            \left(  b_1 \cup b_2 + \frac{1}{2}q(b_1)
            -b_2 \cup B - \frac{1}{2} q(B)
            \right)
    \right)
\end{equation}
from which we can deduce $\mathcal{D}_3 \times \mathcal{D}_3 
\times \mathcal{D}_3 = U(1)_{N} \,\mathcal{C}_{\frac{N}{2}}$.
Combining the above, we conclude $\mathcal{D}_3 \times \mathcal{D}_3 \times \mathcal{D}_3 = U(1)_{N} \,\mathcal{C}_0  =  U(1)_{N} \,\mathcal{C}_{\frac{N}{2}}$.

\end{description}

\subsection{Odd $N$}\label{app:oddN}

Here we derive the fusion rule in Section \ref{Sec:fusion_triality_codim_1_odd_N} for the triality defect ${\cal D}_3$ and the condensation defects ${\cal C}_{0}$ in any 3+1d QFT $\cal Q$ invariant under the $ST$ gauging in the sense of \eqref{Eq:invariance_under_ST} for the odd $N$ case.

\begin{description}[style=unboxed,leftmargin=0cm,font=\normalfont\textbullet\space]
\item[$\overline{\mathcal{D}}_3 \times \mathcal{D}_3$ 
    and $\mathcal{D}_3 \times \overline{\mathcal{D}}_3$ :]

First, we have
\begin{align}
    \begin{split}
        \overline{\mathcal{D}}_3 \times \mathcal{D}_3 \colon
        & \mathcal{L}[b_1] + \frac{2\pi i}{N}\left(
            -b_1 \cup b_2 - \frac{N+1}{2} b_2 \cup b_2
            + b_2 \cup B + \frac{N+1}{2} b_2 \cup b_2
        \right)
        \\ 
        = & \mathcal{L}[b_1] + \frac{2\pi i}{N} \left(
            -b_1 \cup b_2 + b_2 \cup B
        \right)
            \,.
    \end{split}
\end{align}
By flipping the sign of $b_2$, we obtain $\overline{\mathcal{D}}_3 \times \mathcal{D}_3 = \mathcal{C}_0$.
Next,
\begin{align}
    \begin{split}
        \mathcal{D}_3 \times \overline{\mathcal{D}}_3 \colon
        & \mathcal{L}[b_1] + \frac{2\pi i}{N}\left(
            b_1 \cup b_2 + \frac{N+1}{2} b_1 \cup b_1
            - b_2 \cup B - \frac{N+1}{2} B \cup B
        \right)
        \\ 
        = & \mathcal{L}[b_1] + \frac{2\pi i}{N} \left(
            b_1 \cup (b'_2 + \frac{N+1}{2} B)
            - (b'_2 + \frac{N+1}{2} B) \cup B
        \right)
            \,,
    \end{split}
\end{align}
where we have defined $b'_2 \equiv b_2 + \frac{N+1}{2} b_1$.
The resulting defect obeys the same parallel fusion rule with $\eta$ as $\mathcal{C}_0$ (see Section \ref{Sec:eta_fusion_triality_odd_N}). 
Thus, if we ignore the transverse junctions, we can further shift $b'_2$ by $\frac{N+1}{2} B$.
This gives  $\mathcal{D}_3 \times \overline{\mathcal{D}}_3 = \mathcal{C}_0$.

\item[$\mathcal{D}_3 \times \mathcal{D}_3$ :]

We have
\begin{align}
    \begin{split}
        \mathcal{D}_3 \times \mathcal{D}_3 \colon
        & \mathcal{L}[b_1] + \frac{2\pi i}{N}\left(
            b_1 \cup b_2 + \frac{N+1}{2} b_1 \cup b_1
            + b_2 \cup B + \frac{N+1}{2} b_2 \cup b_2
        \right)
        \\ 
        = & \frac{2\pi i}{N} \frac{N+1}{2} (\tilde{b}+B)^2
        +
        \left(
            \mathcal{L}[b] + \frac{2\pi i}{N} \left(
            -b \cup B
            - \frac{N+1}{2} B \cup B
        \right)
        \right)
            \,,
    \end{split}
\end{align}
where we have defined $b \equiv b_1$ and $\tilde{b} \equiv b_1 + b_2$.
The first term is a 3+1d invertible field theory living in $x>0$ with Dirichlet boundary condition $\tilde b|=0$ imposed at $x=0$. 
This gives  the 2+1d $SU(N)_{-1}$ Chern-Simons theory living on the defect \cite{Hsin:2018vcg}.
The remaining terms define $\overline{\mathcal{D}}_3$.
Thus, we have $\mathcal{D}_3 \times \mathcal{D}_3 = SU(N)_{-1} \,\overline{\mathcal{D}}_3$.
From this and the earlier fusion rule, we also find
$\mathcal{D}_3 \times \mathcal{D}_3 \times \mathcal{D}_3 = SU(N)_{-1} \,\mathcal{C}_0$.

\item[$\mathcal{D}_3 \times \mathcal{C}_0$ 
    and $\mathcal{C}_0 \times \mathcal{D}_3$ :]

We have
\begin{align}
    \begin{split}
        \mathcal{D}_3 \times \mathcal{C}_0 \colon
        & \mathcal{L}[b_1] + \frac{2\pi i}{N}\left(
            b_1 \cup b_2 + \frac{N+1}{2} b_1 \cup b_1
            + b_2 \cup b_3 - b_3 \cup B
        \right)
        \\ 
        = & \frac{2\pi i}{N} (b_2 - B) \cup b'_3
        +
        \left(
            \mathcal{L}[b] + \frac{2\pi i}{N} \left(
            b \cup B
            + \frac{N+1}{2} b \cup b
        \right)
        \right)
            \,,
    \end{split}
\end{align}
where $b \equiv b_1$ and $b'_3 \equiv b_3 + b_1$.
The first term gives us the 2+1d $(\mathcal{Z}_N)_0$ gauge theory living on the defect, and the remaining terms define $\mathcal{D}_3$.
Thus, $\mathcal{D}_3 \times \mathcal{C}_0 = (\mathcal{Z}_N)_0 \, \mathcal{D}_3$.

Similarly, we have
\begin{align}
    \begin{split}
        \mathcal{C}_0 \times \mathcal{D}_3 \colon
        & \mathcal{L}[b_1] + \frac{2\pi i}{N}\left(
            b_1 \cup b_2 - b_2 \cup b_3
            + b_3 \cup B + \frac{N+1}{2} b_3 \cup b_3
        \right)
        \\ 
        = & \frac{2\pi i}{N} (b'_2 + B) \cup b'_3
        +
        \left(
            \mathcal{L}[b] + \frac{2\pi i}{N} \left(
            b \cup B
            + \frac{N+1}{2} b \cup b
        \right)
        \right)
            \,,
    \end{split}
\end{align}
where $b \equiv b_1$, $b'_2 \equiv \frac{N+1}{2}b_1 -b_2 +\frac{N+1}{2}b_3$, and $b'_3 \equiv b_3 - b_1$.
We see that $\mathcal{C}_0 \times \mathcal{D}_3 = (\mathcal{Z}_N)_0 \, \mathcal{D}_3$.

\end{description}

\section{$N$-ality Defects  From Mixed Anomalies}

The Kramers-Wannier duality defect in 1+1d Ising model can be obtained from the invertible
chiral symmetry defect in Majorana fermion by gauging the fermion parity symmetry \cite{Thorngren:2018bhj,Ji:2019ugf}. The chiral
symmetry has a mixed anomaly with the fermion parity symmetry, and thus gauging the fermion
parity symmetry ``extends'' the chiral symmetry to be a non-invertible symmetry generated by
the Kramers-Wannier duality defect. Conversely, the Majorana fermion can be obtained from the
Ising model by gauging a $\mathbb{Z}_2$ zero-form symmetry with a local counterterm \cite{Karch:2019lnn,Ji:2019ugf,Hsieh:2020uwb}.

We would like to understand whether the non-invertible $N$-ality defects discussed here can be obtained from invertible zero-form symmetry by gauging a one-form symmetry
in a ``parent theory", generalizing  the algorithm in \cite{Kaidi:2021xfk}. 
 In other words, we would like to understand whether the non-invertible
symmetry defect becomes an invertible symmetry when we gauging a one-form symmetry with
suitable local counterterm.
As we will show in this appendix, this is in general not the case. Thus the non-invertible
defect discussed here in general cannot be obtained from an invertible zero-form symmetry that has
mixed anomaly with one-form symmetry by gauging the one-form symmetry. The condition that
the non-invertible symmetry can be realized as an invertible symmetry upon gauging a discrete
symmetry coincides with the condition  that such non-invertible defect can be realized by invertible
phases discussed in Section \ref{sec:dynamical}.

We define the codimension-one non-invertible defect in the same way as in Section \ref{Sec:define_defects} by gauging in half of the spacetime. 
On one side of the defect the theory has partition function $Z[B]$ with the background   two-form gauge field $B$. 
On the other side the partition function is $\sum_b Z[b]e^{iS[b,B]}$, where\footnote{For notational simplicity, we suppress the normalization factor from  gauging of the one-form symmetry in this appendix. }
\ie
S[b,B]=
\begin{cases}
{2\pi\over N}\int b\cup B+\frac{p}{2}q(b) & \text{even }N\\
\frac{2\pi}{N}\int b\cup B+ 2^{-1}p \, b\cup b & \text{odd }N
\end{cases}
\fe
where $2^{-1}=(N+1)/2$ is the inverse in $\mathbb{Z}_N$ for odd $N$.

Next, we gauge the one-form symmetry by promoting $B$ to be dynamical, and add an additional local counterterm $S'[B,B']$ with a new background two-form gauge field $B'$,
\begin{equation}
S'[B,B']=
\left\{\begin{array}{ll}
\frac{2\pi}{N}\int r B\cup B'+\frac{p'}{2}q(B)& \text{even }N\\
\frac{2\pi}{N}\int r B\cup B'+ p' B\cup B & \text{odd }N
\end{array}
\right.~.
\end{equation}
The theories on the two sides of the defect are now
\begin{align}
&\text{LHS}: \sum_B Z[B] e^{iS'[B,B']}
\cr
&\text{RHS}:\sum_{b,B}Z[b]e^{iS[b,B]+iS'[B,B']}~.
\end{align}
If there is a choice of $S'$ such that we can integrate out $B$ on the right hand side to obtain
$\sum_b Z[b]e^{iS'[b,\ell B']}$ for some integer $\ell$ that is coprime with $N$ and up to a local counterterm of the
background field $B'$, then the defect becomes an invertible defect that generates zero-form symmetry in the theory with partition
function $\sum_B Z[B]e^{iS'[B,B']}$. 
More specifically, this invertible symmetry acts on the dual $\mathbb{Z}_N$ one-form symmetry by $n\rightarrow n\ell$ for $n\in \mathbb{Z}_N$.
If we gauge the dual symmetry generated by $\exp(i \oint B)$, such invertible zero-form symmetry becomes
the non-invertible  defect.

\subsection{Duality Defect}

Consider the Kramers-Wannier duality defect with $p=0$. We will take $\gcd(p',N)=1$.

\subsubsection{Even $N$}

The partition function $\sum_{b,B}Z[b]e^{iS[b,B]+iS'[B,B']}$ is
\begin{align}
&\sum_{b,B}Z[b]\exp \left({2\pi i\over N}\int \left(b\cup B+r B\cup B'+\frac{p'}{2}q(B)\right)\right)\\ 
&=\hspace{-.1in}\sum_{\tilde b=B+p'^{-1}(b+rB'), b} \hspace{-.3in}Z[b] \exp\left(\frac{2\pi i p'}{2N}\int q(\tilde b)+\frac{2\pi i}{N}\int \left(-\frac{p'^{-1}}{2}q(b)-p'^{-1} r b\cup B'-\frac{p'^{-1}r^2}{2}q(B')\right)\right)~. \notag
\end{align}
Thus we need to solve
\begin{equation}
p'^2=-1\text{ mod }2N\quad (\text{bosonic});\quad p'^2=-1\text{ mod }N\quad (\text{fermionic})~,
\end{equation}
for bosonic and fermionic theories, respectively.

The solutions are discussed in \cite{Choi:2021kmx}, in a different context of invertible phases that possess
Kramers-Wannier duality defect.
For instance, if the theory is bosonic, then the first equation does not have a solution and
the Kramers-Wannier duality defect cannot be obtained from an invertible symmetry by gauging
one-form symmetries. 

On the other hand, if the theory is fermionic, and for instance take $N = 2$, then the equations
have solution $p' = 1$. Thus the Kramers-Wannier duality non-invertible defect can be obtained
from an invertible zero-form symmetry in the theory $\sum_B Z[B]e^{{\pi i\over 2} \int q(B)}$ by gauging the dual symmetry generated by $\exp(i\oint B)$. The dual symmetry and the invertible zero-form symmetry has a mixed anomaly: in the presence of the background gauge field $B'$ for the dual symmetry, the two sides of the defect differ by
\begin{equation}
-2\pi \frac{p'^{-1}r^2}{2N}\int q(B')~.
\end{equation}
The invertible symmetry acts on the one-form symmetry by multiplying the $\mathbb{Z}_N$ element with $\ell=-(p')^{-1}$.

\subsubsection{Odd $N$}

From a similar discussion, we need to solve
\begin{equation}
4p'^2=-1\text{ mod }N~.
\end{equation}
The solutions are discussed in \cite{Choi:2021kmx}, in a different context of invertible phases that possess
Kramers-Wannier duality defect.
When the equation is solvable, the Kramers-Wannier duality defect can be obtained by gauging
one-form symmetry in a theory with invertible symmetry. The invertible symmetry has mixed
anomaly with the dual one-form symmetry: in the presence of background gauge field $B'$ for the dual one-form symmetry, the two sides of the defect differ by
\begin{equation}
-2\pi \frac{2^{-1}(2p')^{-1}r^2}{N}\int B'\cup B'~.
\end{equation}
The invertible symmetry acts on the one-form symmetry by multiplying the $\mathbb{Z}_N$ element with $\ell=-(2p')^{-1}$.

\subsection{Triality Defect}

\subsubsection{Even $N$}

We have $p=1$ and take $\gcd(p',N)=1$. The partition function $\sum_{b,B}Z[b]e^{iS[b,B]+iS'[B,B']}$ is
\begin{align}
&\sum_{b,B}Z[b]\exp\left( {2\pi i\over N}\int \left(\frac{1}{2}q(b)+ b\cup B+r B\cup B'+\frac{p'}{2}q(B)\right)\right)\\ 
&=\sum_{\tilde b=B+p'^{-1}(b+rB'), b}\hspace{-.3in}Z[b]\exp\left(\frac{2\pi i p'}{2N}\int q(\tilde b) +\frac{2\pi i}{N}\int \left(\frac{1-p'^{-1}}{2}q(b)-p'^{-1} r b\cup B'-\frac{p'^{-1}r^2}{2}q(B')\right)\right)~.\notag
\end{align}
Thus we need to solve
\begin{equation}
p'(p'-1)+1=0\text{ mod }2N\quad (\text{bosonic});\quad p'(p'-1)+1=0\text{ mod }N\quad (\text{fermionic})~.
\end{equation}
Since $p'(p'-1)$ is even, and the above equation has definite even/odd parity, there is no solution. Thus the triality defect cannot
be obtained from an invertible symmetry by gauging a one-form symmetry.

\subsubsection{Odd $N$}

Here $p=N+1$ and we take $\gcd(p',N)=1$. The partition function $\sum_{b,B}Z[b]e^{iS[b,B]+iS'[B,B']}$ is
\begin{align}
&\sum_{b,B}Z[b]\exp\,\left( {2\pi i\over N}\int \left(\frac{N+1}{2} (b\cup b)+ b\cup B+r B\cup B'+p'(B\cup B)\right)\right)\cr 
&=\sum_{\tilde b=B+(2p')^{-1}(b+rB'), b}\hspace{-.4in}Z[b]\exp\left(\frac{2\pi ip'}{N}\int (\tilde b\cup \tilde b) \right)\\
&\times \exp\left(\frac{2\pi i}{N}\int \left(\left(\frac{N+1}{2}-2^{-1}(2p')^{-1}\right)(b\cup b)-(2p')^{-1} r b\cup B'-2^{-1}(2p')^{-1}r^2(B'\cup B')\right)\right)~. \notag
\end{align}
Thus we need to solve
\begin{equation}
(2p')^2-2p'+1=0\text{ mod }N~.
\end{equation}
There exits solution if and only if $N\in {\cal X}$ as defined in (\ref{eqn:solsetX}).

When $N\in {\cal X}$ and thus a solution exists, the coupling to $B'$ on the two sides of the invertible
wall is $r$ and $-(2p')^{-1}r$. 
The invertible symmetry has mixed anomaly with the dual symmetry: in
the presence of background $B'$ for the dual symmetry generated by $\exp(i \oint B)$, the two sides of the
defect differ by a local counterterm (take $r =1$)
\begin{equation}
-\frac{\pi(N+1)(2p')^{-1}}{N}\int (B'\cup B')~.
\end{equation}

\bibliographystyle{JHEP}
\bibliography{ref}

\end{document}